\documentclass[12pt,a4paper]{article}
\usepackage{amsfonts,amssymb,amsmath}
\usepackage{mathtools}
\usepackage[
      colorlinks=true,
      linkcolor=blue,
      urlcolor=blue,
      filecolor=blue,
      citecolor=red,
      pdfstartview=FitV,
      linktoc={page}
]{hyperref}

\renewcommand{\baselinestretch}{1.2}
\renewcommand{\thefootnote}{\fnsymbol{footnote}}
\newcommand{\be}{\begin{eqnarray}}
\newcommand{\ee}{\end{eqnarray}}
\newcommand{\nn}{\nonumber}

\begin{document}

\makeatletter \@addtoreset{equation}{section} \makeatother
\renewcommand{\theequation}{\thesection.\arabic{equation}}
\renewcommand{\thefootnote}{\alph{footnote}}

\begin{titlepage}
\begin{flushright}
\hfill {\tt arXiv:1206.6781\\
 KIAS-P12033}\\
\end{flushright}
\begin{center}

\vspace{1.5cm}

{\Large\bf 5-dim Superconformal Index \\ with Enhanced $E_n$ Global Symmetry}

\vspace{2.0cm}

{\large\it
Hee-Cheol Kim, Sung-Soo Kim, and Kimyeong Lee}

\vspace{0.2cm}

\text{\large   Korea Institute for Advanced Study,
Seoul 130-722, Korea}

\vspace{0.0cm}

E-mails: {\tt heecheol1@gmail.com, sungsoo.kim@kias.re.kr, klee@kias.re.kr}

\end{center}

\vspace{0.8cm}

\begin{abstract}
The five-dimensional ${\cal N}=1$ supersymmetric gauge theory with $Sp(N)$ gauge group and $SO(2N_f)$ flavor symmetry describes the physics on $N$ D4-branes with $N_f$ D8-branes on top of a single O8 orientifold plane in Type I$'$ theory. This theory is   known to be superconformal at the strong coupling limit with the enhanced global symmetry $E_{N_f+1}$ for $N_f\le 7$.  In this work we calculate the superconformal index on $S^1\times S^4$ for  the   $Sp(1)$ gauge theory by the localization method and confirm such enhancement of the global symmetry  at the superconformal limit for $N_f\le 5$  to a few leading orders in the chemical potential. Both perturbative and (anti)instanton contributions are present in this calculation. For $N_f=6,7$ cases   some issues related the pole structure of the instanton calculation could not be resolved   and here we could provide only some suggestive answer for the leading contributions to the index. For the  $Sp(N)$  case,  similar issues related to the pole structure appear.
\end{abstract}
\end{titlepage}

\renewcommand{\thefootnote}{\arabic{footnote}}
\setcounter{footnote}{0}
\renewcommand{\baselinestretch}{1.0}
\tableofcontents
\renewcommand{\baselinestretch}{1.2}

\section{Introduction}

Seiberg has proposed sometime ago several classes of 5-dim ${\cal N}=1$ superconformal field theories \cite{Seiberg:1996bd}. Especially an interesting class is the $Sp(N)$ gauge theories with $N_f$ fundamental hypermultiplets and one antisymmetric hypermultiplet which appear naturally on $N$ D4-branes near the $N_f$ D8 branes on top of a single O8 orientifold plane in the so-called type I$'$ string theory. The infinite coupling limit of the gauge theory in the symmetric phase is the superconformal fixed point with superconformal group $F(4)$ whose bosonic part consists of $SO(2,5)$ conformal symmetry and  $SU(2)$ R-symmetry. These theories with $SO(2N_f)$ flavor symmetry and $U(1)$ for   instanton number are expected to have  enhanced global symmetries $E_{N_f+1}$ for $N_f\le 7$ at the superconformal point \cite{Seiberg:1996bd,Morrison:1996xf,Douglas:1996xp,Ganor:1996pc,Intriligator:1997pq,Danielsson:1997kt}. While the Coulomb phase moduli of these theories are the coordinates for the positions of D4 branes away from the orientifolds, the Higgs phase of these theories are known to be the center moduli space of $N$ $E_{N_f+1}$ instantons.

In this work we set up the superconformal index calculation of these gauge theories on $S^1\times S^4$ and evaluate it by the localization method with suitable chemical potentials turned on. This index has both perturbative contribution and nonperturbative instanton and anti-instanton contribution.
In our index calculation for the $Sp(1)$ theory  to three and four instanton contributions for $N_f\le 5$, the chemical potentials for the  $SO(2N_f)$ flavor symmetry and the $U(1)$ instanton charge are merged to  the  characters for the enhanced $E_{N_f+1}$.  However   the difficult pole structure appears in calculation of the instanton contributions for $N_f=6,7$ and so in this case we suggest a few leading order expression for the index based on the general pattern.    Similar obstacle exists for $N\ge 2$.

These 5d ${\cal N}=1$ supersymmetric gauge field theories are   nonrenormalizable with Yang-Mills kinetic term and ultra-violet incomplete. However each of them may be regarded as a relevant perturbation of a 5d UV-complete superconformal field theory which corresponds the infinite  gauge coupling   limit of the gauge theory  interacting with hypermultiplets. One should not be wary of the infinite coupling   limit where the Yang-Mills action can be ignored. A most typical such example is the $CP(N)$ model which is written with one auxiliary gauge field. That can be regarded as the infinite coupling limit of the abelian Higgs model with the multiple flavor and FI term turned on. The gauge kinetic term gives   more weight to the smooth field configurations in the path integral. In the infinite gauge coupling limit, all the gauge field configurations contribute with the equal weight. The result is the constraint leading to the $CP(N)$ model. For our $Sp(1)$ gauge theory at the conformal limit also, one has to sum over all gauge fields with equal weight.

In 5d gauge theory with Yang-Mills action, the inverse of the gauge coupling constant $1/g_{YM}^2$ has the mass dimension which is also the mass scale of  instanton solitons.  While the BPS instantons form a massive tensor multiplet in the maximally supersymmetric ${\cal N}=2$ case due to the gaugino zero modes, they form a massive hypermultiplet in the ${\cal N}=1$ case. For the ${\cal N}=1$ $Sp(1)$ gauge theory with fundamental $N_f$ flavor, there would be additional zero modes due to the spinors in the fundamental representation. These zero modes produce no spin but just flavor charge.
Instantons and anti-instantons appear as chiral ${\bf 2}^{N_f-1}_{+1}$ and its complex conjugate $\bar{\bf 2}^{N_f-1}_{-1}$ representations of flavor group $SO(2N_f)$ and the $U(1)$ instanton number, respectively.

We choose the mass for the hypermultiplet to vanish. The enhancement of the global symmetry occurs at the conformal point.  At the infinite coupling limit, instantons also become massless. Seiberg has argued that    the flavor symmetry $SO(2N_f)$ and the instanton number charge $U(1)_I$ get merged into   a   global symmetry $E_{N_f+1}\supset SO(2N_f)\times U(1)_I$ \cite{Seiberg:1996bd}. 
Besides the exceptional groups $E_8, E_7, E_6$, the remaining ones are $E_5=Spin(10), E_4=SU(5),E_3=SU(3)\times SU(2), E_2=SU(2)\times U(1)$ and $E_1=SU(2)$.

The 5d ${\cal N}=1$ supersymmetric gauge theories can have many origins. One is the theory on D4 branes with other branes, say D8 and O8, like in our setting.
Further exploration of 5d gauge theories and D4 on D8 branes and O8 orientifolds has shown one can have $ \tilde E_1$ with $U(1)$ symmetry and $E_0$ without any global symmetry. All these class of theories appear naturally on the M-theory compactification on Calabi-Yau 3-fold with the contracted del Pezzo surfaces \cite{Morrison:1996xf, Douglas:1996xp, Intriligator:1997pq}.

The superconformal index on $S^1\times S^4$ is a tool to examine the enhancement of the global symmetry. A method for calculating the index is by evaluating the path integral by the localization. We are turning on all chemical potential allowed  by our choice of the supercharges which define the index. The contributions consist of perturbative and nonperturbative parts. They are localized at north and south poles. With our convention at south pole instantons contribute and at north pole anti-instantons contribute. While the calculation of the perturbative part is somewhat straightforward, the  instanton and anti-instanton contributions can be obtained from somewhat indirect approach, which is to evaluate the D0 branes contribution on D4 branes with the orientifold $O8$ and $D8$ branes  via the ADHM method.

The gauge group on $k$ D0 branes turns out to be $O(k)$ which consists of the connected subgroup $O_+(k)=SO(k)$ and the disconnected part $O_-(k)$ of minus one determinant. The D0 brane contributions are composed of these two parts whose detail identification need some effort. Once one obtains the D0-brane contribution to the index, one has to integrate over the loop-variables along the circle to get the gauge invariant expression. Here we are turning on all the chemical potentials for the flavor group $SO(2N_f)$ which can be regarded as the mass parameter for each hypermultiplet and the fugacity for the instanton number. The evaluation of the instanton contribution goes well for the cases $N_f\le 5$.
While we do not know the closed form of the index, our method here leads to a series expansion whose coefficients are expressed in terms of the characters of the enhanced global symmetry $E_{N_f+1}$. Especially the leading nontrivial contribution is given by the character of the adjoint representation of $E_{N_f+1}$. We reach, however, some obstacles for $N_f=6,7$ whose solution is not obvious to us at this moment. For $N_f=6,7$, the characters for the adjoint representation of $E_7,E_8$ have the contribution from two (anti)instantons, besides the perturbative parts and the single (anti)instanton contribution.

The rest of the paper is organized as follows. In Section \ref{sec:5dSCFT} we review the 5d ${\cal N}=1$ gauge theories and their superconformal limit. Global symmetry enhancement raining at the conformal fixed point is discussed. 
In Section \ref{sec:SCindex} we define the superconformal index and approach it with localization method.
It has both perturbative and nonperturbative contributions. The superconformal index for $Sp(N)$ and $U(N)$ are presented as an holonomy integral form. In Section \ref{enhancement} we explicitly calculate the index for $Sp(N)$ and show the global symmetry enhancement by showing they are expressed in terms of the characters of the enhanced symmetry. In Section \ref{sec:discussions} we conclude with some perspective and remarks. In Appendices we collect various related formulas.

\section{5d superconformal theories}\label{sec:5dSCFT}

We review the salient features of the 5d $\mathcal{N}=1$ supersymmetric gauge theory with focus made on the theory in the infinite gauge coupling limit. This theory has the vector multiplet and the hypermultiplet.
The vector multiplet consists of a gauge field $A_\mu$, a real scalar $\phi$, and a symplectic-Majorana fermion $\lambda^A$
(where $A$ denotes the $SU(2)_R$ R-symmetry doublet index) in the adjoint representation of gauge group $G$.
The Lagrangian for the vector multiplet $\Phi$ is encoded in the prepotential $\mathcal{F}$.
For generic gauge group $G$, the classical prepotential is given by \cite{Seiberg:1996bd,Intriligator:1997pq}
\begin{eqnarray}
    \mathcal{F} = \frac{1}{2g_{cl}^2} {\rm tr} \Phi^2 + \frac{\kappa}{6}{\rm tr} \Phi^3,
\end{eqnarray}
where $g_{cl}^2$ is the classical gauge coupling and $\kappa$ is a real number which is quantized.
The first term gives rise to the usual 5d Yang-Mills term which drops out in the infinite coupling limit.

The cubic term leads to the Chern-Simons term and its supersymmetric completions \cite{Seiberg:1996bd,
Intriligator:1997pq,Kugo:2000af,Bergshoeff:2001hc,Bergshoeff:2002qk}
\be\label{conformal-Lagrangian}
\mathcal{L}_{cubic} \!\!&\!\!=\!\!&\!\! \mathcal{L}_{cs} +\mathcal{L}_\kappa \,, \nn \\
\mathcal{L}_{cs} \!\!&\!\!=\!\!&\!\! \frac{\kappa}{24\pi^2}{\rm tr}\,\bigg[A\wedge F\wedge F+\frac{i}{2}A\wedge A\wedge A \wedge F -\frac{1}{10}A\wedge A\wedge A\wedge A\wedge A  \nn \\
&& \qquad \qquad  -3\bar\lambda\gamma^{\mu\nu}\lambda F_{\mu\nu}+6i \bar\lambda\sigma^I{\rm D}^I\lambda \bigg] \,, \nn \\
\mathcal{L}_\kappa \!\!&\!\!=\!\!&\!\! \frac{\kappa}{2\pi^2}{\rm tr}\,\big[\phi \mathcal{L}_{YM}\big] \nn \\
\!\!&\!\!=\!\!&\!\! \frac{\kappa}{2\pi^2}{\rm tr}\,\phi\bigg[-\frac{1}{2}F_{\mu\nu}F^{\mu\nu} -D_\mu \phi D^\mu\phi+\frac{i}{2}D_\mu\bar\lambda\gamma^\mu\lambda-\frac{i}{2}\bar\lambda \gamma^\mu D_\mu \lambda + {\rm D}^I {\rm D}^I +i\bar\lambda[\phi,\lambda] \bigg],\qquad
\ee
where ${\rm D}^I\ (I=1,2,3)$ are auxiliary scalars transforming as a triplet of $SU(2)_R$, and
$D_\mu = \partial_\mu-iA_\mu$ is the covariant derivative. See Appendix A for the detail notation.
The trace is taken over the gauge indices of the adjoint matrices $T^a$ and it follows that the cubic terms are proportional to the totally symmetric structure constant of the gauge group $G$
\be
    d^{abc} = \frac{1}{2}{\rm tr}\,T^a\{T^b,T^c\}.
\ee
We note that $d^{abc}$ is nonzero only for $SU(N)$ with $N\ge3$, and thus one neglects this classical cubic term for other gauge groups than $SU(N)$. (Of course, the cubic term exists for abelian case.)

The theory is invariant under the supersymmetry (SUSY) transformations
\be\label{SUSY-transformation1}
\delta A_\mu &=& i\bar\lambda\gamma_\mu \epsilon \,, \nn \\
\delta \phi &=& \bar\lambda\epsilon \,, \nn \\
\delta \lambda &=& \frac{1}{2}F_{\mu\nu}\gamma^{\mu\nu}\epsilon -i D_\mu\phi\gamma^\mu \epsilon  +i{\rm D}^I\sigma^I\epsilon \,, \nn \\
\delta \bar\lambda &=& -\frac{1}{2}F_{\mu\nu}\bar\epsilon\gamma^{\mu\nu} -i \bar\epsilon\gamma^\mu D_\mu \phi -i\bar\epsilon\sigma^I {\rm D}^I\,, \nn \\
\delta {\rm D}^I &=& D_\mu\bar\lambda \gamma^\mu\sigma^I\epsilon -[\phi,\bar\lambda]\sigma^I\epsilon,
\ee
where $\sigma^I$ are the usual Pauli matrices and the $R$-symmetry indices are contracted as $\bar \lambda \epsilon\equiv \bar\lambda_A\epsilon^A$. The supersymmetry parameters $\epsilon$ are symplectic-Majorana spinors defined as
\begin{align}
\bar\epsilon_A = (\epsilon^T)^B\varepsilon_{BA}\Omega,
\end{align}
where $\varepsilon_{AB}$ is the invariant tensor of $SU(2)_R$ while $\Omega$ is the invariant tensor of $Sp(2)$\footnote{To be more precise, $Sp(2)$ should be $USp(2,2)=SO(1,4)$.} which corresponds to the Lorentz rotation in five dimensions.
More precisely, the SUSY parameter $\epsilon^A_m$ transforms as the doublet of $SU(2)_R$  ($A=1,2$) and also
as the spinor of $Sp(2)$  ($m=1,2,3,4$).

The hypermultiplet consists of a complex scalar $q^A$ (an $SU(2)_R$ doublet) and a complex fermion $\psi$ in a representation of the gauge group.
With the matter coupling, the prepotential receives quantum corrections \cite{Witten:1996qb}.
For example, when the $SU(N)$ gauge theory is coupled to a fundamental matter field with real mass $m$,
the quantum contribution to the prepotential is given by \cite{Witten:1996qb}
\be
    \mathcal{F}_{quantum} = - sgn(m)\frac{1}{2}{\rm tr}\,\Phi^3,
\ee
which comes from one-loop computations.
Since the prepotential is at most locally cubic, the one-loop correction is exact.
One can regard  the classical cubic term 
as a quantum mechanically induced prepotential by integrating out the massive fundamental hypermultiplets. 
We note that
the gauge invariance restricts the coefficient $\kappa$ of the cubic terms to be \cite{Intriligator:1997pq,Witten:1996qb}
\be
    \kappa_{eff} = \kappa - \sum_{i}sgn(m_i)\frac{1}{2} \in \mathbb{Z},
\ee
Here, the sum is taken over all the hypermultiplets coupled to the vector multiplet.
The classical Chern-Simons level $\kappa$ is therefore quantized: integer or half-integer depending on even or odd number of fundamental matters, respectively.
In particular, when the number of matter hypermultiplets is odd,
the classical Chern-Simons level $\kappa$ cannot be zero. This means that the classical cubic term   always exists and thus parity symmetry is broken.

The gauge theories have the Coulomb branch and the Higgs branch.
The vacuum expectation value of the real scalar field $\phi$ in the vector multiplet parametrizes the Coulomb branch of the classical vacua of the low energy theory.
The effective theory at low energy is described by the diagonal elements of $\phi$  in the Cartan subalgebra of $G$.
The exact prepotential for the arbitrary gauge group $G$ with various flavors of masses $m_i$
 is expressed as  \cite{Intriligator:1997pq}
\be\label{one-loop-correction}
    \mathcal{F} = \frac{1}{2g_{cl}^2} {\rm tr}\, \phi^2 + \frac{\kappa}{6}{\rm tr}\, \phi^3
    +\frac{1}{12}\left(\sum_{\bf R}|{\bf R}\cdot\phi|^3 - \sum_i\sum_{{\bf w}\in {\bf W}_i}|{\bf w}\cdot\phi+m_i|^3\right),
\ee
where ${\bf R}$ are the roots of $G$ and ${\bf W}_i$ is the weight space of $G$ in the representation for $i$-th hypermultiplet. 
This prepotential is obtained by integrating out all the massive fields at generic point of Coulomb branch where
the charged matters acquire additional masses from the gauge coupling to $\phi$.
The last two terms are the quantum corrections arising from the massive vector multiplet
and the massive hyper multiplets, respectively.

The Higgs branch is where the scalar fields in the hypermultiplet take nonzero vacuum expectation value.
 The Higgs branch moduli space is a hyper-K\"{a}hler manifold. 
From the point of view of D4\mbox{--}\!\! D8  system, $N$ D4 branes  on $N_f$ D8 branes appear  as $N $ instantons in $SU(N_f)$ gauge theory,  and thus the Higgs branch of $D4$ is the   moduli space of $N$ instantons in $SU(N_f)$ gauge theory.   The $Sp(N)$ theories on D4-branes   with an additional $O8$ orientifold  have $SO(2N_f)$ global flavor symmetry where they are interact with $N_f$ hypermultiplets, which corresponds to $N_f$ $D8$ brans in the $D$-brane picture. The Higgs branch in this case is the moduli space of $N$ instantons in the $SO(2N_f)$ gauge theory.

It should be stressed that instantons are associated with the $U(1)_I$ current
\begin{align}
    J = *\,{\rm tr}\,(F\wedge F),
\end{align}
which is topological and always conserved \cite{Seiberg:1996bd}. This $U(1)_I$ charge corresponds to the instanton number. The instantons in 5d maximally supersymmetric theory play the role of Kaluza-Klein (KK) modes from a circle compactification of 6d $(2,0)$ theory and its mass is identified with KK momentum \cite{
Aharony:1997an,Douglas:2010iu,Lambert:2010iw,Lambert:2011gb,Kim:2011mv}.
In our case, however, the instanton solitons in 5d gauge theory form massive hypermultiplet and participate in the enhancement of global symmetry at the conformal point.
 The $U(1)_I$ provides an extra Cartan for the enhanced global symmetry $E_{N_f+1}$   \cite{Seiberg:1996bd, Morrison:1996xf, Ganor:1996pc}.
The way the instantons contribute the symmetry enhancement is one of our main points of the paper and will be explained throughout the paper, especially in Section \ref{enhancement}.

\subsection{Conformal limit}

A non-trivial conformal field theory emerges in the infinite coupling limit of the 5d gauge theory.
The $\mathcal{N}=1$ superconformal theory in five dimensions enjoys
$F(4)$ superconformal symmetry whose bosonic part is $SO(2,5)\times SU(2)_R$ where $SO(2,5)$ is conformal group and $SU(2)_R$ are R-symmetry group \cite{Nahm:1977tg,Romans:1985tw}.

When the gauge group is $U(1)$ or $SU(N) \, (N \ge 3)$, the Lagrangian describing the conformal fixed point is the Chern-Simons action given in (\ref{conformal-Lagrangian}).
This Lagrangian preserves 8 Poincar\'e and 8 conformal supersymmetries.
The supersymmetry transformation then extends from \eqref{SUSY-transformation1} to
\be\label{SUSY-transformation2}
\delta A_\mu &=& i\bar\lambda\gamma_\mu \epsilon \,, \nn \\
\delta \phi &=& \bar\lambda\epsilon \,, \nn \\
\delta \lambda &=& \frac{1}{2}F_{\mu\nu}\gamma^{\mu\nu}\epsilon -i D_\mu\phi\gamma^\mu \epsilon  +i{\rm D}^I\sigma^I\epsilon -\frac{2i}{5}\phi\gamma^\mu D_\mu\epsilon\,, \nn \\
\delta \bar\lambda &=& -\frac{1}{2}F_{\mu\nu}\bar\epsilon\gamma^{\mu\nu} -i \bar\epsilon\gamma^\mu D_\mu \phi -i\bar\epsilon\sigma^I {\rm D}^I -\frac{2i}{5}D_\mu\bar\epsilon\gamma^\mu\phi\,, \nn \\
\delta {\rm D}^I &=& D_\mu\bar\lambda \gamma^\mu\sigma^I\epsilon -[\phi,\bar\lambda]\sigma^I\epsilon-\frac{1}{5}\bar\lambda\sigma^I\gamma^\mu D_\mu\epsilon\ ,
\ee
where the SUSY parameters are $\epsilon=\epsilon_q+x\cdot\gamma \epsilon_s$ with the constant spinors $\epsilon_q$ and $\epsilon_s$.
The commutator of two SUSY transformations leads to the superconformal algebra
\be
\left[\delta_1,\delta_2\right]\!A_\mu &=& \xi^\nu\partial_\nu A_\mu +\partial_\mu\xi^\nu A_\nu + D_\mu \Lambda , \nn \\
\left[\delta_1,\delta_2\right]\!\phi &=& \xi^\mu\partial_\mu\phi +i[\Lambda,\phi]  + \rho\phi,\nn \\
\left[\delta_1,\delta_2\right]\!\lambda &=& \xi^\mu \partial_\mu\lambda +\frac{1}{4}\Theta_{\mu\nu}\gamma^{\mu\nu}\lambda+i[\Lambda,\lambda] +\frac{3}{2}\rho\lambda +\frac{3}{4}R^{IJ}\sigma^{IJ}\lambda,\nn\\
\left[\delta_1,\delta_2\right]\!{\rm D}^I &=& \xi^\mu\partial_\mu {\rm D}^I + i[\Lambda,{\rm D}^I] +2\rho {\rm D}^I+3R^{IJ}{\rm D}^J.
\ee
where the parameters are defined by
\be
\xi^\mu &=& -2i\bar\epsilon_1\gamma^\mu\epsilon_2,\nn \\
\Lambda &=& 2i\bar\epsilon_1\gamma^\mu\epsilon_2A_\mu+2\bar\epsilon_1\epsilon_2\phi, \nn \\
\Theta^{\mu\nu} &=& D^{[\mu}\xi^{\nu]} +\xi^\lambda\omega_\lambda^{\mu\nu},\nn \\
\rho &=& -\frac{2i}{5}D_\mu(\bar\epsilon_1\gamma^\mu\epsilon_2), \nn \\
R^{IJ} &=& -\frac{2i}{5}(\bar\epsilon_1\gamma^\mu\sigma^{IJ}D_\mu\epsilon_2-D_\mu\bar\epsilon_1\gamma^\mu \sigma^{IJ}\epsilon_2).
\ee
The Killing vector $\xi^\mu$ generates the $SO(2,5)$ conformal transformation and $R^{IJ}$ is the $SU(2)_R$ R-symmetry generator,
and $\Lambda$ is the gauge transformation parameter. This is a part of 
superconformal $F(4)$ algebra which is expected for the fixed point theories.

For the matter hypermultiplet, the canonical Lagrangian is already superconformal invariant.
The matter Lagrangian is given by
\be\label{matter-lagrangian}
    \mathcal{L}_{matter} =|D_\mu q|^2 -i\bar\psi\gamma^\mu D_\mu\psi +\bar{q}\phi^2q - q\sigma^I\bar{q} {\rm D}^I  -\sqrt{2}\bar\psi\lambda q +\sqrt{2}\bar{q}\bar\lambda \psi -i\bar\psi\phi\psi,
\ee
which is invariant under the SUSY transformation
\be\label{SUSY-transformation3}
    \delta q^A &=& \sqrt{2} i \bar\epsilon^A\psi, \nn \\
    \delta \bar{q}_A &=& \sqrt{2} i \bar\psi\epsilon_A, \nn \\
    \delta\psi &=& \sqrt{2}(- D_\mu q^A \gamma^\mu\epsilon_A +\phi q^A\epsilon_A-\frac{3}{5}q^A\gamma^\mu D_\mu\epsilon_A),\nn \\
    \delta\bar\psi &=& \sqrt{2}(\bar\epsilon^A\gamma^\mu D_\mu\bar{q}_A+\bar\epsilon^A\bar{q}_A\phi+\frac{3}{5}D_\mu\bar\epsilon^A\gamma^\mu\bar{q}_A).
\ee
The supersymmetry algebra closes on-shell
\begin{align}
[\delta_1,\delta_2]q^A&=\xi^\mu\partial_\mu q^A + i \Lambda q^A
+\frac32 \rho \,q^A + \frac{3}{4} R^{IJ} (\sigma^{IJ} q )^A,\cr
[\delta_1,\delta_2]\psi &= \xi^\mu\partial_\mu \psi +\frac14\Theta_{\mu\nu}\gamma^{\mu\nu} \psi+ i \Lambda \psi+2 \rho \,\psi +{\rm (e.o.m.)},
\end{align}
where the e.o.m. term is given by
\begin{align}
\big ( \bar\epsilon_2\epsilon_1-(\bar\epsilon_2\gamma^\nu \epsilon_1)\gamma_\nu  \big) \big(i\gamma^\mu D_\mu \psi + i\phi\psi-i\sqrt2 q \lambda\big).
\end{align}

When the gauge group is $Sp(N)$, unlike the $SU(N)$ case, the Lagrangian for the vector multiplet at the conformal fixed point does not exist. It is because there is no Chern-Simons term for $Sp(N)$. Nevertheless, the theory itself exists and
it is believed that such a theory flows to non-trivial interacting fixed point as well. Moreover, at the strong coupling limit,
the conformal theories for $Sp(N)$ also enjoy the same superconformal symmetry $F(4)$.
The supersymmetry transformation rules are presumably the same as those of the $SU(N)$ case
\eqref{SUSY-transformation2}. On the other hand, the canonical Lagrangian of the hypermultiplets remains invariant under the same SUSY transformation (\ref{SUSY-transformation3}).

The physical observables of the conformal field theory we are interested in are the gauge invariant operators.
Once the superconformal algebra is given, the gauge invariant operators are classified according to their representations of the superconformal algebra. These operators correspond to the physical states in a radially quantized theory. We will be interested in radially quantized theories on $\mathbb{R}\times {S}^4$. Physical states are then labeled by the charges of the Cartan generators of the bosonic subalgebra $SO(2,5)\times SU(2)_R$: the energy $\epsilon_0$ is the dilatation of the original theory, the angular momenta $j_1,j_2$ are the charges of $SU(2)_1\times SU(2)_2 \subset Sp(2)\cong  SO(5)$, and $j_R$ is the $SU(2)_R$-charge.

The supercharge $Q_m^A$ and the conformal supercharge $S^m_A$ are conjugate each other in the radially quantized theories and
they have the dilatation charge $+\frac{1}{2}$ and $-\frac{1}{2}$, respectively.
Their commutator gives a multiplet shortening condition often called BPS bound and the multiplet satisfying such a shortening condition is called the short (or BPS) multiplet. The commutator of $Q_m^A$ and $S^m_A$ reads \cite{Minwalla:1997ka,Bhattacharya:2008zy}
\be\label{SUSY-algebra2}
    \{Q_m^A,S^n_B\} = \delta_m^n \,\delta^A_B\, D + 2\,\delta^A_B\, M_m{}^{n} - 3\, \delta_m^n \,R_B{}^{ A} \ ,
\ee
where $D$ is the dilatation, $M_m^{\ \ n}$ are the $SO(5)$ rotations, and $R_B{}^{A}$ are the $SU(2)_R$ R-symmetry generators.
As the BPS states satisfy the BPS bound of (\ref{SUSY-algebra2}) their spectrum is protected by the supersymmetry.
This property allows us to count the exact spectrum of the BPS states (or operators) in the conformal field theory.
We will count the spectrum in Section \ref{sec:SCindex} with the superconformal index which is a kind of partition function counting BPS states.

\subsection{Global symmetry enhancement}\label{sec:GSE}
 
Let us first focus on   the $Sp(1)\cong SU(2)$ gauge theory  with $N_f$ fundamental hypermultiplet matters.  This  is the simplest example revealing non-trivial fixed point in the strong coupling limit.  To see such fixed point, it is instructive to review D-brane descriptions for this.

To begin with, let us consider familiar D-brane configuration of Type I string theory.
Type I theory has 16 D9-branes required to cancel the gravitational anomaly arising from the spacetime filling O9 orientifold plane. The gauge group of this background is $SO(32)$.
We introduce a D5-brane to this system.
The worldvolume theory of the D5-brane is  a six dimensional $SU(2)$ gauge theory coupled to $N_f=16$ hypermultiplets. This Higgs phase of this gauge theory is the center moduli space of a single $SO(32)$ instanton. We compactify the theory on a small circle ${S}^1$ along the D5-brane worldvolume direction and then perform  T-duality transformation.
It gives rise to the brane configuration of a D4-brane on 10d background compactified on ${S}^1/\mathbb{Z}_2$ interval with two orientifolds at the ends of the interval.
The two O8 orientifolds at the tip of the interval  ${S}^1/\mathbb{Z}_2$  are the T-dual of the  O9 orientifold plane in Type I theory.
This theory is often referred to as Type I$'$ theory.
The location of the D4-brane on the ${S}^1/\mathbb{Z}_2$ (from $x^9=0$ to $x^9=\pi$) corresponds to the Coulomb branch of the moduli space of the gauge theory which is parametrized by the scalar component of the vector multiplet.
In Type I picture this corresponds to the $SU(2)$ Wilson line of the D5-brane appearing after the circle compactification.
There are also 16 D8-branes filling the nine dimensions transverse to the compact direction.
They are T-dual of 16 D9-branes in Type I theory and their positions on the ${S}^1/\mathbb{Z}_2$ are again $SO(32)$ Wilson lines
on the compact circle.
For generic $U(1)^{16}$ Wilson lines or equivalently the positions of the D8-branes, the $SO(32)$ symmetry is spontaneously broken to $U(1)^{16}$ subgroup.

Of particular interest is the dynamics near one of the O8 orientifold fixed points, say at $x^9=0$.
We put a D4-brane and $N_f$ D8-branes close to   $x^9=0$ and thus the moduli can be seen as $\mathbb{R}^1/\mathbb{Z}_2$.
This D4-brane worldvolume theory then describes five-dimensional $SU(2)$ gauge theory with $N_f$ fundamental hypermultiplets.
The moduli space of the D4-brane becomes $\mathbb{R}^+$ which coincides with the Coulomb branch of the moduli space of the gauge theory.
The strings stretched between D4- and D8-branes provide $N_f$ hypermultiplets whose nonzero masses correspond to the distances of D8-branes from the fixed point.
The Higgs branch becomes the center moduli space of $N$ instantons in $N_f$ gauge theory.

Now let us analyze the field theory described by the above D-brane configuration.
For $SU(2)$ gauge group, the Coulomb branch is one-dimensional given by $\phi = {\rm diag}(a,-a)$.
Without loss of generality we can take $a\ge 0$ using the unbroken Weyl symmetry of $SU(2)$ gauge group.
The corresponding moduli space is $\mathbb{R}^+$. Along the Coulomb branch the gauge group is broken to $U(1)$.
Clearly there is a singularity at the boundary of the moduli space $a=0$ where additional massless vector fields arise and as a result the $SU(2)$ symmetry is restored.

As mentioned earlier, the classical cubic term vanishes because the totally symmetric structure constant $d^{abc}$ does not exit. According to (\ref{one-loop-correction}) the prepotential for the abelian gauge theory in the Coulomb branch gets corrected by the quantum effect and the effective gauge coupling is
\be
    \frac{1}{g^2_{eff}} = \frac{1}{g^2_{cl}} + 8a-\frac{1}{2}\sum_{i=1}^{N_f}|a-m_i|-\frac{1}{2}\sum_{i=1}^{N_f}|a+m_i|
\ee
where $m_i$ are the masses for $N_f$ fundamental flavors.
For consistency of the theory, the effective coupling should be non-negative on the entire Coulomb branch of the moduli space.
This cannot be guaranteed if $N_f >8$.
There always exists the finite point $a_c$ of the moduli space away from the origin where the effective coupling $g_{eff}$ diverges.
Beyond $a_c$ the effective coupling flips its sign and thus the theory becomes sick.
This reflects the fact that the quantum theories for $N_f>8$ are not renormalizable \cite{Seiberg:1996bd}.

When $N_f\le 8$, on the other hand, the effective gauge coupling is positive everywhere on the Coulomb branch and therefore the Lagrangian description of the 5d gauge theory is still valid even for the quantum level. 
Let us place all $N_f$ D8-branes at the fixed point. It follows that the effective coupling becomes
\be
    \frac{1}{g^2_{eff}} = \frac{1}{g^2_{cl}} + (8-N_f)a.
\ee
As we take the classical gauge coupling $g_{cl}$ to be positive there is no singularity on the Coulomb branch
when $N_f\le 8$.
Moreover, if the classical gauge coupling is taken to the infinity, $g_{cl}\rightarrow\infty$,
there can be a scale invariant fixed point at the origin of the Coulomb branch where $a=0$ \cite{Seiberg:1996bd}.
Hence, the 5d gauge theories with the condition above are reliable field theories over the entire Coulomb branch of the moduli space.
The field theory at the conformal fixed point is very strongly interacting since the effective coupling diverges.
 We note that when $N_f=8$, the infinite coupling limit yields the vanishing of the metric on the entire Coulomb moduli space, and thus the 5d description becomes not meaningful. 

The global flavor symmetry of the theory is $SO(2N_f)$ and  there is conserved $U(1)$   charge for the instanton soliton numbers in 5d $Sp(1)$ gauge theory.  Together, the global symmetry is expected to be enhanced to $E_{N_f+1}$ at the conformal fixed point for $N_f<8 $ \cite{Seiberg:1996bd,Intriligator:1997pq}, which is related to the gauge symmetry enhancement of the heterotic string theory on the self-dual radius. 
We note that the Higgs branch is also enhanced and thus it becomes  the moduli space of $E_{N_f+1}$ instantons \cite{Ganor:1996mu,Seiberg:1996bd}. 

The symmetry enhancement from the point of view of the heterotic string theories is as follows. 
Consider the eleven-dimensional M-theory compactified on an interval ${S^1}/\mathbb{Z}_2$ with a radius $R_{11}$ along  the 11-th direction. The orbifold action $\mathbb{Z}_2$ introduces  ten-dimensional orientifold plane  at each tip  of the interval.   
To cancel gravitational anomalies arising from two hyperplanes, we need to place a 10d $E_8$ gauge theory at each end of the interval.
This theory realizes the strong coupling limit of the $E_8\times E_8$ heterotic string theory in ten dimensions \cite{Horava:1995qa,Horava:1996ma}.
The heterotic string coupling constant $g_h$ is related to the radius of 11d circle by $R_{11}\sim g_h^{2/3}$
and, at small coupling limit, M-theory on a ${S}^1/\mathbb{Z}_2$ reduces to the $E_8\times E_8$ heterotic string theory.

One further compactifies the heterotic string on a circle with the radius $R_{h}$,
which exhibits the gauge symmetry enhancement at the special point of the moduli space.
More explicitly, one can turn on the Wilson lines that break the gauge symmetry to $SO(14)^2 \times U(1)^2\times U(1)^2$
where the later $U(1)^2$ are given by right and left-moving winding modes.
Then, there is symmetry enhancement at the self-dual radius to 
 $E_8\times E_8\times U(1)^2$ \cite{Polchinski:1995df,Polchinski:1996fm}.
One may view the symmetry enhancement in the heterotic string theory 
from the point of view of Type I$'$ string theory using the heterotic/Type I and Type I/Type I$'$
string theory dualities.
The gauge coupling and the radius of the circle $S^1$ with $R_{h}$ are mapped as follows
\be
    R_{I'} = \sqrt{R_h g_h}\ , \quad g_{I'} = \sqrt{R^3_h/g_h} \ ,
\ee
where $g_{I'},R_{I'}$ are the gauge coupling and the radius of the compact circle in 
Type I$'$ string theory, respectively.
Thus, at the self-dual radius $R_h= R_{\rm sd} \sim 1$ and the strong coupling limit $g_h=\infty$ in the heterotic theory, the dual Type I$'$ theory is
at weak coupling region with the infinity radius $R_{I'}$ of the circle ${S}^1/\mathbb{Z}_2$ \cite{Polchinski:1995df,Polchinski:1996fm}.
The dual description of   $SO(14)\times U(1)$ preserving Wilson line is one D8-brane away from a fixed point while 7 D8-branes are
located exactly at the fixed point.

The string coupling $g_{I'}$ which is the dilaton expectation value diverges and the perturbative description breaks down precisely at which we hope to see the symmetry enhancement.
Since D8-branes are the dilaton sources the existence of D8-branes causes the dilaton gradient.
So, as the eighth D8-brane approaches to the fixed point, the string coupling at the orientifold diverges, which implies that D0-branes become massless and
they can provides the massless gauge bosons that are necessary for $E_8$ gauge symmetry enhancement.

From the heterotic string theory point of view, D4-branes on $N_f$ D8-branes at the O8 orientifold fixed point can be interpreted as the instantons of the $E_{N_f+1}$ gauge symmetry in the heterotic theory.
The string duality relates the D4-branes to the NS5-branes in the heterotic theory.
In the heterotic theory, the five-branes couples to the gauge symmetry which corresponds to $SO(2N_f)$ global symmetry at finite coupling of the D4-brane theory.
The Higgs branch of the moduli space in the five-brane worldvolume theory is known to be the moduli space of $SO(2N_f)$ instantons.
One may notice that the $Sp(N)$ gauge symmetry of $N$ D4-brane theory coincides with the gauge symmetry of the $N$ instanton moduli space of $SO(2N_f)$ gauge theories.
As the global symmetry is enhanced to $E_{N_f+1}$ in the D4-brane theory at the infinite coupling limit,
the gauge symmetry of the heterotic theory is also enhanced to $E_{N_f+1}$ gauge symmetry in this regime.
Therefore the Higgs branch of $N$ D4-branes describes the $N$ instanton moduli space of $E_{N_f+1}$ gauge theory.\\

The conformal fixed point can exit for other gauge groups as well \cite{Intriligator:1997pq}.
The condition for the non-trivial fixed point is determined by the matter content of the theories as done in the $Sp(1)=SU(2)$ case, where the matter content is restricted by the positivity of the effective gauge coupling. 
Among other gauge theories of the fixed point, we pay special attention to the $Sp(N)$ gauge theories, as $Sp(N)$ gauge theory describes the worldvolume theory on $N$ coincident D4-branes on the orientifold fixed point. Take the $Sp(N)$ gauge theory with $N_a$ antisymmetric and $N_f$ fundamental hypermultiplets of the gauge group.
The Coulomb branch moduli is again given by the vacuum expectation value of the vector multiplet scalar,
$\phi = {\rm diag}(a_1,\cdots,a_N,-a_1,\cdots, -a_N)$.
We can always choose $a_i\ge 0\ (i=1,\cdots,N)$ using the unbroken Weyl symmetry.
The effective gauge coupling receives one-loop corrections 
\cite{Intriligator:1997pq}
\be
    (g^{-2}_{eff})_{ii} &=& 2\left[(N-i)a_i + \sum_{k=1}^{i-1}a_k\right](1-N_a) + a_i(8-N_f) ,\cr
    (g^{-2}_{eff})_{i<j} &=& 2(1-N_a)a_j ,
\ee
where we have assumed the classical gauge coupling $\frac{1}{g_{cl}^2}=0$.
The eigenvalue of this effective coupling is positive semi-definite only for $N_a=0,\, N_f\le 2N+4$ or $N_a=1,\, N_f \le 7$.
This shows there is an interesting fixed point when the above conditions for the matter content are satisfied.
In particular, $N$ D4-branes on the $\mathbb{Z}_2$ fixed point correspond to $N_a=1$ and $N_f\le 7$ cases. The global symmetry $SO(2N_f)$ is also enhanced to $E_{N_f+1}$ at the conformal fixed point just as the aforementioned
$Sp(1)=SU(2)$ gauge theories. The $Sp(1)$ theories are, in fact, a special example of such enhancement
 where the antisymmetric hypermultiplet is decoupled because it is a gauge singlet.


\section{Superconformal index}\label{sec:SCindex}

In this section, as a tool for counting BPS states, we introduce the superconformal index \cite{Minwalla:1997ka} for the 5d superconformal theories of the gauge group $G$, with various hypermultiplets (here $N_f$ flavors). We choose a supercharge $Q_{m=2}^{A=1}\equiv Q$ among others to define the superconformal index, so that 
we count the BPS states which are annihilated by the supercharge $Q$ and its conjugate supercharge $S=Q^\dagger$. This means that  we count $\frac{1}{8}$ BPS states. 
It follows from \eqref{SUSY-algebra2} that
\be
    \Delta\equiv\{Q,S\} = \epsilon_0 - 2j_1 - 3R,
\ee
where the energy or the dilatation is denoted by $\epsilon_0$, the Cartan generators of $SU(2)_1\times SU(2)_2\subset Sp(2)$ are $j_1$, $j_2$, and the Cartan generator for $R$-symmetry (or  $R$-charge) is $R$. The BPS bound is saturated when $\Delta=0$ or $\epsilon_0 = 2j_1+3R$.
Each of three Cartan generators $\Delta,\,j_1+R,\,j_2$ of $F(4)$ commuting with the supercharges $Q$ and $S$ have the chemical potentials $e^{-\beta},\,x=e^{-\gamma_1},$ and $y=e^{-\gamma_2}$, respectively.    The instanton number  $k$ also commute with the supercharges.   The  Cartan generators $H_i (i=1,2,\cdots N_f)$  of the flavor symmetry and the instanton charge $k$ have the chemical potential $e^{-im_i}$ and $q$,  respectively.   We then use these Cartan generators to label the BPS states.  With these ingredients, we define
the superconformal index
\be\label{index}
I(x,y,m_i,q) = {\rm tr}\Big[(-1)^Fe^{-\beta\{Q,S\}}x^{2(j_1+R)}y^{2j_2}e^{-i\sum_i H_im_i} q^k\Big],
\ee
where $F$ is the fermion number operator.  The trace is taken over the Hilbert space on ${S}^4$ after radial quantization.
 It is easy to see that 
the index counts only the number of BPS states ($\Delta=0$)  because 
the states with non-zero $\Delta$ pairwise cancel out due to $(-1)^F$.  As a result,
the index does not depend on the chemical potential $e^{-\beta}$.

The index (\ref{index}) admits a functional integral representation.
Let us consider radial quantization of a conformal theory and compactification along the Euclidean time direction after Wick rotation $x^0=-i\tau$.
It is equivalent to put the conformal theory on ${S}^1\times {S}^4$.
The index then can be expressed as a path integral of the Euclidean action on ${S}^1\times {S}^4$:
\be
    I(x,y,m_i,q) = \int_{{S}^1\times {S}^4}\mathcal{D}\Psi e^{-S_E[\Psi]}.
\ee
The factor $(-1)^F$ enforces
both bosonic fields and fermionic fields satisfy periodic boundary conditions
along the time circle ${S}^1$ of radius $\beta$. The insertion of the chemical potentials leads to 
the twisted boundary condition 
\be
    \Psi(\tau+\beta) = e^{-(-2j_1-3R)\beta -2(j_1+R)\gamma_1 -2j_2\gamma_2-iH_im_i}\Psi(\tau).
\ee
Equivalently, we can shift the time derivatives in the action
\begin{align}
\label{time-shift}
    \partial_\tau \rightarrow \partial_\tau +\frac{2\beta-2\gamma_1}{\beta}j_1 - \frac{2\gamma_2}{\beta}j_2 + \frac{3\beta-2\gamma_1}{\beta}R
    -i\frac{m_i}{\beta}H_i ,
\end{align}
with the usual periodic boundary condition. From here on, we regard all the time derivatives as this shifted one \eqref{time-shift}.


\subsection{Localization}\label{sec:localization}

The localization technique \cite{Pestun:2007rz} is very powerful in 
 evaluating the superconformal index.
The superconformal index is independent not only of the parameter $\beta$ but also of any continuous deformation of the theory as long as the deformation preserves the chosen supercharge $Q$.
This means that under deformation of the Lagrangian 
with arbitrary $Q$-exact terms and a continuous parameter $t$
\begin{align}
 \mathcal{L}\rightarrow \mathcal{L}+t\{Q,V \},
\end{align}
the result of the path integral is not altered. In particular, when we take $t$ to infinity, the path integral is localized around a set of the classical solutions to the saddle point equation $\{Q,V \}=0$.
In this limit the Gaussian integral over the quadratic fluctuations near the saddle points yields the exact result of the superconformal index.

To apply the localization,
let us choose a Killing spinor $\epsilon$ parametrizing the SUSY transformation of $Q+S$
\be\label{SUSY-parameter}
    \epsilon \equiv \epsilon_q + \epsilon_s =  e^{\frac{1}{2}\theta_1\gamma^{51}}e^{\frac{1}{2}\theta_2\gamma^{12}}e^{\frac{1}{2}\theta_3\gamma^{23}}e^{\frac{1}{2}\theta_4\gamma^{34}}\epsilon_0^q + \gamma^5 e^{\frac{1}{2}\theta_1\gamma^{51}}e^{\frac{1}{2}\theta_2\gamma^{12}}e^{\frac{1}{2}\theta_3\gamma^{23}}e^{\frac{1}{2}\theta_4\gamma^{34}}\epsilon_0^s,
\ee
where $\epsilon^q_0$ and $\epsilon^s_0$ are the constant spinors corresponding to $Q$ and $S$, respectively.
The spinor $\epsilon$ satisfies the Killing spinor equation
\be
    \nabla_\mu\epsilon = \frac{1}{2}\gamma_\mu\gamma^5\tilde\epsilon \ ,
\ee
where another Killing spinor $\tilde\epsilon$ is given as $\tilde\epsilon = -\epsilon_q+\epsilon_s$.
Here the twist of the time derivatives \eqref{time-shift} should be understood.
The supercharge $Q$ has definite charges under the bosonic symmetries of the superconformal algebra: $\epsilon_0=\frac12$, $j_1=-\frac12$ and $R=\frac12$.  It is then easy to see that 
the constant spinor $\epsilon^q_0$ obeys the
projection condition
\be
    \gamma^5\epsilon^q_0 = i\gamma^{12}\epsilon^q_0 = \sigma^3\epsilon^q_0 = -1.
\ee
It follows from $S=Q^\dagger$ that
\begin{align}
   (\epsilon^{q*}_0)^m_A = \varepsilon_{AB}\Omega^{mn} (\epsilon^s_0)_n^B.
\end{align}
(See Appendix \ref{appendixA} and \ref{appendixB} for the gamma matrix convention and the metric of ${S}^1\times{S}^4$ in detail.)
The square of this supercharges $Q+S$ takes the form
\be\label{cohomology}
    \delta_\epsilon^2 = -iL_\tau +i\Lambda+ i \frac{2\gamma_1}{\beta} (j_1+R) +i \frac{2\gamma_2}{\beta}j_2 - \frac{m_i}{\beta}H_i,
\ee
where $L_\tau$ is the Lie derivative along the ${S}^1$ (or time) direction and the symmetry generators $j_1,j_2,R$ and $H_i$
appear due to the twist of the time derivative \eqref{time-shift}. Here, $\Lambda$ generates the gauge transformation and its action on the fields is determined by the representations of the fields.
Later we will see that $\Lambda$ is replaced by the holonomy variable $A_\tau$ which is the remaining moduli
of the superconformal index after localizing all field configurations.

We now deform the Lagrangian by adding the $Q$-exact term
\be\label{Q-exact}
    \delta\mathcal{L}= t\,\delta_\epsilon\!\left((\delta_\epsilon \lambda)^\dagger \lambda\right),
\ee
where $\delta_\epsilon\lambda$ is the gaugino variation with respect to the SUSY parameter $\epsilon$.
When we take $t\rightarrow +\infty$ limit, this deformation localizes the vector multiplet part of the path integral
near the critical points of the bosonic potential given by
\be\label{bosonic-potential}
    V_b \!\!&\!\!=\!\!&\!\! (\delta_\epsilon\lambda)^\dagger \delta_\epsilon \lambda \nn \\
    \!\!&\!\!=\!\!&\!\!F_{\tau\mu}F^{\tau\mu} + \cos^2\frac{\theta_1}{2}(F^-_{ij}-\omega^-_{ij}\phi)^2+ \sin^2\frac{\theta_1}{2}(F^+_{ij}- \omega^+_{ij}\phi)^2  +(\nabla_\mu\phi)^2-{\rm D}^2,
\ee
where $F^\pm_{ij} = \frac{1}{2}\left[F_{ij} \mp *_4F_{ij}\right]$ are the self/anti-self dual part of the field strength on ${S}^4$ ($*_4$ is the Hodge star operator on ${S}^4$)
 and
\be
    && \omega^+_{ij} = \frac{i}{2\sin^2\frac{\theta_1}{2}}\bar{\tilde\epsilon}^R\gamma^5\gamma^{ij}\epsilon^R \ , \quad \omega^-_{ij} = \frac{i}{2\cos^2\frac{\theta_1}{2}}\bar{\tilde\epsilon}^L\gamma^5\gamma^{ij}\epsilon^L
    \ , \quad \omega^+_{ij}\omega^{+ij} =\omega^-_{ij}\omega^{-ij}=1 \,, \nn \\
    && \gamma^5\epsilon^R = \epsilon^R \ , \quad \gamma^5\epsilon^L = -\epsilon^L.
\ee
The deformed potential is obviously positive semi-definite apart from the D-term potential which comes with minus sign in front.
The D-term potential, in fact, is positive semi-definite as well. It is because when we put the theory on the Euclidean space through analytic continuation,
the field D is also analytically continued to be pure imaginary.
Therefore, the critical points of the bosonic potential are those at which all the square terms in the potential vanish.
This is very similar to the $Q$-exact deformation done in \cite{Pestun:2007rz} for the localization of the ${S}^4$ partition function. Except for the time derivatives in the kinetic terms of \eqref{bosonic-potential}, the rest bosonic potential terms are exactly the same as those of the vector multiplet part in the ${S}^4$ partition function computation \cite{Pestun:2007rz}.

We can evaluate the path integral exactly using the saddle point approximation around the classical solutions to the bosonic potential (\ref{bosonic-potential}).
Let us first analyze the classical saddle points of the deformed potential.
The first term and the last two terms in (\ref{bosonic-potential}) imply that both $F_{\tau i}$ and ${\rm D}^A$ vanish and $\phi$ is covariantly constant everywhere on ${S}^4$ at the critical point.
It follows from the Bianchi identity that the remaining terms, the second and third terms, yield the classical solution $F_{ij}=0$ and $\phi=0$ away from the north and south poles \cite{Pestun:2007rz}.
Thus we can only turn on the holonomy $\alpha$, which take values in the Cartan subalgebra of the gauge group $G$, along the time circle as a smooth classical solution to the saddle point equations.
(Other terms in the vector multiplet are set to zero.)

We, however, note that singular instanton solution can be localized exactly at the south pole ($\theta_1=\pi$) of ${S}^4$ and  singular anti-instanton solution can be localized at the north pole ($\theta_1=0$) \cite{Pestun:2007rz}.
At the north or south pole, the constraint on the field strength $F^+=F^-=0$ can be relaxed because either $\cos\frac{\theta_1}{2}$ or $\sin\frac{\theta_1}{2}$ becomes zero.
Since the anti-self-dual condition $F^-=0$ solves the saddle point equations,
the singular configuration corresponding to point-like anti-instantons can be located at the north pole.
The self-dual condition $F^+=0$ also solves the saddle point equations, but this yields that
point-like instantons are located at south pole. \\

To localize the integral over the matter fields in the hypermultiplets, we shall use the original matter Lagrangian (\ref{matter-lagrangian}).
This follows from the fact that the superconformal index does not depend on the continuous parameter of the theory.
We can always pull out a continuous parameter, say $t$, from the matter Lagrangian by the field redefinition, and use $t$ as a deformation parameter.
As in the vector multiplet localization, we take the limit $t\rightarrow \infty$ and localize the matter field integral around
the critical point of the bosonic potential.
Due to the conformal mass terms associated with the coupling to the curvature of ${S}^4$,
there is no zero modes from the matter fields.
Thus we set the classical values of all bosonic fields in the hypermultiplets to zero.

In summary, the path integral receives the perturbative contributions with holonomy $\alpha$ on the entire ${S}^4$ and, in addition,
the instanton and anti-instanton contributions localized at the south pole and the north pole on ${S}^4$, respectively.
Indeed, the one-loop perturbative contribution can also be localized to the south and north poles of ${S}^4$ \cite{Gomis:2011pf}
at which the equivariant rotations $j_1$ and $j_2$ of 
(\ref{cohomology}) become singular.
One then deduce that the superconformal index becomes the holonomy integral of the product of contributions from the south/north poles,
\be
    I(x,y,m_i,q)=\int [d\alpha] \  I_{south}(\alpha,x,y,m_i,q)I_{north}(\alpha,x,y,m_i,q^{-1})
\ee
where the integral is taken over the holonomy $\alpha$ and $[d\alpha]$ involves the Haar measure of the gauge group $G$.
The integrand $I_{south}$ and $I_{north}$ contain both one-loop perturbative contributions and instanton contributions\footnote{
The classical Chern-Simons term modifies instanton dynamics through induced Chern-Simons term on the instanton moduli space, which affects instanton contributions.
More detail is explained in Section \ref{instantioncontri}}
from the south and north poles,
\be
    &&I_{south}(\alpha,x,y,m_i,q) = I^{1-loop}_{south} \times I^{inst}_{south} , \nn \\
    &&I_{north}(\alpha,x,y,m_i,q^{-1}) = I^{1-loop}_{north} \times I^{inst}_{north},
\ee
Notice that $I_{south}$ is a function of $q$ reflecting that the instantons are localized at the south pole while $I_{north}$ is a function of $q^{-1}$ reflecting that the anti-instantons are localized at the north pole.

\subsection{One-loop contribution}
The one-loop perturbative contribution can be computed using the Atiyah-Singer equivariant index theorem \cite{Atiyah}. Recall the ${S}^4$ partition function in four dimensions was obtained from the index theorem \cite{Pestun:2007rz,Gomis:2011pf}. 5d calculation is not much different from the 4d calculation.
We have already noted that the algebra \eqref{cohomology} and the $Q$-exact deformation \eqref{bosonic-potential} are analogous to those in \cite{Pestun:2007rz,Gomis:2011pf}.
The only difference arises from the momentum modes along the time circle in 5d calculation.
With this in mind, we will follow each step of 4d perturbative computation  in \cite{Pestun:2007rz,Gomis:2011pf} and extend it to the 5d case with the insertion of the time circle dependence whenever necessary.

Let us start with the cohomological formulation of the supersymmetry transformation by $Q$.\footnote{
Before evaluating the path integral, we need to gauge fix the theory. We introduce the standard ghost fields and BRST transformations.
The supercharge $Q$ here is a linear sum of the original supercharge $Q+S$ and BRST operator.
From now on, we also assume the $Q$-exact term \eqref{Q-exact} includes the gauge fixing terms.
It is known \cite{Pestun:2007rz} that the saddle points of the gauge fixed $Q$-exact term remains the same as those of \eqref{Q-exact}.}
The supercharge $Q$ behaves as an equivariant differential operator on a supermanifold
formed by the bosonic and fermionic fields including the ghost fields in the gauge fixed theory.
The bosonic and fermionic fields can be regarded as differential forms on a supermanifold such that they form the $Q$-complex as 
\be
    Q\Psi_{b,f} = \Psi_{f,b}' \ , \quad Q\Psi'_{f,b} = \mathcal{H}\Psi_{b,f},
\ee
where $\Psi_b$ and $\Psi_f$ are the bosonic and fermionic fields respectively. It follows from \eqref{cohomology} that $\mathcal{H}$ is given by 
\be
    Q^2 =\mathcal{H} \equiv L_\tau -i\frac{\alpha}{\beta} - \frac{2\gamma_1}{\beta} (j_1+R) - \frac{2\gamma_2}{\beta}j_2 -i \frac{m_i}{\beta}H_i.
\ee
Therefore $Q^2$ produces a combination of $U(1)$ symmetry transformations and the gauge rotation by the holonomy $\alpha$.
The $Q$ is nilpotent only in the subspace of $\mathcal{H}$-invariant fields and the cohomology of the $Q$ is the $\mathcal{H}$-equivariant cohomology of the supermanifold.

We expand the gauge fixed $Q$-invariant terms to quadratic order in the field fluctuations $\Psi_b$ and $\Psi_f$
and evaluate the Gaussian integral for the quadratic terms.
There occurs  huge pairwise cancellation between the bosonic and the fermionic fluctuations, as they are paired by the $Q$-complex.
The remaining contribution arises only from the kernel and cokernel spaces of the operator $D$
that is the quadratic operator acting on $\Psi_b$ and $\Psi_f$ in the $Q$-exact term \eqref{Q-exact}. 
The one-loop determinant becomes \cite{Pestun:2007rz}
\be
    I^{1-loop}=\left(\frac{{\rm det}_{{\rm coker}D}\mathcal{H}|_f}{{\rm det}_{{\rm ker}D}\mathcal{H}|_b}\right)^{1/2}.
\ee
The fermionic contribution comes to the numerator while the bosonic contribution comes to the denominator due to their statistics.
This one-loop determinant is the equivariant Euler class of the normal bundle whose sections are 
$\Psi_b$ and $\Psi_f$.
It is given by the product of the weights of irreducible representations with respect to the group action $\mathcal{H}$.
We can compute the weights of the representations from the equivariant index (or the equivariant Chern character) of the operator $D$.
The equivariant index is expressed as the sum over weights and
one can easily convert it into a product of weights such as
\be
    {\rm ind}D_{\mathcal{H}}\equiv \sum_{i}\epsilon_ie^{w_i} \rightarrow \prod_i w_i^{-\epsilon_i},
\ee
where $w_i$ is a weight and $\epsilon_i$ is the multiplicity of $w_i$ representation.

To compute the equivariant index of the operator $D$ we use the equivariant Atiyah-Singer index theorem.
The Atiyah-Singer index theorem is defined as follows:
Let $E$ and $F$ be vector bundles over a manifold $M$, and $\Gamma(E)$,  $\Gamma(F)$ be the space of sections. A differential operator $D$ is a map of sections, $D : \Gamma(E) \rightarrow\Gamma(F)$.
For a compact Lie group $\mathcal{G}$ acting on $M$, let $T=U(1)^n$ be the maximal torus
 of $\mathcal{G}$. We can define the $\mathcal{G}$-equivariant index with respect to $t\in T$
\be
    {\rm ind}D(t) = {\rm tr}_{{\rm ker}D}t-{\rm tr}_{{\rm coker}D}t.
\ee
The equivariant index is an alternating sum over the cohomologies constructed by the differential $D$, which
receives the contributions only from the kernel and the cokernel of $D$ as mentioned above. The index is not altered under small deformation and thus topological. It turns out that the index can be expressed as a sum over the contributions from the fixed points of the group action $T$.
Then the Atiyah-Singer index theorem states that \cite{Atiyah}
\be\label{Atiyah-Singer}
    {\rm ind}D(t)=\sum_{{\rm fixed \, point}\, p}\frac{{\rm tr}_{E(p)}t-{\rm tr}_{F(p)}t}{{\rm det}_{TM_p}(1-t)},
\ee
where we have assumed there are discrete number of fixed points $p$.
Therefore when there are fixed points of the action $T$ the index reduces to the summation of
the equivariant index around each fixed point.

The fixed points of the group action $\mathcal{H}$ are the south and north poles of ${S}^4$,
which are the critical points of the Lorentz rotation $j_1$ and $j_2$.
We consider the equivariant index around these two fixed points.
Since the operator $D$ does not mix the vector multiplet and hypermultiplet complexes,
we compute the equivariant indices for these two multiplets independently.
Let us first compute the equivariant index of the vector multiplet.
Near the south pole, the operator $D$ for the vector multiplet is isomorphic to the self-dual complex $(d,d^+)$ \cite{Pestun:2007rz,Gomis:2011pf}
\be
    \Omega^0 \overset{d}{\rightarrow} \Omega^1 \overset{d^+}{\rightarrow} \Omega^{2+},
\ee
where $\Omega^0$, $\Omega^1$, and $\Omega^{2+}$ denote zero-forms, one-forms and self-dual two-forms, respectively, and $d^+$ is the self-dual projection.
On the other hand, this operator is isomorphic to the anti-self-dual complex $(d,d^-)$ near the north pole
\be
    \Omega^0 \overset{d}{\rightarrow} \Omega^1 \overset{d^-}{\rightarrow} \Omega^{2-},
\ee
where $\Omega^{2-}$ stands for anti-self-dual two-forms, and $d^-$ involves the anti-self-dual projection.

In the neighborhood of the south pole on ${S}^4$, we take a local manifold $\mathbb{R}^4=\mathbb{C}^2$. The $\mathcal{H}$ acts on the local coordinate as $(z_1,z_2)\rightarrow (e^{i\epsilon_1}z_1,e^{i\epsilon_2}z_2)$
where $z_1,z_2$ are the $\mathbb{C}^2$ coordinates.
These  $U(1)_{\epsilon_1}\!\times\!U(1)_{\epsilon_2}$ rotations correspond to the diagonal and off-diagonal combinations of the Cartan generators in $SU(2)_1\!\times\!SU(2)_2\!=\!SO(4)$ respectively.
The rotation parameters $\epsilon_{1},\,\epsilon_{2}$ are related to the chemical potentials $\gamma_{1},\,\gamma_{2}$
by
\be
\epsilon_1=i\frac{\gamma_1+\gamma_2}{\beta}\ ,\qquad \epsilon_2 = i\frac{\gamma_1-\gamma_2}{\beta}.
\ee
Recall that $\Omega^0$, $\Omega^1$ and $\Omega^{2+}$ in the self-dual complex
take the representations of $SO(4)$ as $({\bf 0},{\bf 0})$, $({\bf \frac{1}{2}},{\bf \frac{1}{2}})$ and $({\bf 1}, {\bf 0})$, respectively.
Therefore the fields consisting of the self-dual complex also rotate under the $U(1)^2$ action according to their representations. As an example, let us compute the equivariant index of the self-dual complex with the torus $T=U(1)_{\epsilon_1}\!\times\! U(1)_{\epsilon_2}$.
The Atiyah-Singer equivariant index theorem (\ref{Atiyah-Singer}) tells us that the index of the self-dual complex on $\mathbb{C}^2$ is expressed as \cite{Gomis:2011pf}
\be\label{self-dual-index1}
    {\rm ind}(D_{SD}) \!\!&\!\!=\!\!&\!\! \frac{(2+e^{i\epsilon_1+i\epsilon_2}+e^{-i\epsilon_1-i\epsilon_2})-(e^{i\epsilon_1}+e^{-i\epsilon_1}+e^{i\epsilon_2}+e^{-i\epsilon_2})}{(1-e^{i\epsilon_1})(1-e^{-i\epsilon_1})(1-e^{i\epsilon_2})(1-e^{-i\epsilon_2})} \nn \\
    \!\!&\!\!=\!\!&\!\!\frac{1+e^{i\epsilon_1+i\epsilon_2}}{(1-e^{i\epsilon_1})(1-e^{i\epsilon_2})}
\ee
The numerator in the first line is the Chern character of the vector bundle associated with the self-dual complex.
One can easily read off this from the representations of $U(1)^2$ action on the form fields given above.
The denominator in the first line is from $\mathbb{C}^2$ coordinate dependence of the sections of the vector bundle,
where $U(1)^2$ acts on the coordinate as $(z_{1,2},\bar{z}_{1,2})\rightarrow (e^{i\epsilon_{1,2}}z_{1,2},e^{-i\epsilon_{1,2}}\bar{z}_{1,2})$.

Let us then compute the equivariant index with $\mathcal{H}$ action which includes not only the $U(1)_{\epsilon_1}\!\times \!U(1)_{\epsilon_2}$ but also additional $U(1)$ actions
such as the ${S}^1$ translation, the gauge transformation, and the flavor rotation. 
We can consider the ${S}^1$ circle as a $U(1)$ line bundle over $\mathbb{C}^2$ and expand the elements of the self-dual complex
by the eigenmodes of the circle momentum as $\Psi = \sum_{n\in \mathbb{Z}}\Psi_n e^{\frac{2\pi in}{\beta}}$.
As the self-dual complex is formed by the fields in the vector multiplet, $\Psi_n$ are in the adjoint representation of the gauge group
and the flavor rotation will not act on them.
From (\ref{self-dual-index1}) one can obtain the equivariant index for the vector multiplet
\be\label{self-dual-index2}
    {\rm ind}(D_{vec}) = -\frac{1+e^{i\epsilon_1+i\epsilon_2}}{2(1-e^{i\epsilon_1})(1-e^{i\epsilon_2})}\sum_{{\bf R}}e^{-i{\bf R}\cdot\frac{\alpha}{\beta}}\sum_{n\in \mathbb{Z}}e^{\frac{2\pi i n}{\beta}},
\ee
where ${\bf R}$ is the roots of the gauge group.
Here we have summed over all the eigenmodes of $L_\tau$.
The factor $-\frac{1}{2}$ follows from both the consideration of the statistics and the fact that the self-dual complex for the vector multiplet is real.
To read the weights and the degeneracies of the representations for $\mathcal{H}$ action,
we need to expand this index in a power series of $e^{i\epsilon_{1,2}}$ but there are various ways of expansion
depending on whether $|e^{i\epsilon_{1,2}}|<1$ or $|e^{i\epsilon_{1,2}}|>1$.
It turns out \cite{Pestun:2007rz} that we should expand the index (\ref{self-dual-index2}) in positive powers
of $e^{i\epsilon_{1,2}}$ at the north pole,
while we should expand the index in negative powers of $e^{i\epsilon_{1,2}}$ at the south pole, or equivalently
we flip all the signs of the chemical potentials other than $\epsilon_{1},\,\epsilon_{2}$.
This argument also holds for the equivariant index of the hypermultiplets which we will compute shortly.
For the adjoint vector multiplet or hypermultiplets in the real representation, two indices from the north and south poles yield the same results since
the roots and the weight space of the gauge group are invariant under the sign flip.

Then the one-loop perturbative contribution at the south pole from the vector multiplet reads
\begin{align}\label{1-loopvec}
&I^{1-loop}_{vec,south}\\
&    \!\!\! =\!\!\prod_{n=-\infty}^{\infty}\prod_{n_1,n_2=0}^{\infty}\prod_{{\bf R}}\left[\frac{2\pi n}{\beta}+n_1\epsilon_1+n_2\epsilon_2-\frac{{\bf R}\cdot\alpha}{\beta}\right]^{\frac12}
        \left[\frac{2\pi n}{\beta}+(n_1+1)\epsilon_1+(n_2+1)\epsilon_2-\frac{{\bf R}\cdot\alpha}{\beta}\right]^{\frac12}  \cr
&   \!\!\! =\!\!\prod_{n_1,n_2=0}^{\infty}\!\!\prod_{{\bf R}}\sinh\left[\frac{(n_1\!+\!n_2)\gamma_1\!+\!(n_1\!-\!n_2)\gamma_2\!+\!i{\bf R}\cdot\alpha}{2}\right]^{\frac12}\!\!
\sinh\left[\frac{(n_1\!+\!n_2\!+\!2)\gamma_1\!+\!(n_1\!-\!n_2)\gamma_2\!+i{\bf R}\cdot\alpha}{2}\right]^{\frac12}\!. \nn
\end{align}
In the second line, we factor out the divergent constant which is independent of chemical potentials and set it to unity \cite{Aharony:2003sx}.
The contribution from the north pole differs from that of the south pole only by the signs of chemical potentials, which means that the contribution at the north can be obtained from complex conjugation of the the contribution at the south pole
\be
    I^{1-loop}_{vet,north}(\gamma_1,\gamma_2,\alpha) = I^{1-loop}_{vet,south}(\gamma_1,\gamma_2,-\alpha) = \left(I^{1-loop}_{vet,south}\right)^*.
\ee
It is more convenient to rewrite the one-loop result in terms of the single letter index as
\be
    \tilde{I}^{1-loop}_{vec} = x^{\epsilon_0}{\rm exp}\left[\sum_{n=1}^\infty \frac{1}{n}\tilde{f}_{vec}(x^n,y^n, n \alpha)\right]\ , \quad
    \tilde{f}_{vec}(x,y,\alpha) = -\frac{1+x^2}{(1-xy)(1-x/y)}\sum_{{\bf R}}e^{-i{\bf R}\cdot\alpha} \ , \quad
\ee
where $\epsilon_0$ is the Casimir energy and we can regularize it to be unity using the procedure in Appendix B.3 of \cite{Kim:2009wb}.
However, this is not the appropriate 1-loop result which we want to compute.
This is because the single letter index $\tilde{f}_{vec}$ includes the contribution corresponding to Haar measure on the gauge group manifold,
which we already factored out in front in the measure of the path integral $[d\alpha]$.
Therefore, we have to subtract Haar measure contribution $-\sum_{{\bf R}}e^{-i{\bf R}\cdot\alpha}$ from $\tilde{f}_{vec}$ and obtain the proper one-loop determinant for the vector multiplet
\be
    I^{1-loop}_{vec}\!&\!\!=\!\!&\ {\rm exp}\left[\sum_{n=1}^\infty \frac{1}{n}f_{vec}(x^n,y^n, n \alpha)\right] , \nn \\
    f_{vec}(x,y,\alpha) \!&\!\!=\!\!&\!\tilde{f}_{vec}+\sum_{{\bf R}}e^{-i{\bf R}\cdot\alpha} =-\frac{x(y+1/y)}{(1-xy)(1-x/y)}\sum_{{\bf R}}e^{-i{\bf R}\cdot\alpha} .
\ee
One may notice that the single letter index for the vector multiplet $\tilde{f}_{vec}$ is very similar to
the equivariant index of the self-dual complex (\ref{self-dual-index2}) if we ignore the $U(1)$ line bundle contribution.
This similarity can also be found in the hypermultiplet determinant. 

In a similar way we can compute the one-loop determinant for a matter hypermultiplet.
The bosonic and fermionic fields in the hypermultiplet are the sections of the spin bundles of positive and negative chiralities on $\mathbb{C}^2$ respectively.
The differential operator acting on these fields is the Dirac operator that is a map from the spin bundle of positive chirality to that of negative chirality,
\be\label{Dirac-complex}
    D_{Dirac} : \Omega^{(\frac{1}{2},0)} \rightarrow \Omega^{(0,\frac{1}{2})},
\ee
which implies that the field content of a hypermultiplet forms a Dirac complex associated with the operator $D_{Dirac}$.
It is straightforward to
compute the equivariant index of the Dirac complex (\ref{Dirac-complex})
with respect to $U(1)_{\epsilon_1}\!\times\! U(1)_{\epsilon_2}$ action using Atiyah-Singer index theorem.
The result is given by \cite{Gomis:2011pf}
\begin{align}
    {\rm ind}(D_{Dirac}) &=
    \frac{(e^{i(\epsilon_1+\epsilon_2)/2}+e^{-i(\epsilon_1+\epsilon_2)/2})-(e^{i(\epsilon_1-\epsilon_2)/2}+e^{-i(\epsilon_1-\epsilon_2)/2})}
    {(1-e^{i\epsilon_1})(1-e^{-i\epsilon_1})(1-e^{i\epsilon_2})(1-e^{-i\epsilon_2})} \cr
    &=\frac{e^{i(\epsilon_1+\epsilon_2)/2}}{(1-e^{i\epsilon_1})(1-e^{i\epsilon_2})}.
\end{align}

We now compute the equivariant index with $\mathcal{H}$ action from this result.
We again expand all the fields in the Dirac complex by the eigenmodes of $L_\tau$, and also consider the gauge and flavor symmetry acting on the fields.
For the hypermultiplet in the fundamental representation of the gauge group with $N_f$ flavor symmetry,
the equivariant index has the form
\begin{align}
    {\rm ind}(D_{mat})=\frac{e^{i(\epsilon_1+\epsilon_2)/2}}{(1-e^{i\epsilon_1})(1-e^{i\epsilon_2})}\sum_{{\bf w}\in{\bf W}}\sum_{i=1}^{N_f}e^{-i\frac{{\bf w}\cdot\alpha+m_i}{\beta}}\sum_{n\in \mathbb{Z}}e^{\frac{2\pi i n}{\beta}}.
\end{align}
Just as the vector multiplet case,  we perform a series expansion in positive power of $e^{i\epsilon_{1,2}}$ at the south pole,
and read off the one-loop determinant contributions of $N_f$ matter hypermultiplet at the south pole
\be\label{1-loophyp}
   I^{1-loop}_{mat,south} \!=\!\!\!&&\!\!\!\prod_{n=-\infty}^{\infty}\prod_{n_1,n_2=0}^{\infty}\prod_{{\bf w}\in{\bf W}}\prod_{i=1}^{N_f}\left[\frac{2\pi n}{\beta}+\big(n_1+\frac{1}{2}\big)\epsilon_1+\big(n_2+\frac{1}{2}\big)\epsilon_2-\frac{{\bf w}\cdot\alpha+ m_i}{\beta}\right]^{-1} \\
   \!\!\! =\!\!\!&&\!\!\!\prod_{n_1,n_2=0}^{\infty}\prod_{{\bf w}\in{\bf W}}\prod_{i=1}^{N_f}\sinh\left[\frac{(n_1+n_2+1)\gamma_1+(n_1-n_2)\gamma_2+i{\bf w}\cdot\alpha+i m_i}{2}\right]^{-1}\ . \nn
\ee
As mentioned earlier, the contribution from the north pole differs from that of the south pole only by the signs of chemical potentials $\alpha,m_i$.
Therefore, we see that two results are related to each other by complex conjugation
\be
    I^{1-loop}_{mat,north}(\gamma_1,\gamma_2,\alpha,m_i) = I^{1-loop}_{mat,south}(\gamma_1,\gamma_2,-\alpha,-m_i) = \left(I^{1-loop}_{mat,south}\right)^*.
\ee
Combining two pole contributions and rewriting them in terms of the single letter index, we find that the one-loop determinant for the fundamental hypermultiplet is given by
\be\label{one-loop-hyper}
    I^{1-loop}_{mat}\!&\!\! =\!\!&\!x^{-\epsilon_0}{\rm exp}\left[\sum_{n=1}^\infty \frac{1}{n}f_{mat}(x^n,y^n, n \alpha, n m)\right] , \nn \\
    f_{mat}(x,y,\alpha) \!&\!\! =\!\!&\! \frac{x}{(1-xy)(1-x/y)}\sum_{{\bf w}\in {\bf W}}\sum_{i=1}^{N_f}(e^{-i{\bf w}\cdot\alpha-im_i}+e^{i{\bf w}\cdot\alpha+i m_i}).
\ee
The Casimir energy $\epsilon_0$ is again regularized to be zero.
The result for the hypermultiplets in other representation $R$ can be obtained by
replacing the weight space ${\bf W}$
for the fundamental representation by the corresponding weight space ${\bf W}_R$.

One can also evaluate the one-loop determinant for the hypermultiplet by computing the single letter partition function.
In other words, the path integral for the hypermultiplet in the trivial background can be evaluated with the canonical action \eqref{matter-lagrangian} at the weak coupling limit.
This is possible because there is no zero modes in the hypermultiplet.
The single letter index is defined as a trace over the single letter operators and its derivatives saturating the BPS bound $\epsilon_0-2j_1-3R=0$ modulo the equation of motion.
 (The charges of the single letters in the hypermultiplet under the superconformal symmetry are listed in the Table \ref{tab:letters}.)
Since only the BPS operator $q^{A=1}\,(=\!q^{A=+})$ and the arbitrary number of derivatives $\partial_{+\pm}$ acting on it contribute to the letter index, one can easily read off the single letter index for the hypermultiplet
\be
    f_{matter} = {\rm tr\,}_{{\rm letters}}\left[(-1)^F x^{2(j_1+R)}y^{2j_2}\right] = \frac{x}{(1-xy)(1-x/y)}.
\ee
One then evaluates the single letter index with the holonomy $\alpha$ and the chemical potentials for the flavor symmetry, which indeed reproduces the same letter index in \eqref{one-loop-hyper} computed using the index theorem.

\begin{table}[t!]
$$
\begin{array}{c|c|c|c}
    \hline {\rm letter}& \epsilon_0 & (j_1,j_2) & R \\
    \hline q^A & \frac{3}{2} & (0,0) & \pm \frac{1}{2} \\
     \psi& 2 & (\pm \frac{1}{2},0)\oplus(0,\pm \frac{1}{2}) & 0 \\
    \hline \partial_\mu & 1 & (\pm\frac{1}{2},\pm \frac{1}{2})\oplus (0,0) & 0 \\ \hline
\end{array}
$$
\caption{The ``letters" in a hypermultiplet and the derivatives acting on them.}
\label{tab:letters}
\end{table}


\subsection{Instanton contribution}\label{instantioncontri}
We have shown in Section \ref{sec:localization} that the path integral of the superconformal index localizes on the space of
the instanton $F^+\!=\!0$ and the anti-instanton $F^-\!=\!0$ solutions at the south and north poles of the four-sphere, respectively.
Near one of the fixed points, the spacetime manifold looks like a product space ${S}^1\times\mathbb{R}^4$
and the path integral over the solution space of the instanton equation $F^+({\rm or}\,F^-)\!=\!0$
reduces to Nekrasov's instanton partition function \cite{Nekrasov:2002qd,Nekrasov:2003rj,Nakajima:2003pg} of 5d theory on a compact circle,
which has the meaning of the Witten index counting BPS instanton particles living on the 5d theory \cite{Kim:2011mv, Nekrasov:2002qd}.
The product of these 5d instanton partition functions (or the indices) from two fixed points will give the instanton contributions to the superconformal index.

It is known \cite{Nekrasov:2002qd,Nekrasov:2003rj} that the instanton partition function can be computed by putting the theory in the $\Omega$-background
and thus using the equivariant localization technique.
Or, equivalently, one can compute the Witten index of the 1d quantum mechanics on the instanton moduli space,
which can be understood as the Higgs branch of D0-brane worldvolume theory describing D0-D4 brane bound states, using the localization technique as in \cite{Kim:2011mv}.

5d instanton index is related to 4d Nekrasov's partition function by the dimensional reduction along the compact circle upon the suitable identification of the parameters in 4d partition function such as the $\Omega$-deformation parameters with the chemical potentials in 5d index.
This is possible because the our supercharge $Q$ near the fixed points is identical to the supercharges used in the computation of the 4d partition function.
Our strategy to obtain the 5d instanton index is to use
the known results of 4d instanton partition functions to convert them into 5d instanton index by carefully considering all KK-modes along the compact circle.
4d instanton partition functions for the classical groups that we are mainly interested in have been computed in various literature \cite{Nekrasov:2002qd,Nekrasov:2003rj,Nekrasov:2004vw,Shadchin:2005mx}.

In the presence of the classical Chern-Simons term, the instanton carry nonzero electric charge and thus the moduli space dynamics should be modified.
The Chern-Simons term in 5d theory induces the Chern-Simons term of the Lagrangian in the 1d instanton quantum mechanics \cite{Kim:2008kn,Collie:2008vc}
\be\label{1d-CS}
    \mathcal{L}_{CS}^{1d} = \kappa\int dt\ {\rm tr}(A_t-\phi),
\ee
where $A_t$ and $\phi$ are the gauge field and the scalar component of the vector multiplet in the adjoint representation of the gauge group $U(k)$ in the 1d quantum mechanics, respectively.
This term preserves half of the supersymmetries in the instanton moduli space which of course includes our supercharge $Q$. 
This allows us to recycle the localization technique used for the case of no Chern-Simons term, but we need to take into account the classical contribution from (\ref{1d-CS}).

Introducing the fugacity $q$ for labeling the instanton number, we get the instanton contribution from the south pole
\be
    I_{south}^{inst}(\gamma_1,\gamma_2,\alpha,m_i,q) = \sum_{k=0}^\infty q^k I^{k}(\gamma_1,\gamma_2,\alpha,m_i),
\ee
where $I^k$ is the instanton index with charge $k$ and $I^{k=0}=1$.
As the anti-instanton index can be obtained by the sign flip of the chemical potentials $\alpha,m_i$ of the instanton index, we find the instanton contribution from the north pole as
\be
    I_{north}^{inst}(\gamma_1,\gamma_2,\alpha,m_i,q) =  \sum_{k=0}^\infty q^{-k} I^{k}(\gamma_1,\gamma_2,-\alpha,-m_i).
\ee
Here the instanton sum is given in negative power of $q$ because the instanton charge for the anti-instantons is negative. $I_{north}^{inst}(\gamma_1,\gamma_2,\alpha,m_i,q)$ is basically the complex conjugation of  $I_{south}^{inst}(\gamma_1,\gamma_2,\alpha,m_i,q)$
where the complex conjugation exchanges $(\alpha,m_i,q)$ to $(-\alpha,-m_i,q^{-1})$.


\subsection{$U(N)$ gauge theories}
As a simple example that one can apply the localization technique, we compute the superconformal index for $U(N)$ gauge theories with $N_f$ fundamental flavor hypermultiplets.
Classical Chern-Simons term exists for all $N$ unlike $SU(N)$ for which it exists when $N\ge 3$.
We include the Chern-Simons term of level $\kappa$ in superconformal index calculations. After localization the Chern-Simons term does not contribute to perturbative part, but contributes nonperturbative part as explained.
The gauge invariance demands the quantization condition $\kappa+{N_f}/{2} \in \mathbb{Z}$.

The the superconformal index takes the form
\begin{align}
    I_{U(\!N\!)_\kappa}^{N_f} (x,y,m_i,q,\kappa) \!= \int[d\alpha]{\rm PE}\Big[f_{mat}(x,y,e^{i\alpha},e^{im})\!+\!f_{vec}(x,y,e^{i\alpha})\Big]\Big|I^{inst}(x,y,e^{i\alpha},e^{im},q,\kappa)\Big|^2
\end{align}
where the invariant Haar measure is given by
\begin{align}
    \left[d\alpha\right]=\frac{1}{N!}\left[\prod_{i=1}^N\frac{d\alpha_i}{2\pi}\right]\prod_{i<j}\left[2\sin\left(\frac{\alpha_i-\alpha_j}{2}\right)\right]^2.
\end{align}
The PE in the integrand is
the Plethystic exponential
\be
    {\rm PE}\Big[f(\cdot)\Big] = {\rm exp}\Big[\sum_{n=1}^\infty \frac{1}{n} f(\cdot ^n)\Big] ,
\ee
which is used to obtain the multi particle index from a free single particle index $f(\cdot)$.
Here $f_{vec}$ and $f_{mat}$ are the single particle indices for the vector and hypermultiplets respectively 
\begin{align}
    f_{vec} &=-\frac{x(y\!+\!1/y)}{(1\!-\!xy)(1\!-\!x/y)}\sum_{i, j}^Ne^{-i\alpha_i+i\alpha_j},\cr
    f_{mat} &= \frac{x}{(1\!-\!xy)(1\!-\!x/y)}\sum_{i=1}^N\sum_{l=1}^{N_f}\left(e^{-i\alpha_i-im_l}\!+\!e^{i\alpha_i+i m_l}\right) .
\end{align}
The instanton index $I^{inst}$ for $U(N)$ gauge group can be read off from \cite{Nekrasov:2004vw,Shadchin:2005mx} in which
the 4d Nekrasov's partition functions are written as the contour integral formula
over the Cartan subalgebra of $U(k)$ gauge group of the instanton moduli space.
As mentioned in the previous section, the 4d instanton partition function
can be uplifted to 5d instanton index by taking into account the full KK modes along the time circle.
In addition, we need to insert the classical Chern-Simons contribution $e^{\kappa\phi_I}$
in the matrix integral 
\cite{Tachikawa:2004ur}.
This is from the induced Chern-Simons term on the instanton moduli space.
For $k$ instantons, the integral formula of the $U(N)$ instanton index is given by
\begin{equation}\label{contour-integral}
    \hspace{-.3cm}I^{k} =\frac{(2i)^{k(N_f-2N-1)}}{k!}\oint \prod_{I=1}^k\left(\frac{d\phi_I}{2\pi}\frac{e^{i\kappa\phi_I}\prod_{l=1}^{N_f}\sin\frac{\phi_I+m_l}{2}}
    {\prod_{i=1}^N\sin\frac{\phi_I-\alpha_i-i\gamma_1}{2}\sin\frac{-\phi_I+\alpha_i-i\gamma_1}{2}}\right)
    \frac{\prod_{I\neq J}\sin\frac{\phi_{IJ}}{2}\prod_{I,J}\sin\frac{\phi_{IJ}-2i\gamma_1}{2}}{\prod_{I,J}\sin\frac{\phi_{IJ}-i\gamma_1-i\gamma_2}{2}\sin\frac{\phi_{IJ}-i\gamma_1+i\gamma_2}{2}}
\end{equation}
where $\phi_I$ ($I=1,\cdots, k$) is the $U(k)$ gauge transformation parameter that takes a value in $U(1)^k$ Cartan subalgebra, and $\phi_{IJ} = \phi_I -\phi_J$. The $U(k)$ gauge invariance is attained by integration over $\phi_I$.
We briefly comment on the sine factors in \eqref{contour-integral}: the sine factors in the numerator in the bracket come from the fermion zero modes of $N_f$ fundamental hypermultiplets,
and the rest of sine factors is the contributions from the ADHM data in the instanton moduli space of the pure $U(N)$ Yang-Mills theory.

Using the residue theorem this integral can be explicitly evaluated.
Let us define $z_I =e^{i\phi_I}$ and consider all poles enclosed by the contour around unit circles $|z_I| <1$,
with the assumption $|e^{-\gamma_1}|\!\ll\! |e^{-\gamma_2}|\! \ll \! 1$.
Note that irrelevant poles at $z_I=0$ can appear when $N_f\ge 2N$.
This pole is unphysical as it corresponds to $\phi_I = i\infty$.
The contributions from the poles $z_I=0$ should be excluded from the result.
In addition, 
more irrelevant poles apart from $z_I=0$ appear when we introduce other types of hypermultiplets, for example, an adjoint hypermultiplet which we will compute shortly. 
The contributions from such 
poles must also be excluded otherwise we will get the wrong index.

For $U(k)$ gauge theories, we can avoid this notorious pole problems with the help of 
the localization on the instanton quantum mechanics as noted in 
\cite{Kim:2011mv}.
The path integral of the 1d quantum mechanics on the instanton moduli space can be completely localized
around the set of classical saddle points by turning on the FI parameter $\zeta$ as well as other chemical potentials.
At the saddle points, the $U(k)$ gauge symmetry is completely broken and
there is no remaining gauge transformation parameter to be integrated.
In this way,
we can compute the path integral without the contour description.
See \cite{Kim:2011mv} for more details.
Though two prescriptions for the instanton index may look different, they give the same result.
Moreover, one can find the one-to-one map between the physical poles in the contour integral
and the classical fixed points of the path integral of the 1d quantum mechanics,
which also supports the fact that the poles at $z_I=0$ as well as poles from hypermultiplet contributions are
irrelevant since there is no corresponding fixed point in the second prescription.

The poles of the contour integral can be classified by N-colored Young diagram $\{Y_1,Y_2,\cdots,Y_N\}$ \cite{Nekrasov:2002qd,Nekrasov:2003rj}.
Each Young diagram $Y_i$ contains $k_i$ boxes and the total number of boxes in Young diagrams is $k=\sum_ik_i$.
We denote the position in the $i$-th Young diagram by $s=(m,n)(\in Y_i)$, where $m$ and $n$ are the vertical and horizontal position from the upper-left corner of the Young diagram $Y_i$, respectively.
The corresponding pole is given by
\be
    \phi(s) = \alpha_i+i\gamma_1+i(m-1)(\gamma_1+\gamma_2)+i(n-1)(\gamma_1-\gamma_2) \ .
\ee
For a given colored Young diagram, we can fully evaluate the contour integral and write it in the simple form \cite{Nekrasov:2002qd,Nekrasov:2003rj,Bruzzo:2002xf}
\be
    I_{\{Y_1,Y_2,\cdots,Y_N\}} = \prod_{i}^N\prod_{s\in Y_i}\frac{e^{i\kappa\phi(s)}\prod_{l=1}^{N_f}2i\sin\frac{\phi(s)+m_l}{2}}{\prod_{j=1}^N(2i)^2\sin\frac{E_{ij}}{2}\sin\frac{E_{ij}+2i\gamma_1}{2}} \ ,
\ee
where
\be
    E_{ij}=\alpha_i-\alpha_j +i(\gamma_1+\gamma_2)h_i(s) -i(\gamma_1-\gamma_2)(v_j(s)+1) \ .
\ee
Here $h_i(s)$ and $v_j(s)$ are 
 the distance from $s \in Y_i$ to the right end of the $i$-th Young diagram and the bottom end of the $j$-th Young diagram respectively.
To obtain the index for $k$ instantons, we need to sum over all possible Young diagram configurations with total $k$ boxes.
Then the full instanton index is given by
\be
    I^{inst}_{U(N)} = \sum_{k=0}^\infty q^kI^k\ , \quad I^k = \sum_Y I_{\{Y_i(k_i)\}}\ , \quad \sum_i k_i=k \ ,
\ee
where $I^{k=0}=1$.
This is the instanton index at the south pole on ${S}^4$.
The complex conjugation of the instanton index gives the anti-instanton index at the north pole.

Note that the Chern-Simons term provides the coupling of instantons with the holonomy variables $\alpha_i$,
which reflects the fact that the instantons carry  $U(N)$ electric charges.
As discussed in Section \ref{sec:5dSCFT}, the Chern-Simons term can be induced by integrating out the massive fundamental hypermultiplets.
This can be seen from our instanton index by taking large mass limit $m_l \rightarrow i\infty$ for $n_f$ hypermultiplet.
Provided that the divergence arising from the large mass limit is regularized to unity, one sees an additional Chern-Simons term
is induced and the Chern-Simons level is shifted as
\be
    \kappa_{eff} = \kappa -\frac{n_f}{2}\ ,
\ee
which agrees with the expected result from \cite{Witten:1996qb}.

Now we consider the $U(N)$ gauge theory with one adjoint hypermultiplet (corresponding to so-called $\mathcal{N}=2^*$ theory in 4d).
The single particle index for the matter hypermultiplet in the adjoint representation of $U(N)$ is given by
\be
    f_{mat}=\frac{x}{(1\!-\!xy)(1\!-\!x/y)}(e^{im}+e^{-im})\sum_{i, j}e^{-i\alpha_i+i\alpha_j} \ ,
\ee
while that for the vector multiplet remains the same as before.
Here $m$ is the chemical potential of $U(1)$ flavor symmetry which is enhanced to $SU(2)$ as the hypermultiplet is in the real adjoint representation of the gauge group.
In 4d limit, this chemical potential becomes the mass parameter of the adjoint hypermultiplet.

The 5d instanton index for $\mathcal{N}=2^*$ theory was computed in \cite{Kim:2011mv}.
Adding the Chern-Simons term to the result of \cite{Kim:2011mv} is straightforward and we get   \footnote{Chemical potentials here and those in \cite{Kim:2011mv} are related as $(\gamma_1,\gamma_2,\alpha_i,m)_{{\rm here}} \rightarrow (i\gamma_R,-i\gamma_1,i\alpha_i,\gamma_2)_{{\rm }}$ of \cite{Kim:2011mv}.}
\be\label{instanton-adj}
    I_{\{Y_1,Y_2,\cdots,Y_N\}} = \prod_{i,j=1}^N\prod_{s\in Y_i}e^{\kappa\phi(s)}\frac{\sin\frac{E_{ij}+i(\gamma_1+m)}{2}\sin\frac{E_{ij}+i(\gamma_1-m)}{2}}{\sin\frac{E_{ij}}{2}\sin\frac{E_{ij}+2i\gamma_1}{2}}.
\ee
Two sine factors in the numerator correspond to the adjoint hypermultiplet.
This result is derived directly from the contour integral formula, Eq.\! \!(5.15) in \cite{Kim:2011mv}.
Note that the adjoint hypermultiplet introduces many irrelevant poles that should be discarded from the contour.
 The instanton index for $U(N)$ theory with an adjoint hypermultiplet, (\ref{instanton-adj}), is obtained
by considering only the residues for the relevant poles. 


\subsection{$Sp(N)$ gauge theories}\label{sec:sp-gaugegr}
We now compute the index for  $Sp(N)$ gauge theories.
These theories are of our main interest as they exhibit intriguing global symmetry enhancements at the conformal fixed point. Firstly, we consider $Sp(N)$ gauge theory with $N_f$ fundamental flavors and
later add one additional hypermultiplet in the antisymmetric representation of $Sp(N)$
which is required to see the global symmetry enhancement when $N\ge 2$.
In this case, there is no classical Chern-Simons terms because the symmetric structure constant $d^{abc}$
is identically zero.

Considering the perturbative part and the instanton part all together we find the superconformal index for the $Sp(N)$ gauge theory
with $N_f$ flavors as follows
\be\label{I4spN}
    I_{Sp(\!N\!)}^{N_f} (x,y,m_i,q) \!\!&\!\!=\!\!&\!\! \int[d\alpha]\ {\rm PE}\Big[f_{mat}(x,y,e^{i\alpha},e^{im})+f_{vec}(x,y,e^{i\alpha})\Big]\Big|I^{inst}(x,y,e^{i\alpha},e^{im},q)\Big|^2 \ , \nn \\
    \left[d\alpha\right]\!\!&\!\!=\!\!&\!\!\frac{2^N}{N!}\left[\prod_{i=1}^N\frac{d\alpha_i}{2\pi}\sin^2\alpha_i\right]\prod_{i<j}\left[2\sin\left(\frac{\alpha_i-\alpha_j}{2}\right)\right]^2
    \left[2\sin\left(\frac{\alpha_i+\alpha_j}{2}\right)\right]^2 \ ,
\ee
where the single letter indices for the vector and hypermultiplets are given by
\be
    f_{vec} \!\!&\!\!=\!\!&\!\! -\frac{x(y\!+\!1/y)}{(1\!-\!xy)(1\!-\!x/y)}\left[\sum_{i<j}^N\left(e^{-i\alpha_i-i\alpha_j}\!+\!e^{-i\alpha_i+i\alpha_j}\!+\!e^{i\alpha_i-i\alpha_j}\!+\!e^{i\alpha_i+i\alpha_j}\right)
    \!+\!\sum_{i=1}^N\left(e^{-2i\alpha_i}\!+\!e^{2i\alpha_i}\right)\!+\!N\right] \ , \nn \\
    f_{mat} \!\!&\!\!=\!\!&\!\! \frac{x}{(1\!-\!xy)(1\!-\!x/y)}\sum_{i=1}^N\sum_{l=1}^{N_f}\left(e^{-i\alpha_i-im_l}\!+\!e^{i\alpha_i-i m_l}\!+\!e^{-i\alpha_i+i m_l}\!+\!e^{i\alpha_i+i m_l}\right)\ ,
\ee
after taking into account the root and fundamental weight of the Lie algebra of $Sp(N)$ gauge symmetry.

For the instanton part $I^{inst}$, we borrow the result of \cite{Nekrasov:2004vw,Shadchin:2005mx} in which
the 4d  Nekrasov's partition functions for $Sp(N)$ gauge theories was computed. These 4d instanton partition functions are obtained from  the ADHM construction of the instanton moduli spaces. It was noticed \cite{Nekrasov:2004vw} that the dual gauge group $G_D$ on the instanton moduli space of $Sp(N)$ gauge theory is $O(k)$ whereas the instanton calculus is done only for $SO(k)$ dual gauge group in \cite{Nekrasov:2004vw,Shadchin:2005mx}. $O(k)$ group is different from $SO(k)$ group by $\mathbb{Z}_2$ factor and the consideration of this difference is crucial to obtain the correct instanton index.

One needs to carefully consider the $O(k)$ group action on the instanton moduli space.
The $O(k)$ group has two components. One component contains the group elements whose determinants are $+1$ and the other contains the elements of the determinant $-1$. 
The former component forms a group itself and is called $SO(k)$ group, which we denote by  $O(k)_+$, and the latter component does not form a group itself, which we denote by $O(k)_-$ for convenience.
The torus action of the dual gauge group is generated by the following elements: for $O(k)_+$ ,
\be\label{gauge-parameter1}
    e^{i\phi_+} = \left\{
    \begin{array}{l}{\rm diag}(e^{i\sigma_2\phi_1},\cdots,e^{i\sigma_2\phi_{n}})\ \ {\rm for \ even} \ k\ ,  \\
    {\rm diag}(e^{i\sigma_2\phi_1},\cdots,e^{i\sigma_2\phi_n},1)\ \ {\rm for \ odd} \ k ,\end{array} \right.
\ee
and for $O(k)_-$,
\be\label{gauge-parameter2}
    e^{i\phi_-} = \left\{
    \begin{array}{l}{\rm diag}(e^{i\sigma_2\phi_1},\cdots,e^{i\sigma_2\phi_{n-1}},\sigma_3)\ \ {\rm for \ even} \ k\ ,  \\
    {\rm diag}(e^{i\sigma_2\phi_1},\cdots,e^{i\sigma_2\phi_n},-1)\ \ {\rm for \ odd} \ k, \end{array} \right.
\ee
where $k=2n+\chi\, (\chi =0\ {\rm or}\ 1)$.%
\footnote{For $-1$ and $\sigma_3$ that appear in the $e^{i\phi_-}$ action, they should be understood as the operators, $e^{i\pi}$ and  $\sigma_3={\rm diag}(1,e^{i\pi})$ acting on representations. For example, for $e^{i\pi}\in O(1)$ acting on the fundamental representation ${\bf f}$, $e^{i\pi}{\bf f}^n = (-1)^n{\bf f}^n$.} 
As there are two disjoint torus actions for $O(k)$ gauge group,
there are two corresponding disjoint contour integral formulas for the instanton index: $I^k_+$ and $I^k_-$ which come from the torus actions $e^{i\phi_+}$ and $e^{i\phi-}$, respectively.
The correct way of imposing $O(k)$ gauge singlet constraint is to take an average of these indices $I_+$ and $I_-$ after performing contour integration.

Let us first obtain the contour integral representation of the instanton index with $O(k)_+=SO(k)$ dual gauge group.
Taking into account the torus action on the ADHM data, we obtain
\be\label{integral-formula1}
    I^k_{+}\!\!&\!\!\!=\!\!\!&\!\!(2i)^{k(N_f-2N-2)-n}i^{n+2\chi}\oint[d\phi]\left[\frac{\prod_{l=1}^{N_f}\sin\frac{m_l}{2}}{\sinh\frac{\gamma_1\pm\gamma_2}{2}
    \prod_{i=1}^N\sin\frac{i\gamma_1\pm\alpha_i}{2}}
    \prod_{I=1}^n\frac{\sin(\frac{\phi_I\pm2i\gamma_1}{2})}
    {\sin\frac{\phi_I\pm i\gamma_1\pm i\gamma_2}{2}}\right]^\chi \nn \\
    &&\times \prod_{I=1}^n\left[\frac{\sinh\gamma_1}
    {\sinh\frac{\gamma_1\pm\gamma_2}{2}\sin\frac{2\phi_I\pm i\gamma_1\pm i\gamma_2}{2}}\frac{\prod_{l=1}^{N_f}\sin\frac{m_l\pm\phi_I}{2}}{\prod_{i=1}^N \sin\frac{\phi_I\pm\alpha_i\pm i\gamma_1}{2}}\right]
    \prod_{I< J}^n\left[\frac{\sin\frac{\phi_I\pm\phi_J\pm2i\gamma_1}{2}}
    {\sin\frac{\phi_I\pm\phi_J\pm i\gamma_1\pm i\gamma_2}{2}}\right],\qquad
\ee
where $[d\phi]$ is the Haar measure for $SO(k)$ and we used a succinct notation, $\sin(a \pm b)\equiv \sin(a + b)\sin(a - b)$ and so on.\footnote{For example,
$\sin\frac{\phi_I\pm\alpha_i\pm i\gamma_1}{2}\equiv
\sin\frac{\phi_I+\alpha_i+ i\gamma_1}{2}\sin\frac{\phi_I+\alpha_i- i\gamma_1}{2}\sin\frac{\phi_I-\alpha_i+ i\gamma_1}{2}
\sin\frac{\phi_I-\alpha_i- i\gamma_1}{2}$.
}
This is the 5d version of Nekrasov's partition function for $Sp(N)$ gauge theories with $N_f$ matter hypermultiplets (the corresponding 4d partition function was computed in \cite{Nekrasov:2004vw,Marino:2004cn,Fucito:2004gi}).
We can decompose this formula into the contributions from the vector multiplet and the $N_f$ fundamental hypermultiplets separately.
The fundamental hypermultiplet contribution is
\be
    z_{fund}^{N_f} = \left[\prod_{l=1}^{N_f}2i\sin\frac{m_l}{2}\right]^\chi\prod_{I=1}^n\prod_{l=1}^{N_f}2i\sin\frac{m_l\pm \phi_I}{2}
\ee
which comes from the fermion zero modes in the fundamental representation of $O(k)$.
Here we have considered the mass shift $m_l\rightarrow m_l+i\gamma_1$ in the hypermultiplet contribution which was first noticed in \cite{Okuda:2010ke}.
The remaining factors are the vector multiplet contribution.

The integrations are taken over the $SO(k)$ algebra elements $\phi_I$.
As in the previous $U(N)$ instanton case, we define $z_I=e^{i\phi_I}$ and take contours around unit circles.
We assume $|e^{-\gamma_1}|\!\ll\!|e^{-i\gamma_2}|\!\ll\!1$ and only keep the residues from the poles inside the unit circles on $z_I$ planes.
Then it provides the clear pole prescription for this contour integral.
The relevant poles for $z_I$ are located at
\be
    z_I=e^{-\gamma_1\pm i\alpha_i}\,,\ e^{-\gamma_1\pm\gamma_2}\, , \ e^{\frac{-\gamma_1\pm\gamma_2}{2}} \, ,\ -e^{\frac{-\gamma_1\pm\gamma_2}{2}} ,
\ee
which are from the denominators of the first and second brackets in \eqref{integral-formula1}.
The poles from the last bracket are determined by the relative size of radii of $|z_J|$ and $|e^{\pm\gamma_1\pm\gamma_2}|$.

There also exist unphysical poles at $z_I=0$ when $N_f \ge 6$.
The residues from these irrelevant poles
should not be considered in our computation.
We will subtract the contributions from the residues of the irrelevant poles from our instanton index.
However it turns out that the naive subtraction of the irrelevant pole contributions seems not to give sensible answer in this case.
We will explicitly evaluate these integrals for lower $k$'s in the next section
and discuss subtleties arising from irrelevant poles.

Now we turn to the instanton index with the $O(k)_-$ torus action.
As the torus actions for odd $k$ and even $k$ are different, we have to treat them separately as shown in Appendix \ref{App:ECc}.
For odd $k$, the contour integral formula of the instanton index with $O(k)_-$ is
\be\label{integral-formula2}
    I_{-}^{k:odd}\!\!&\!\!\!=\!\!\!&\!\!\frac{(2i)^{k(N_f-2N-2)-n}}{\,i^{N_f-2N-n-2}}\oint[d\phi] \left[\frac{\prod_{l=1}^{N_f}\cos\frac{m_l}{2}}{\sinh\frac{\gamma_1\pm\gamma_2}{2}
    \prod_{i=1}^N\cos\frac{i\gamma_1\pm\alpha_i}{2}}
    \prod_{I=1}^n\frac{\cos(\frac{\phi_I\pm2i\gamma_1}{2})}
    {\cos\frac{\phi_I\pm i\gamma_1\pm i\gamma_2}{2}}\right] \nn \\
    &&\times \prod_{I=1}^n\left[\frac{\sinh\gamma_1}
    {\sinh\frac{\gamma_1\pm\gamma_2}{2}\sin\frac{2\phi_I\pm i\gamma_1\pm i\gamma_2}{2}}
    \frac{\prod_{l=1}^{N_f}\sin\frac{m_l\pm\phi_I}{2}}{\prod_{i=1}^N \sin\frac{\phi_I\pm\alpha_i\pm i\gamma_1}{2}}\right]
    \prod_{I< J}^n\left[\frac{\sin\frac{\phi_I\pm\phi_J\pm2i\gamma_1}{2}}
    {\sin\frac{\phi_I\pm\phi_J\pm i\gamma_1\pm i\gamma_2}{2}}\right] \ ,\qquad
\ee
and the formula for even $k$ is
\be\label{integral-formula3}
    \hspace{-.6cm}I_{-}^{k:even}\!\!&\!\!\!=\!\!\!&\!\!(2i)^{(k-1)(N_f-2N)-\frac{5}{2}k}i^{n+4}\oint[d\phi]
    \left[\frac{\cosh\gamma_1}{\cosh\frac{\gamma_1\pm\gamma_2}{2}\sinh^2\frac{\gamma_1\pm\gamma_2}{2}}\frac{\prod_{l=1}^{N_f}\sin m_l}{ \prod_{i=1}^N \sin(i\gamma_1\!\pm\!\alpha_i)}\right]
      \\
     &&\times
    \prod^{n-1}_{I=1}
    \!\left[
    \frac{\sinh\gamma_1\sin(\phi_I\!\pm\!2i\gamma_1)}{\sinh\frac{\gamma_1\pm\gamma_2}{2}\sin\frac{2\phi_I\pm i\gamma_1\pm i\gamma_2}{2}\sin(\phi_I\!\pm \!i\gamma_1\!\pm \!i\gamma_2)}
    \frac{\prod_{l=1}^{N_f}\sin\frac{m_l\pm\phi_I}{2}}{\prod_{i=1}^N\sin\frac{\phi_I\pm\alpha_i\pm i\gamma_1}{2}} \right]
    \prod_{I< J}^{n-1}\!\left[\frac{\sin\frac{\phi_I\pm\phi_J\pm2i\gamma_1}{2}}
    {\sin\frac{\phi_I\pm\phi_J\pm i\gamma_1\pm i\gamma_2}{2}}\right]. \nn
\ee
Here $[d\phi]$'s denote the Haar measures for $O(k)_-$ whose explicit expressions are listed in Appendix E.
We can evaluate these contour integrations using the pole prescription discussed above.
The physical poles whose residues give nontrivial contributions to the instanton index appear at
\be
    z_I=e^{-\gamma_1\pm i\alpha_i}\, ,\ -e^{-\gamma_1\pm\gamma_2}\, ,\ e^{\frac{-\gamma_1\pm\gamma_2}{2}} \, ,\ -e^{\frac{-\gamma_1\pm\gamma_2}{2}}\ ,
\ee
for odd $k$ and
\be
    z_I=e^{-\gamma_1\pm i\alpha_i}\, ,\ -e^{-\gamma_1\pm\gamma_2}\, ,\  e^{-\gamma_1\pm\gamma_2} \,, \ e^{\frac{-\gamma_1\pm\gamma_2}{2}} \, ,\ -e^{\frac{-\gamma_1\pm\gamma_2}{2}}\ ,
\ee
for even $k$, respectively.
We note that the poles from $\sin\frac{\phi_I\pm\phi_J\pm i\gamma_1\pm i\gamma_2}{2}$ factors in the denominator must be chosen
if they are inside unit circles $|z_I| < 1$.

Unlike $U(N)$ gauge theories, there is neither proper Young diagram correspondence for the physical poles in the contour integration
nor closed form of the index after performing all the contour integral.
We have to evaluate the contour integrations case by case using the pole prescription given above.
Evaluating all the contour integrations the final instanton index contribution at the south pole is given by
\be
    I^{inst}_{Sp(N)} = \sum_{k=0}^\infty q^k I^k \ , \quad I^k=\frac{1}{2}\left[I^k_++I^k_-\right].
\ee
The anti-instanton contribution at the north pole can be obtained from this instanton index by complex conjugation which
flips the sign of chemical potentials, $\alpha_i\rightarrow -\alpha_i$ and $m_l\rightarrow -m_l$, and reverses $q$ to $q^{-1}$.

Finally let us introduce the antisymmetric hypermultiplet of $Sp(N)$.
The single letter index of the matter hypermultiplets in the perturbative part is modified by this antisymmetric hypermultiplet contribution which is given by
\be
     f_{mat}^{asym} \!\!&\!\!=\!\!&\!\! \frac{x}{(1\!-\!xy)(1\!-\!x/y)}(e^{im}+e^{-im})\!\!\left[\sum_{i<j}^N\left(e^{-i\alpha_i-i\alpha_j}\!+\!e^{-i\alpha_i+i\alpha_j}\!+\!e^{i\alpha_i-i\alpha_j}\!+\!e^{i\alpha_i+i\alpha_j}\right)+N\right],\qquad\quad
\ee
where $m$ is the chemical potential for $Sp(1)$ global symmetry acting on the antisymmetric matter.

The introduction of the antisymmetric hypermultiplet modifies the field content of the instanton moduli space or the instanton quantum mechanics.
It provides, for example, four $k\times k$ symmetric bosonic fields and their superpartners describing the positions of D0-branes transverse to the D4-branes (but on D8-branes).
The field content of the D0-brane quantum mechanics with D4-branes on the Type I$'$ system is listed in \cite{Douglas:1996uz, Aharony:1997pm} and in Appendix \ref{instanton-QM}.
These additional zero modes contribute to the instanton moduli space integral.
The equivariant index for the antisymmetric hypermultiplet can be read off from symmetry properties of these zero modes,
or it can also be read off from Eq.(5.14) in \cite{Shadchin:2005mx}.
With the $SO(k)$ dual gauge group, the additional terms induced by the antisymmetric hypermultiplet are
\begin{equation}\label{formula-anti1}
    \hspace{-.8cm}z_{asym,+}\!\!=\!(2i)^{2k(N\!-\!1)}\!\!\left[\frac{\prod_{i=1}^N\sin\frac{m\pm\alpha_i}{2}}{\sin\frac{m\pm i\gamma_1}{2}}
    \prod_{I=1}^n\!\frac{\sin\frac{\phi_I\pm i\gamma_2\pm m}{2}}{\sin\frac{\phi_I\pm i\gamma_1\pm m}{2}}\right]^\chi
    \prod_{I=1}^n\!\left[\frac{\sin\frac{m\pm i\gamma_2}{2}\prod_{i=1}^N\sin\frac{\phi_I\pm\alpha_i \pm m}{2}}{\sin\frac{m\pm i\gamma_1}{2}\sin\frac{2\phi_I\pm i\gamma_1 \pm m}{2}}\right]
    \!\prod_{I<J}^n\!\frac{\sin\frac{\phi_I\pm\phi_J\pm i\gamma_2\pm m}{2}}{\sin\frac{\phi_I\pm\phi_J\pm i\gamma_1\pm m}{2}}\quad
\end{equation}
The supplementary terms with $O(k)_-$ are
\begin{equation}\label{formula-anti2}
   \hspace{-1cm} z_{asym,-}^{k:odd}\!\!=\!\frac{(2i)^{2k(N-1)}}{i^{2N}}\!\!\left[\frac{\prod_{i=1}^N\cos\frac{m\pm\alpha_i}{2}}{\sin\frac{m\pm i\gamma_1}{2}}
    \prod_{I=1}^n\!\frac{\cos\frac{\phi_I\pm i\gamma_2\pm m}{2}}{\cos\frac{\phi_I\pm i\gamma_1\pm m}{2}}\right]
    \prod_{I=1}^n\!\left[\frac{\sin\frac{m\pm i\gamma_2}{2}\prod_{i=1}^N\sin\frac{\phi_I\pm\alpha_i \pm m}{2}}{\sin\frac{m\pm i\gamma_1}{2}\sin\frac{2\phi_I\pm i\gamma_1 \pm m}{2}}\right]
    \!\prod_{I<J}^n\!\frac{\sin\frac{\phi_I\pm\phi_J\pm i\gamma_2\pm m}{2}}{\sin\frac{\phi_I\pm\phi_J\pm i\gamma_1\pm m}{2}}\quad
\end{equation}
for odd $k$ and
\begin{equation}\label{formula-anti3}
   \hspace{-1.2cm} z_{asym,-}^{k:even}\!\!=\!(2i)^{2(kN\!-\!k\!-\!N)}\!\frac{\cos\!\frac{m\pm i\gamma_2}{2}\!\prod_{i=1}^N\!\sin(m\!\pm\!\alpha_i)}{\cos\frac{m\pm i\gamma_1}{2}\sin^2\left(\frac{i\gamma_1 \pm m}{2}\right)}
   \!
    \prod_{I=1}^{n-1}\!\!\left[\!\frac{\sin\!\frac{m\pm i\gamma_2}{2}\sin(\!\phi_I\!\pm \!i\gamma_2\!\pm\! m\!)\!\prod_{i=1}^N\!\sin\!\frac{\phi_I\pm\alpha_i \pm m}{2}}{\sin\frac{m\pm i\gamma_1}{2}\sin\frac{2\phi_I\pm i\gamma_1 \pm m}{2}\sin(\phi_I\!\pm\! i\gamma_1\!\pm\! m)}\!\right]
    \!\!\prod_{I<J}^{n-1}\!\frac{\sin\!\frac{\phi_I\pm\phi_J\pm i\gamma_2\pm m}{2}}{\sin\!\frac{\phi_I\pm\phi_J\pm i\gamma_1\pm m}{2}}
\end{equation}
for even $k$, respectively.
The contour integral of the full instanton index including the antisymmetric hypermultiplet contribution should be evaluated with these $z_{asym}$ terms
in the integrand as well as the vector and $N_f$ fundamental hypermultiplet terms.
It seems that these additional factors provide extra poles apart from the poles from the vector multiplet factors.
However, as we saw from the contour integration for the instanton index of $\mathcal{N}=2^*$ $U(N)$ gauge theory, these extra poles are all irrelevant.
New poles from the antisymmetric matter factors $z_{asym}$ are always irrelevant poles and thus their contribution should not be included in the instanton index computation.
However, as for theories without the antisymmetric hyper matter, we have encountered an obstacle that the naive elimination of these irrelevant contributions does not seem to yield a sensible answer.
One needs to develop an appropriate prescription for this.

Let us briefly comment about the role of the antisymmetric hypermultiplet.
The scalar field in the antisymmetric hypermultiplet represents the fluctuation of a D4-brane along the transverse directions to its worldvolume on the orientifold plane.
For $Sp(1)$, the antisymmetric representation is trivial, so it decouples from the gauge theory and we expect it does not affect the instanton dynamics.
However we can see that if we keep the antisymmetric hyper even though it decouples from the theory, then its instanton index contribution becomes non-trivial. This would imply that
the instanton index seems to contain extra information when the antisymmetric hypermultiplet is involved.
It turns out that it captures the D0-D8 brane bound state information. 

As there is no integral variable in the integral formulas for $k=1$, we do not need to consider the subtleties arising from the pole description.
Thus the one instanton results from the above indices give the correct result without pole ambiguity for any $N$ and $N_f$.
The one instanton index of the $Sp(1)$ gauge theory with the antisymmetric hypermultiplet shows interesting factorization structure as
\be
    &&\hspace{-.6cm}\frac{1}{32i^2}\left[\frac{\sin\frac{m\pm\alpha_1}{2}\prod_{l=1}^{N_f}2i\sin\frac{m_l}{2}}{\sinh\frac{\gamma_1\pm\gamma_2}{2}\sin\frac{m\pm i\gamma_1}{2}\sin\frac{i\gamma_1\pm\alpha_1}{2}}
    +\frac{\cos\frac{m\pm \alpha_1}{2}\prod_{l=1}^{N_f}2\cos\frac{m_l}{2}}{\sinh\frac{\gamma_1\pm\gamma_2}{2}\sin\frac{m\pm i\gamma_1}{2}\cos\frac{i\gamma_1\pm\alpha_1}{2}}\right]  \\
    &\hspace{-1cm}=&\hspace{-.6cm}\frac{1}{32}\left[\frac{\prod_{l=1}^{N_f}2i\sin\frac{m_l}{2}}{i^2\sinh\frac{\gamma_1\pm\gamma_2}{2}\sin\frac{i\gamma_1\pm\alpha_1}{2}}
    +\frac{\prod_{l=1}^{N_f}2\cos\frac{m_l}{2}}{\sinh\frac{\gamma_1\pm\gamma_2}{2}\cos\frac{i\gamma_1\pm\alpha_1}{2}}\right]
    +\frac{\prod_{l=1}^{N_f}2i\sin\frac{m_l}{2}+\prod_{l=1}^{N_f}2\cos\frac{m_l}{2}}{32i^2\sinh\frac{\gamma_1\pm\gamma_2}{2}\sin\frac{m\pm i\gamma_1}{2}}. \nn
\ee
The first two terms in the bracket agrees with the one instanton index of the $Sp(1)$ theory without the antisymmetric hypermultiplet.
The last term is the extra contribution from the antisymmetric hypermultiplet.
This extra index is exactly the same as the index for a single D0-brane floating freely on $N_f$ D8-brane background with the orientifold
\begin{equation}\label{D0-index}
    I_{{\rm D0}}^{k=1} =\frac{\prod_{l=1}^{N_f}2i\sin\frac{m_l}{2}+\prod_{l=1}^{N_f}2\cos\frac{m_l}{2}}{32i^2\sinh\frac{\gamma_1\pm\gamma_2}{2}\sin\frac{m\pm i\gamma_1}{2}}.
\end{equation}
Let us review the partition function of a single D0-brane sitting at the orientifold plane to compare to this index.
The partition function corresponding to the 8 translational zero modes of a single D0-brane is given by
\begin{equation}
    \frac{1}{(1-e^{\pm \gamma_1 \pm \gamma_2})(1-e^{\pm\gamma_1 \pm im})} \sim \frac{1}{\sinh^2\frac{\gamma_1\pm\gamma_2}{2} \sin^2\frac{m\pm i\gamma_1}{2}}
\end{equation}
Here the chemical potential $m$ for $Sp(1)$ global symmetry is identified with that for $SU(2)_{R'}$ symmetry in $SU(2)_1\times SU(2)_2\times SU(2)_{R}\times SU(2)_{R'} \subset SO(4)_1\times SO(4)_2$
where $SO(4)_1 \ (SO(4)_2)$ is the rotation symmetry along (transverse to) the D4-branes.
The broken supersymmetries by the presence of D0- and D8-branes provide the fermionic oscillators on the D0-brane quantum mechanics.
The corresponding 8 fermion zero modes transform in the $({\bf 1},{\bf2},{\bf 2},{\bf1})$ and $({\bf 2},{\bf1},{\bf 1},{\bf2})$ representations of the symmetry group and, after quantizing the zero modes,
their partition function is
\be
    \big(e^{\frac{\gamma_1 \pm \gamma_2}{2}}-e^{-\frac{\gamma_1 \pm \gamma_2}{2}}\big)\big(e^{\frac{\gamma_1 \pm im}{2}}-e^{-\frac{\gamma_1 \pm im}{2}}\big)
    \sim \sinh\frac{\gamma_1\pm\gamma_2}{2}\sin\frac{m\pm i\gamma_1}{2}.
\ee
Multiplying these partition functions together, we obtain the denominator of the D0-brane index (\ref{D0-index}).
The numerator comes from the open strings connecting one D0- and $N_f$ D8-branes.
As the lightest states of the 0-8 strings are in the Ramond-sector, we have the $N_f$ fermions in the fundamental representation of $O(k)$ ($O(1)=\mathbb{Z}_2$ in this case).
Therefore the quantization of $N_f$ fermion zero modes gives ${\bf 2}^{N_f-1}$ and its conjugate representation of $SO(2N_f)$, and $O(1)$ gauge constraint
leaves only ${\bf 2}^{N_f-1}$ representation which is the spinor representation of $SO(2N_f)$ group.
The numerator in (\ref{D0-index}) yields the expected spectrum of ${\bf 2}^{N_f-1}$ states.

The instanton states we are interested in are the D0-D4 bound states which are the degrees of freedom in the interior of the Higgs branch in the D0-brane quantum mechanics.
The factorization of the above instanton index indicates that our instanton calculation captures the states in both the Coulomb branch and the Higgs branch.
We are not able to restrict it only to the Higgs branch. This might be related to the fact that the non-commutativity of the instanton moduli space cannot be turned on for the simple gauge groups.
We have to remove the Coulomb branch index by hand.

The extra D0-brane index is also included for higher rank gauge group $Sp(N)$ with the antisymmetric hypermultiplet.
Therefore we argue that the correct instanton index is obtained by subtracting the extra index \eqref{D0-index}.
Therefore, the one instanton index for $Sp(N)$ is given by
\begin{equation}
    I^{k=1}_{Sp(N)}\!=\! \frac{1}{32i^2}\!\!\left[\frac{\prod_{l=1}^{N_f}2i\sin\frac{m_l}{2}\prod_{i=1}^N2i\sin\frac{m\pm\alpha_i}{2}}{\sinh\frac{\gamma_1\pm\gamma_2}{2}\sin\frac{m\pm i\gamma_1}{2}\prod_{i=1}^N2i\sin\frac{i\gamma_1\pm\alpha_i}{2}}\!
    +\!\frac{\prod_{l=1}^{N_f}2\cos\frac{m_l}{2}\prod_{i=1}^N2\cos\frac{m\pm \alpha_i}{2}}{\sinh\frac{\gamma_1\pm\gamma_2}{2}\sin\frac{m\pm i\gamma_1}{2}\prod_{i=1}^N2\cos\frac{i\gamma_1\pm\alpha_i}{2}}\right]\!-\! I_{D0}^{k=1}
\end{equation}
For higher instantons, the factorization structure of the Coulomb and Higgs branch indices is unclear because irrelevant poles start to appear when $k>1$.
So we have no good prescription to factor out the free D0-brane index from our result for higher instanton cases at the moment.


\section{Enhanced global symmetry}\label{enhancement}
In the worldvolume theory on a D4-brane, the $E_{N_f+1}$ symmetry enhancement occurs at the origin of the Coulomb branch when the classical gauge coupling constant diverges \cite{Seiberg:1996bd}. With $N_f<8$ massless hypermultiplets, the global flavor symmetry of the 5d gauge theory is $SO(2N_f)\times U(1)_I$. 
The instantons of $Sp(1)$ gauge group, which are D0 branes on the D4 brane,  play a crucial role for this symmetry enhancement. The instantons have masses proportional to the inverse gauge coupling as $1/g^{2}_{YM}$ and they become massless at the infinite coupling limit. They mix with the excitations of the elementary fields and form certain representations of the enhanced global symmetry $E_{N_f+1}$. The instanton charge behaves as the $U(1)_I$ Cartan generator of the $SO(14)\times U(1)_I$ subgroup of $E_{N_f+1}$ symmetry group. In this way, at the conformal point, the global symmetry is enhanced to $E_{N_f+1}$ symmetry: $E_8,E_7,E_6,E_5=SO(10),E_4=SU(5),E_3=SU(3)\times SU(2),E_2=SU(2)\times U(1)$ and $E_1=SU(2)$.


\subsection{Superconformal index for $Sp(1)$}
We now present the results of the index computations for $Sp(1)$. As explained earlier, there are perturbative part and instanton part in the superconformal index. For the perturbative part, we obtain merely the spectrum for the global symmetry $SO(2N_f)$. This is the case where the $U(1)_I$ charge is zero. For the instanton part which is associated with non-zero $U(1)_I$ charges, we find the spectrum for the enhanced symmetry $E_{N_f+1}$. The $U(1)_I$ provides an extra Cartan and thus leads to symmetry enhancements from $SO(2N_f)\times U(1)_I$ to $E_{N_f+1}$.

An instructive example is the case with $N_f=3$. The global symmetry for this case is $SO(6)$.
It follows from \eqref{I4spN} that,
dropping the instanton part $I^{inst}$, we find the lowest energy states appear at $x^2$ order in the superconformal index
\begin{align}
I_{\rm pert}= 1+ \big(e^{- im_1 -i m_2}+\cdots+e^{im_2+i m _3}+3+1\big) x^2 + {\cal O}(x^{3}),
\end{align}
where the constant $1$ is a singlet of the global symmetry, and the chemical potentials $m_i$ are arranged to form the (15-dim) adjoint representation of $SO(6)$, $e^{-i m_1-i m_2}+\cdots+e^{im_2+im_3}+3$. In terms of the character, we express it as 
\begin{align}
I_{\rm pert}= 1+ \big(1+ \chi^{SO(6)}_{\rm 15} \big) x^2 + \cdots,
\end{align}
where $SO(6)$ characters are denoted by $\chi^{SO(6)}_{irrep}$ with dimension of irreducible representations written in the subscript.
If we take into account the instantons, we find that the superconformal index contains extra contributions coming from the instantons
\begin{align}\label{eq:IforE5}
I= 1+ \Big(1+\chi^{SO(6)}_{\rm 15}+ q\,\chi^{SO(6)}_{4} +q^{-1}\,\chi^{SO(6)}_{\overline{4}}\Big) x^2 + \cdots,
\end{align}
where the power of the fugacity $q$ represents the $U(1)_I$ charges or instanton number. It is then clear that the first two characters of $x^2$ are from the perturbative part and the last two, which are the spinor representations of $SO(6)$, are from the instanton part with the opposite $U(1)_I$ charges, i.e., one instanton contribution arises in one of two spinor representations; the anti-instanton contribution arises in the other spinor representation. This shows how symmetry enhancement takes place.
On top of the adjoint representation of $SO(6)$, the instanton contributions come into play in symmetry enhancement to $E_4$ providing an extra Cartan generator in the form of fugacity $q$ attached to the spinor representations of $SO(6)$. In other words,
it follows from
the embedding of $SO(6)$
\begin{align}
SU(5) &\supset SO(6)\times U(1)_I\cr
{\bf 24}&= {\bf 1}_{0}+{\bf 15}_{0}+ {\bf 4}_{1}+ {\bf \overline 4}_{-1},
\end{align}
where the subscript denotes $U(1)$ charges, that the $SO(6)$ characters in $x^2$ of \eqref{eq:IforE5} are captured in the character of the (24-dim) adjoint representation of $E_4=SU(5)$, and hence one is allowed to write \eqref{eq:IforE5} as
\begin{align}
I&=1+ \big ( \chi^{SO(6)}_{\bf 1}+\chi^{SO(6)}_{\bf 15}+ q \,\chi^{SO(6)}_{\bf 4}+ q^{-1}\,\chi^{SO(6)}_{\bf \overline 4} \big) x^2 + \cdots\cr
&\equiv 1+ \chi^{E_4}_{\bf 24} \,x^2 + \cdots.
\end{align}
Higher $U(1)_I$ charges (or instanton number) appear as we go along with  higher powers of $x$ (or higher energies in the sense that the power of $x$ in the index is  $2j_1+2R$ that is roughly proportional to the energy $\epsilon_0$ for the BPS states\footnote{At least, $2(j_1+R) =\frac{2}{3}\epsilon_0$  at the leading order of $x$, because the single letter $q^1(=q^+)$ for the hypermultiplet does not carry $j_1$ charges.}). For this case, two instantons start to contribute from order $x^4$, and three instantons from order $x^6$, and so  on.

The pattern of the symmetry enhancement for other cases is very similar except for $E_7$ and $E_8$.  For a given $N_f$, the perturbative part of the index is
the superconformal index takes the form
\begin{align}
I_{N_f} = 1+ \chi^{E_{N_f+1}}_{\bf adj} x^2 + \cdots,
\end{align}
with the following generic embedding
\begin{align}
E_{N_f+1} &= SO(2N_f)\times U(1)_I,\cr
{\bf adj}^{E_{N_f+1}} &\supset {\bf 1}^{SO(2N_f)}_0 + {\bf adj}^{SO(2N_f)}_0 +{\bf 2}^{N_f-1}_1 +{\bf 2'}^{N_f-1}_{-1},
\end{align}
where ${\bf 2}^{N_f-1}_1$ and ${\bf 2'}^{N_f-1}_{-1}$ are two spinor representations denoted by their dimensions (they can be conjugate or self-conjugate depending of $N_f$), and the subscripts are the $U(1)$ charges. The following table summarizes the relevant embeddings:\\
\begin{flalign*}
\qquad N_f=2:\quad& E_3 =SU(3)\times SU(2) \supset SO(4)\times U(1)_I 
&
\underset{\mathclap{}}{\circ} \quad\qquad\quad~
\\
&~\qquad SU(3)\supset SU(2)\times U(1)_I
&
\underset{\mathclap{}}{\circ}
-\!\!\!- \underset{\mathclap{I}}{\bullet}
\\
&~~~~~~~\qquad {\bf 8}={\bf 1}_0+{\bf 3}_0+{\bf 2}_1+{\bf 2}_{-1}.
&
\end{flalign*}
\begin{flalign*}
\qquad N_f=3:\quad &E_4 =SU(5)\supset SU(4)\times U(1)_I&
\underset{\mathclap{}}{\overset{\overset{\textstyle\circ_{\mathrlap{}}}{\textstyle\vert}}{\circ}} -\!\!\!-\underset{\mathclap{}}{\circ}
-\!\!\!- \underset{\mathclap{I}}{\bullet}
\\
&~~~~~~\qquad{\bf 24}={\bf 1}_{0}+ {\bf 15}_{0}+{\bf 4}_{1}+ \overline{\bf 4}_{-1} &
\end{flalign*}
\begin{flalign*}
\qquad N_f=4:\quad &E_5 = SO(10)\supset SO(8)\times U(1) &
\underset{\mathclap{}}{\circ} -\!\!\!-
\underset{\mathclap{}}{\overset{\overset{\textstyle\circ_{\mathrlap{}}}{\textstyle\vert}}{\circ}} -\!\!\!-\underset{\mathclap{}}{\circ}
-\!\!\!- \underset{\mathclap{I}}{\bullet}
\\
&~~\qquad\qquad{\bf 45}= {\bf 1}_{0}+{\bf 28}_{0}+{\bf 8}_{-1}+ {\bf 8}_{1}&
\end{flalign*}
\begin{flalign*}
\qquad N_f=5:\quad &E_6\supset SO(10)\times U(1)
&
\underset{\mathclap{}}{\circ}-\!\!\!-
\underset{\mathclap{}}{\circ} -\!\!\!-
\underset{\mathclap{}}{\overset{\overset{\textstyle\circ_{\mathrlap{}}}{\textstyle\vert}}{\circ}} -\!\!\!-\underset{\mathclap{}}{\circ}
-\!\!\!- \underset{\mathclap{I}}{\bullet}
\\
&{\bf 78}={\bf 1}_0 +{\bf 45}_{0}+{\bf 16}_{-1}+ \overline{\bf 16}_{1}
&
\end{flalign*}
\begin{flalign*}
\qquad N_f=6:\quad &E_7\supset SO(12)\times U(1)_I&
 \underset{\mathclap{}}{\circ}-\!\!\!-
\underset{\mathclap{}}{\circ}-\!\!\!-
\underset{\mathclap{}}{\circ} -\!\!\!-
\underset{\mathclap{}}{\overset{\overset{\textstyle\circ_{\mathrlap{}}}{\textstyle\vert}}{\circ}} -\!\!\!-\underset{\mathclap{}}{\circ}
-\!\!\!- \underset{\mathclap{I}}{\bullet}
\\
&{\bf 133}={\bf 1}_0+{\bf 66}_0 + {\bf 32}_1+ {\bf 32}_{-1} + {\bf 1}_2+ {\bf 1}_{-2} &
\end{flalign*}
\begin{flalign*}
\qquad N_f=7:\quad &E_8 \supset SO(14)\times U(1)_I
&
 \underset{\mathclap{}}{\circ} -\!\!\!-
 \underset{\mathclap{}}{\circ}-\!\!\!-
\underset{\mathclap{}}{\circ}-\!\!\!-
\underset{\mathclap{}}{\circ} -\!\!\!-
\underset{\mathclap{}}{\overset{\overset{\textstyle\circ_{\mathrlap{}}}{\textstyle\vert}}{\circ}} -\!\!\!-\underset{\mathclap{}}{\circ}
-\!\!\!- \underset{\mathclap{I}}{\bullet}
\\
&{\bf 248}={\bf 1}_0 +{\bf 91}_0+ {\bf  64}_1+\overline{\bf 64}_{-1}+{\bf 14}_{2}+{\bf 14}_{-2} &
\end{flalign*}
The Dynkin diagram which are made out of the empty nodes represent the Dynkin diagram for $SO(2N_f)$ and the filled node $\underset{\mathclap{I}}{\bullet}$ denotes the extra Cartan stems from the instanton contributions which is connected to the node associated with a spinor representation of $SO(2N_f)$, and thus all together the nodes account for enhanced $E_{N_f+1}$ Dynkin diagrams.

In other words, we see, by tracing the structure of the index, that the perturbative part and the instanton part together form a single state in the adjoint representation of $E_{N_f+1}$ at the leading order of $x$, at order $x^2$. The perturbative part comprises two states in the representations of $SO(2N_f)$: the singlet and the adjoint representation. One instanton and anti-instanton parts, on the other hand, both provide two states in the spinor and its conjugate representations of $SO(2N_f)$. These states altogether make the $SO(2N_f)\times U(1)_I$ decomposition of the adjoint representation of $E_{N_f+1}$ for $N_f<6$, at order $x^2$. 

Notice that, in the above embedding, the $N_f=6, 7$ cases do contain the extra representations of higher instanton charges. This implies that unlike the lower $N_f$ (up to 5) cases, two instantons start to contribute non-trivially to the leading power of $x$ in the superconformal index. For the moment, we do not have a clear understanding of why the enhancements to $E_7$ and $E_8$ is slightly different from the cases for $N_f\le 5$, regarding the instanton contributions.  We see, at least, complications in the pole structures for such cases that appear in the contour integral formula for the index we have discussed in section \ref{sec:sp-gaugegr}, which may reflect such difference. In the next subsections, we present the superconformal index result that we computed to order $x^6$, and then discuss the complications in the $N_f=6, 7$ cases.


\subsection{Index for $N_f\le 5$}\label{sec:Indexless5}
To proceed with higher powers of $x$, we first restrict ourselves to lower $N_f$ that is $N_f\le 5$. The goal of this subsection is to present the superconformal index result  to order $x^6$ (or $x^8$ for $N_f=3, 4$) and to show how the multi-instantons contribute to the global symmetry enhancement.

\noindent $\bullet$ For $N_f=0$, the global symmetry is $U(1)_I$ and one expects that this symmetry is enhanced to $E_1 = SU(2)$
\begin{align}
SU(2) \supset U(1)_I.
\end{align}
The superconformal index for this is given by
\begin{align}\label{I4e1}
I&=1 + \chi^{E_1}_{\bf 3} x^2 +\chi_2(y)\big[1 + \chi^{E_1}_{\bf 3}\big] x^{3} + \Big(\chi_3(y)\big[1 +\chi^{E_1}_{\bf 3}\big] +1+\chi^{E_1}_{\bf 5}\Big) x^4 \cr
& + \Big(\chi_4(y) \big[1+\chi^{E_1}_{\bf 3}\big] +\chi_2(y)\big[1+\chi^{E_1}_{\bf 3}  + \chi^{E_1}_{\bf 5}\big]\Big) x^{5} \cr
&+ \Big(\chi_5(y) \big[1+\chi^{E_1}_{\bf 3} \big] + \chi_3(y)\big[1 + \chi^{E_1}_{\bf 3} + \chi^{E_1}_{\bf 5}  + \chi^{E_1}_{\bf 3}\chi^{E_1}_{\bf 3}\big] +\chi^{E_1}_{\bf 3} + \chi^{E_1}_{\bf 7}-1\Big)x^6\cr
&+\Big(\chi_6(y) \big[1+\chi^{E_1}_{\bf 3}] +
\chi_4(y) \big[2+4\chi^{E_1}_{\bf 3}+2\chi^{E_1}_{\bf 5}\big]
+ \chi_2(y)\big[ 1+3\chi^{E_1}_{\bf 3}+2\chi^{E_1}_{\bf 5}+\chi^{E_1}_{\bf 7}\big]
\Big) x^7\cr
&+\Big(\chi_7(y)\big[1+\chi^{E_1}_{\bf 3}] +\chi_5(y)\big[3\chi^{E_1}_{\bf 5}+5\chi^{E_1}_{\bf 3} +4\big]
+\chi_3(y)\big[2\chi^{E_1}_{\bf 7}+3\chi^{E_1}_{\bf 5}+7 \chi^{E_1}_{\bf 3}+2\big]\cr
&+\chi^{E_1}_{\bf 9}+2\chi^{E_1}_{\bf 5}+2\chi^{E_1}_{\bf 3}+3\Big)x^8+{\cal O}(x^{9}).
\end{align}
where $\chi_n(y)$ stands for $SU(2)$ character for fugacity $y$ with dimension $n$. For instance, 2-dim representation is given by $\chi_2(y)= y+1/y$.
Here, $\chi^{E_1}_n$ is the character for $E_1=SU(2)$ with dimension $n$, whose formula is given in Appendix \ref{app:character}. 
 As higher dimensional representations are associated to higher instanton numbers, we see that the two instantons start to contribute to the index at order $x^4$, through $\chi^{E_1}_{\bf 5}$. Likewise, the three instantons contributes at order $x^6$ through $\chi^{E_1}_{\bf 7}$, and so on. From the point of view of $D0$ branes, this case is when there is no $D8$ branes involved, and a $D4$ probes the theory and thus, $D0$s are bounded to the $D4$. We note that $-1$ at order $x^6$ implies that a fermionic contribution arise as a singlet of $SU(2)$.\\

\noindent $\bullet$ For $N_f=1$, the global symmetry is $SO(2)\times U(1)_I$ and one expects that this symmetry is enhanced to $E_2 = SU(2)\times U(1)$, where the Cartan generator for $SU(2)$ comes form a linear combination of the Cartan generators of $SO(2)$ and $U(1)_I$
\begin{align}
 SU(2)_{\frac{1}{2}(m_1+w)}\times U(1)_{\frac{1}{2}(7m_1-w)} &\supset SU(2)_{m_1}\times U(1)_I{}_{w},
\end{align}
where $w$ is the chemical potential for $U(1)_I$ identified as $q=e^{i\frac{w}{2}}$ and $m_1$ is the chemical potential associated withe $SO(2)$.  
The superconformal index for $N_f=1$ is then given by
\begin{align}\label{I4e2}
I&=1 + \chi^{E_2}_{\bf 4} x^2 +\chi_2(y)\big[1 + \chi^{E_2}_{\bf4}\big] x^{3} + \Big(\chi_3(y)\big[1 +\chi^{E_2}_{\bf 4}\big] +1+\chi^{SU(2)}_{\bf 5}-\chi_{\bf 4}(f)\Big) x^4 \\
& + \Big(\chi_4(y) \big[1+\chi^{E_2}_{\bf 4}\big] 
+\chi_2(y)\big[\chi^{E_2}_{\bf 4}  + \chi^{SU(2)}_{\bf 3}+\chi^{SU(2)}_{\bf 5}-\chi_{\bf 4}(f)\big]\Big) x^{5} \cr
&+ \Big(\chi_5(y) \big[1+\chi^{E_2}_{\bf 4} \big] + \chi_3(y)\big[
4 \chi^{E_2}_{\bf 4} + 2\chi^{SU(2)}_{\bf 5}-\chi_{\bf 4}(f) \big] 
+\chi^{SU(2)}_{\bf 7} +3 \chi^{SU(2)}_{\bf 3}+1\Big)x^6\cr
&+\Big(\chi_6(y) \big[1+\chi^{E_2}_{\bf 4}] +
\chi_4(y) \big[5\chi^{E_2}_{\bf 4}+2\chi^{SU(2)}_{\bf 3}+2\chi^{SU(2)}_{\bf 5}-\chi_{\bf 4}(f)\big]\cr
&
+ \chi_2(y)\big[6\chi^{E_2}_{\bf 4}+2\chi^{SU(2)}_{\bf 5}+\chi^{SU(2)}_{\bf 7}-\chi^{SU(2)}_{\bf 3}\chi_{\bf 4}(f)\big]
\Big) x^7\cr
&+\Big(\chi_7(y)\big[1+\chi^{E_2}_{\bf 4}] 
+\chi_5(y)\big[9\chi^{E_2}_{\bf 4}+3\chi^{SU(2)}_{\bf 5} -\chi_{\bf 4}(f) \big]
+\chi_3(y)\big[9\chi^{E_2}_{\bf 4}+2\chi^{SU(2)}_{\bf 7} +4\chi^{SU(2)}_{\bf 5}\cr
&
+2 \chi^{SU(2)}_{\bf 3}-(\chi^{E_2}_{\bf 4}+\chi^{SU(2)}_{\bf 3})\chi_{\bf 4}(f)\big]
+3\chi^{E_2}_{\bf 4}+\chi^{SU(2)}_{\bf 9}+2\chi^{SU(2)}_{\bf 5}+2-\chi^{E_2}_{\bf 4}\chi_{\bf 4}(f)\Big)x^8+{\cal O}(x^{9}),\nn
\end{align}
where $\chi^{E_2}_{\bf 4}=1+\chi^{SU(2)}_{\bf 3}$ is the adjoint representation of $E_2$ and 
$\chi_{\bf 4}(f)=(e^{i \frac{\rho}{2}}+e^{-i \frac{\rho}{2}})\chi^{SU(2)}_{\bf 2}$ 
with $U(1)$ charge $\rho$. This index shows that states are in $SU(2)$ and $U(1)$ representations, and it is clear that the pattern that multi-instantons appear follows that of the $N_f=0$ case. On the other hand, we observe that the fermionic contribution (with the negative sign in front) appears quite differently compared with the $N_f=0$ case. 
The fermionic contribution appears not as a singlet but as $\chi_{\bf 4}(f)$ that is the fundamental representations of $SU(2)$ with the opposite $U(1)$ charges.  Moreover, it continues to appear in higher powers of $x$.
It is not clear for us, for the moment, how instanton contributions give rise to such fermionic contributions in the process of the global symmetry enhancement for $N_f=0,1$. 
\\

\noindent $\bullet$ For $N_f=2$, the global symmetry is $SO(4)\times U(1)_I\cong SU(2)\times SU(2)\times U(1)_I$ and one expects that this symmetry is enhanced to $E_3 = SU(3)\times SU(2)$.
The symmetry enhancement is understood from  the embedding
\begin{align}
SU(3)\times SU(2) &\supset SU(2)_{m_1}\times SU(2)_{m_2}\times U(1)_I,
\end{align}
where $m_1, m_2$  the chemical potentials of two $SU(2)$s in the right hand side properly arrange themselves  to yield the enhancement to $E_3$
\begin{align}
E_3 =SU(3)\times SU(2)_{\frac{1}{2}(m_1-m_2)}
\end{align}
with the following $SU(2)$ and $U(1)$ charges
\begin{align}
SU(3) &\supset SU(2)_{\frac{1}{2}(m_1+m_2)}\times U(1)_I\cr
{\bf 8}&={\bf 1}_0+{\bf 3}_0+{\bf 2}_1+{\bf 2}_{-1}.
\end{align}
For our convenience, we write this $E_3$ decomposition as
\begin{align}
E_3= SU(3)\times SU(2)&\supset   SU(2)\times SU(2)\times U(1)_I\cr
({\bf 8},{\bf 1})&=({\bf 1},{\bf 1})_0+({\bf 3},{\bf 1})_0+({\bf 2},{\bf 1})_1+({\bf 2},{\bf 1})_{-1}.
\end{align}
The superconformal index for $N_f=2$ is then given by
\begin{align}
I&=1 + \big(\chi^{E_3}_{\bf (8,1)+ (1,3)}\big) x^2 + \chi_2(y)\big(1 + \chi^{E_3}_{\bf (8,1)+(1,3)}\big) x^{3} \\
&+ \Big(\chi_3(y)\big[1 + \chi^{E_3}_{\bf (8,1)+(1, 3)}\big]+1 + \chi^{E_3}_{\bf (27,1)+(1, 5)}  \Big) x^4 \cr
&+ \Big(\chi_4(y) \big[1 + \chi^{E_3}_{\bf (8,1)+(1, 3) }\big] + \chi_2(y)
\big[1 +  \chi^{E_3}_{\bf (8,1)+(1, 3)}+\chi^{E_3}_{\bf (27,1)+ (1, 5)} + \chi^{E_3}_{\bf (8, 3) +(10,1)+(\overline{10},1)} \big]\Big) x^{5}\cr
&+ \Big(\chi_5(y)\big[1+\chi^{E_3}_{\bf (8,1)+(1,3)}\big]+
\chi_3(y)\big[
1+\chi^{E_3}_{\bf (8,1)+(1, 3)} + \chi^{E_3}_{\bf (27,1)+(1, 5)} +\chi^{E_3}_{\bf [(8,1)+ (1,3)]\otimes[(8,1)+ (1,3)]} \big]\cr
&
+2\chi^{E_3}_{\bf (8,1)+(1, 3) }+
\chi^{E_3}_{\bf (8, 3) +(10,1)+(\overline{10},1)}+\chi^{E_3}_{\bf (64,1)+(1,7)}
\Big)x^6\cr
&+ \Big(\chi_6(y)\big[1+\chi^{E_3}_{\bf (8,1)+(1,3)}\big]
+\chi_4(y)\big[3+2\chi^{E_3}_{\bf (27,1)+(1,5)}+2\chi^{E_3}_{({\bf 10,1})+({\overline{\bf 10},1})}
+3\chi^{E_3}_{({\bf8, 3})}+\chi^{E_3}_{({\bf 8,1})}
\cr
& +4\chi^{E_3}_{({\bf 8,1})+({\bf 1,3})}  \big]
+\chi_2(y)\big[\chi^{E_3}_{\bf (64,1)+(1,7)+({\bf 35,1})+({\overline{\bf 35},1})}+3\chi^{E_3}_{\bf (8,1)+(1,3)}\cr
&+\chi^{E_3}_{\bf [(8,1)+ (1,3)]\otimes[(8,1)+ (1,3)]} +
\chi^{E_3}_{\bf (27,1)+(1,5)}+1\big]\Big)x^7\cr
&+ \Big(\chi_7(y)\big[1+\chi^{E_3}_{\bf (8,1)+(1,3)}\big]+ \chi_5(y)\big[2+\chi^{E_3}_{\bf (27,1)+(1,5)}
+2\chi^{E_3}_{\bf [(8,1)+ (1,3)]\otimes[(8,1)+ (1,3)]}
+3\chi^{E_3}_{\bf (8,1)+(1,3)}\big]\cr
&
+\chi_3(y)\big[2+2\chi^{E_3}_{\bf (64,1)+(1,7)+(27,1)+(1,5)+(35,1)+(\overline{35},1)+(10,1)+(\overline{10},1)}
+\chi^{E_3}_{\bf (27,3)+(27,1) +(8,5)+(8,3)}\cr
&
+7\chi^{E_3}_{\bf (8,1)+(1,3)} +\chi^{E_3}_{\bf [(8,1)+ (1,3)]\otimes[(8,1)+ (1,3)]} \big]
+2+\chi^{E_3}_{\bf (35,1)+({\overline{\bf 35},1})+(27,1)+(10,3)+(\overline{10},3)} \cr
&
+\chi^{E_3}_{\bf [(8,1)+ (1,3)]\otimes[(8,1)+ (1,3)] + (8,1)+ (1,3)+(27,1)+(1,5)}
+\chi^{E_3}_{\bf (125,1)+(1,9)} \Big)x^8
+{\cal O}(x^{9}),\nn
\end{align}
where we used the following shorthand notation
\begin{align}
\chi^{E_3}_{\bf (8,1)+ (1,3)} &= \chi^{E_3}_{\bf (8,1)}+\chi^{E_3}_{\bf (1,3)},
\end{align}
and tensor product of $(\bf 8,1)+ (\bf 1,3)$ is given by
\begin{align}
\chi^{E_3}_{[({\bf 8,1})+ ({\bf 1,3})]\otimes[({\bf 8,1})+ ({\bf 1,3})]}
&=\chi^{E_3}_{[({\bf 8,1})+ ({\bf 1,3})]\otimes_A[({\bf 8,1})+ ({\bf 1,3})]}+\chi^{E_3}_{[({\bf 8,1})+ ({\bf 1,3})]\otimes_S[({\bf 8,1})+ ({\bf 1,3})]}
\cr
&=\chi^{E_3}_{({\bf 8,1})+ ({\bf 1,3}) + ({\bf8, 3}) +({\bf 10,1})+({\overline{\bf 10},1})} +  \chi^{E_3}_{({\bf 27,1})+({\bf1, 5})+ ({\bf8, 3}) +({\bf 8,1})+2({\bf1,1})}.
\end{align}
Two instanton contributions start to appear at order $x^4$ where the character $\chi^{E_3}_{(\bf 27,1)+(1,5)}$ contain $q^2$ and $q^{-2}$. In a similar way, the three instanton contributions start to appear at order $x^6$ in  $\chi^{E_3}_{(\bf 64,1)+(1,7)}$, and four instantons contributions appear at order $x^8$, $\chi^{E_3}_{\bf (125,1)+(1,9)}$, and so on.
\\

\noindent $\bullet$ For $N_f=3$ case, the global symmetry is $SO(6)\times U(1)$ and one expects the symmetry is enhanced to $E_4 = SU(5)$. The superconformal index for this is given by
\begin{align}\label{I4e4}
I&=1 + \chi^{E_4}_{\bf 24} x^2 +\chi_2(y)\big[1 + \chi^{E_4}_{\bf 24}\big] x^{3} + \Big(\chi_3(y)\big[1 +\chi^{E_4}_{\bf 24}\big] +1+\chi^{E_4}_{\bf 200}\Big) x^4 \cr
& + \Big(\chi_4(y) \big[1+\chi^{E_4}_{\bf 24}\big] +\chi_2(y)\big[1+\chi^{E_4}_{\bf 24}  + \chi^{E_4}_{\bf 126}+\chi^{E_4}_{\overline{\bf 126}}+ \chi^{E_4}_{\bf 200}\big]\Big) x^{5} \cr
&+ \Big(\chi_5(y) \big[1+\chi^{E_4}_{\bf 24} \big] + \chi_3(y)\big[2 + 3 \chi^{E_4}_{\bf 24} + \chi^{E_4}_{\bf 75}  + \chi^{E_4}_{\bf 126}+\chi^{E_4}_{\overline{\bf 126}}+2 \chi^{E_4}_{\bf 200}\big]\cr
& +2\chi^{E_4}_{\bf 24} + \chi^{E_4}_{\bf 126}+\chi^{E_4}_{\overline{\bf 126}}
+ \chi^{E_4}_{\bf 1000}  \Big)x^6\cr
&+\Big(\chi_6(y) \big[1+\chi^{E_4}_{\bf 24}] +
\chi_4(y) \big[2+2\chi^{E_4}_{\bf 200}+2\chi^{E_4}_{\bf 126}+2\chi^{E_4}_{\overline{\bf 126}}+\chi^{E_4}_{\bf 75}+5\chi^{E_4}_{\bf 24}\big]\cr
&+ \chi_2(y)\big[ \chi^{E_4}_{\bf 1000}+\chi^{E_4}_{\bf 1050}+\chi^{E_4}_{\overline{\bf 1050}}+ 2\chi^{E_4}_{\bf 200}+\chi^{E_4}_{\bf 126} +\chi^{E_4}_{\overline{\bf 126}}+\chi^{E_4}_{\bf 75}+5\chi^{E_4}_{\bf 24}+2\big]
\Big) x^7\cr
&+\Big(\chi_7(y)\big[1+\chi^{E_4}_{\bf 24}] +\chi_5(y)\big[3\chi^{E_4}_{\bf 200}+2\chi^{E_4}_{\bf 126} +2\chi^{E_4}_{\overline{\bf 126}}+2\chi^{E_4}_{\bf 75}+7\chi^{E_4}_{\bf 24}+4\big]\cr
&+\chi_3(y)\big[2\chi^{E_4}_{\bf 1000}+2 \chi^{E_4}_{\bf 1050}+2\chi^{E_4}_{\overline{\bf 1050}}+ \chi^{E_4}_{\bf 1024}+4 \chi^{E_4}_{\bf 200}+3\chi^{E_4}_{\bf 126}+3\chi^{E_4}_{\overline{\bf 126}}+\chi^{E_4}_{\bf 75}+ 9\chi^{E_4}_{\bf 24}+3\big]\cr
&+\chi^{E_4}_{\bf 3675}+\chi^{E_4}_{\bf 1050}+\chi^{E_4}_{\overline{\bf 1050}}+ \chi^{E_4}_{\bf 224}+\chi^{E_4}_{\overline{\bf 224}}+\chi^{E_4}_{\bf 1024}+ 3 \chi^{E_4}_{\bf 200}+ \chi^{E_4}_{\bf 126}+\chi^{E_4}_{\overline{\bf 126}}+\chi^{E_4}_{\bf 75}+3\chi^{E_4}_{\bf 24}+3
\Big)x^8\cr
&+{\cal O}(x^{9})
\end{align}
where we used the branching rules
\begin{align}
SU(5)&\supset SO(6)\times U(1)\cr
{\bf 24}&={\bf 1}_{0}+{\bf 4}_{1}+ \overline{\bf 4}_{-1} + {\bf 15}_{0}\cr
{\bf 200}&= {\bf 1}_{0}+{\bf 4}_{1}+ \overline{\bf 4}_{-1} +{\bf 10}_{2}+ \overline{\bf 10}_{-2}+{\bf 15}_{0}+{\bf 36}_{1}+ \overline {\bf 36}_{-1}+{\bf 84}_{0}\cr
{\bf 1000}&={\bf 1}_{0}+{\bf 4}_{1}+ \overline{\bf 4}_{-1}+{\bf 10}_{2}+ \overline{\bf 10}_{-2}+{\bf 15}_{0}
+{\bf 20}_{3}+{\bf \overline{20}}_{-3}
+{\bf 36}_{1}+ \overline {\bf 36}_{-1}\cr
&\quad+{\bf 70}_2+{\bf \overline{70}}_{-2}+{\bf 84}_{0}+{\bf {\bf 160}_{1}+\overline{160}}_{-1}+{\bf 300}_0,\cr
{\bf 3765}&={\bf 1}_{0}+{\bf 4}_{1}+ \overline{\bf 4}_{-1}+{\bf 15}_{0} +{\bf 10}_{2}+ \overline{\bf 10}_{-2}
+{\bf 20}_{3}+{\bf \overline{20}}_{-3}
+ {\bf 35}_{4}+ \overline{\bf 35}_{-4}
\cr
&\quad
+{\bf 36}_{1}+ \overline {\bf 36}_{-1}
+{\bf 70}_2+{\bf \overline{70}}_{-2}
+{\bf 84}_{0}+ {\bf 120}_{3}+ \overline{\bf 120}_{-3}
+{\bf {\bf 160}_{1}+\overline{160}}_{-1}\cr
&\quad
+{\bf 300}_0
+ {\bf 270}_{2}+ \overline{\bf 270}_{-2}
+ {\bf 500}_{1}+ \overline{\bf 500}_{-1}
+ {\bf 825}_{0}.
\nn
\end{align}
From $U(1)$ charges, we see that the two instantons start to contribute from $x^4$ order; it is captured in the character in $\chi^{E_4}_{\bf200}$. Three instanton contribution first appears at order $x^6$,  $\chi^{E_4}_{\bf1000}$, and four instanton contribution appears at order $x^8$,  $\chi^{E_4}_{\bf3765}$. \\

\noindent $\bullet$ For $N_f=4$ case, the global symmetry is $SO(8)\times U(1)_I$ and the symmetry is enhanced to $E_5 = Spin(10)$. The superconformal index for this is given by
\begin{align}\label{I4e5}
I&=1 + \chi^{E_5}_{\bf 45} x^2 + \chi_2(y) \big[1 + \chi^{E_5}_{\bf45}\big]x^{3} + \Big(\chi_3(y) \big[1 + \chi^{E_5}_{\bf45}\big] + 1+\chi^{E_5}_{\bf770} \Big) x^4  \\
&+ \Big(\chi_4(y) \big[1+\chi^{E_5}_{\bf45}\big] +\chi_2(y) \big[1+\chi^{E_5}_{\bf45}+ \chi^{E_5}_{\bf770} + \chi^{E_5}_{\bf945} \big]\Big) x^{5} \cr
&+\Big(\chi_5(y) \big[1+\chi^{E_5}_{\bf45}\big]  +\chi_3(y)\big[2+ 2 \chi^{E_5}_{\bf45} + \chi^{E_5}_{\bf54}+  \chi^{E_5}_{\bf210} +2 \chi^{E_5}_{\bf770}+ \chi^{E_5}_{\bf945}\big] +2 \chi^{E_5}_{\bf45} + \chi^{E_5}_{\bf945} +\chi^{E_5}_{\bf7644}\Big) x^6 \cr
& + \Big( \chi_6(y)\big[1+ \chi^{E_5}_{\bf 45}\big] + \chi_4(y)  \big[2\chi^{E_5}_{\bf770} + 2\chi^{E_5}_{\bf 945} + \chi^{E_5}_{\bf54} + \chi^{E_5}_{\bf210} + 4\chi^{E_5}_{\bf45}+ 2\big]
\cr
&+ \chi_2(y)\big[\chi^{E_5}_{\bf7644} +2\chi^{E_5}_{\bf 17920} + \chi^{E_5}_{\bf945}+2\chi^{E_5}_{\bf770}+\chi^{E_5}_{\bf54} + \chi^{E_5}_{\bf210} + 4\chi^{E_5}_{\bf45}+ 2\big]\Big)x^7\cr
&+ \Big( \chi_7(y) \big[1+ \chi^{E_5}_{\bf 45}\big]+\chi_5(y) \big[3\chi^{E_5}_{\bf 770}+2\chi^{E_5}_{\bf 945}+2\chi^{E_5}_{\bf 210}+2\chi^{E_5}_{\bf 54}+5\chi^{E_5}_{\bf 45}+4\big] \cr
&+\chi_3(y)\big[2\chi^{E_5}_{\bf 7644} + 2 \chi^{E_5}_{\bf 17920} + \chi^{E_5}_{\bf 1386} + \chi^{E_5}_{\bf 5940} + 3 \chi^{E_5}_{\bf 945} + 3 \chi^{E_5}_{\bf 770} + \chi^{E_5}_{\bf 210} + \chi^{E_5}_{\bf 54} +
 8 \chi^{E_5}_{\bf 45} + 3\big]\cr
&+ \chi^{E_5}_{\bf 52920}+ \chi^{E_5}_{\bf 17920}+ \chi^{E_5}_{\bf 8085} + \chi^{E_5}_{\bf 4125} + \chi^{E_5}_{\bf 945}+ 3 \chi^{E_5}_{\bf 770} + \chi^{E_5}_{\bf 210}+ \chi^{E_5}_{\bf 54 } + 2 \chi^{E_5}_{\bf 45} + 3 \Big)x^8+ {\cal O}(x^{9}).\nn
\end{align}
where we used the branching rule
\begin{align}
SO(10)&\supset SO(8)\times U(1)\\
{\bf 45}&= {\bf 1}_{0}+{\bf 8}_{1}+ {\bf 8}_{-1}+{\bf 28}_{0} \nn\\
{\bf 770}&= {\bf1}_0+{\bf 8}_{1}+{\bf 8}_{-1} +{\bf 28}_{0} + {\bf 35}_{2}+{\bf 35}_{0}+{\bf 35}_{-2} + {\bf 160}_{1} + {\bf 160}_{-1}+{\bf 300}_{0}  \cr
{\bf 7644}& = {\bf 1}_0+{\bf 8}_{1}+{\bf 8}_{-1}+{\bf 28}_0+{\bf 35}_{2}+{\bf 35}_0+{\bf 35}_{-2}+{\bf 112}_{3}+{\bf 112}_{1}+{\bf 112}_{-1}+{\bf 112}_{-3}\cr
&\quad+{\bf 160}_{1}+{\bf 160}_{-1}+{\bf 300}_0+{\bf 567}_{2}+{\bf 567}_0+{\bf 567}_{-2}+{\bf 1400}_{1}+{\bf 1400}_{-1}+{\bf 1925}_0\cr
{\bf 52920}& = {\bf 1}_0+{\bf 8}_{1}+{\bf 8}_{-1}+{\bf 28}_{0}+{\bf 35}_{2}+{\bf 35}_{0}+{\bf 35}_{-2}
+{\bf 112}_{3}+{\bf 112}_{1}+{\bf 112}_{-1}+{\bf 112}_{-3}\cr
&\quad
+{\bf 160}_{1}+{\bf 160}_{-1}
+{\bf 294}_{4}+ {\bf 294}_{2}+{\bf 294}_{0}+{\bf 294}_{-2}+{\bf 294}_{-4}
+{\bf 300}_{0}\cr
&\quad
+{\bf 567}_{2}+{\bf 567}_{0}+{\bf 567}_{-2}+{\bf 1400}_{1}+{\bf 1400}_{-1}
+ {\bf 1568}_{3} +{\bf 1568}_{1}+{\bf 1568}_{-1}+{\bf 1568}_{-3}
\cr
&\quad
+{\bf 1925}_{0}
+{\bf 4312}_{2}+{\bf 4312}_{0}+{\bf 4312}_{-2}
+{\bf 7840}_{1}+{\bf 7840}_{-1}
+{\bf 8918}_{0}.\nn
\end{align}
Note that even though $SO(8)$ has triality automorphism, the representations $\bf 8$, $\bf 35$ and so on are all higher dimensional spinor representations due to fermionic zero modes to which the instantons couple. The result shows that the two instantons again starts to contribute from $x^4$ order in $\chi^{E_5}_{\bf770}$.  Three instanton contribution is at order $x^6$; it is captured in the character $\chi^{E_5}_{\bf7644}$. Four instanton contribution first appears in the character $\chi^{E_5}_{\bf52920}$. \\

\noindent $\bullet$ For $N_f=5$ case, the global symmetry is $SO(10)\times U(1)$ and the symmetry is enhanced to $E_6 $. The superconformal index for this is given by
\begin{align}\label{I4e6}
I&=1 + \chi^{E_6}_{\bf78} x^2 + \chi_2(y) \big[1 + \chi^{E_6}_{\bf78}\big] x^{3} +
\Big(\chi_3(y)\big[1 + \chi^{E_6}_{\bf 78}\big] +1+ \chi^{E_6}_{\bf2430} \Big) x^4 \cr
&+\Big(\chi_4(y) \big[1+\chi^{E_6}_{\bf78} \big] + \chi_2(y) \big[1 + \chi^{E_6}_{\bf78}+\chi^{E_6}_{\bf2430} + \chi^{E_6}_{\bf2925} \big]\Big) x^{5} \cr
&+ \Big(\chi_5(y) \big[1+\chi^{E_6}_{\bf78}\big] + \chi_3(y) \big[2 + 2 \chi^{E_6}_{\bf78} + \chi^{E_6}_{\bf650} +2 \chi^{E_6}_{\bf2430} + \chi^{E_6}_{\bf2925}\big] 
+2 \chi^{E_6}_{\bf78} +\chi^{E_6}_{\bf2925} + \chi^{E_6}_{\bf43758} \Big) x^6\cr
&+\Big(\chi_6(y) \big[1+\chi^{E_6}_{\bf78} \big] + \chi_4(y) \big[2+ 4\chi^{E_6}_{\bf78}+\chi^{E_6}_{\bf650}+2\chi^{E_6}_{\bf2430} + 2\chi^{E_6}_{\bf2925} \big] \cr
&+ \chi_2(y)\big[
2+ 4\chi^{E_6}_{\bf78}+\chi^{E_6}_{\bf650}+2\chi^{E_6}_{\bf2430} + \chi^{E_6}_{\bf2925}
+ \chi^{E_6}_{\bf43758}+ \chi^{E_6}_{\bf105600}
\big]\Big) x^{7} \cr
&+\Big(\chi_7(y) \big[1+\chi^{E_6}_{\bf78} \big] + \chi_5(y) \big[4+ 5\chi^{E_6}_{\bf78}+2\chi^{E_6}_{\bf650}+3\chi^{E_6}_{\bf2430} + 2\chi^{E_6}_{\bf2925} \big] \cr
&+ \chi_3(y)\big[
3+8\chi^{E_6}_{\bf78}+\chi^{E_6}_{\bf650}+3\chi^{E_6}_{\bf2430} + 3\chi^{E_6}_{\bf2925}
+\chi^{E_6}_{\bf34749}+ 2\chi^{E_6}_{\bf43758}+ 2\chi^{E_6}_{\bf105600}
\big]\cr
&
+3+2\chi^{E_6}_{\bf78}+\chi^{E_6}_{\bf650}+3\chi^{E_6}_{\bf2430} + \chi^{E_6}_{\bf2925}
+ \chi^{E_6}_{\bf70070}+ \chi^{E_6}_{\bf105600}+\chi^{E_6}_{\bf537966}
\Big) x^{8} 
+{\cal O}(x^{9}),
\end{align}
where we used the branching rule
\begin{align}
E_6&\supset SO(10)\times U(1)\cr
{\bf 78}&={\bf 1}_0 +{\bf 16}_{1}+ \overline{\bf 16}_{-1}+{\bf 45}_{0}\cr
{\bf 2430}&={\bf 1}_0+{\bf 16}_{1}+ \overline{\bf 16}_{-1}+{\bf 45}_{0}+{\bf 126}_{-2}+ \overline{\bf 126}_{2}+{\bf 210}_{0}+{\bf 560}_{1}+\overline{\bf 560}_{-1}+ {\bf 770}_{0}\cr
{\bf 43758}&={\bf 1}_0+{\bf 16}_{1}+ \overline{\bf 16}_{-1}+{\bf 45}_{0}+{\bf 126}_{-2}+ \overline{\bf 126}_{2}+{\bf 210}_{0}
+{\bf 560}_{1}+\overline{\bf 560}_{-1}+ {\bf 770}_{0}\cr
&+{\bf 672}_{-3}+ \overline{\bf 672}_{3}+{\bf 1440}_{1}+\overline{\bf 1440}_{-1}+{\bf 3696}'_{-2}+\overline{\bf 3696}'_{2}+{\bf 5940}_{0}+{\bf 7644}_{0}\cr
&+{\bf 8064}_{1}+\overline{\bf 8064}_{-1}\cr
{\bf 537966}&= {\bf 1}_{0}+{\bf 16}_{1}+\overline{\bf 16}_{-1}+{\bf 45}_{0}+{\bf 126}_{-2}+\overline{\bf 126}_{2}+{\bf 210}_{0}+{\bf 560}_{1}+\overline{\bf 560}_{-1}\cr
&+\overline{\bf 672}_{3}+{\bf 672}_{-3}+{\bf 770}_{0}+{\bf 1440}_{1}+\overline{\bf 1440}_{-1}+{\bf 2772}_{-4}+\overline{\bf 2772}_{4}+{\bf 3696}'_{-2}+\overline{\bf 3696}'_{2}\cr
&
+{\bf 5940}_{0}+{\bf 6930}'_{-2}+\overline{\bf 6930}'_{2}+{\bf 7644}_{0}+{\bf 8064}_{1}+\overline{\bf 8064}_{-1}+{\bf 8910}_{0}
\cr
&
+\overline{\bf 17280}_{3}+{\bf 17280}_{-3}+{\bf 34992}_{1}+\overline{\bf 34992}_{-1}+{\bf 46800}_{-2}+\overline{\bf 46800}_{2}+{\bf 52920}_{0}
\cr
&
+{\bf 70560}_{1}+\overline{\bf 70560}_{-1}+{\bf 73710}_{0}. 
\end{align}
The result shows that the two instantons again starts to contribute from $x^4$ order in $\chi^{E_6}_{\bf2430}$. Three instanton contribution is at order $x^6$; it is captured in the character $\chi^{E_6}_{\bf43758}$, and  four instanton contribution appear at order $x^8$ with the character $\chi^{E_6}_{\bf537966}$.


\subsection{Index for $N_f=6,7$}\label{Nf67}
As stated before, the way that instanton contributions for $N_f=6, 7$ cases appear in the index is different from the lower $N_f$ cases. Namely, two instanton contributions for such cases do show up in lower powers of $x$, starting at order $x^2$.

To better understand these case, recall how the symmetry enhancement works for $N_f<6$.
By tracing the structure of the index, we deduce that the perturbative part and the instanton part together form a single state in the adjoint representation of $E_{N_f+1}$ at the leading order of $x$, at order $x^2$. The perturbative part comprises two states in the representations of $SO(2N_f)$: the singlet and the adjoint representation. One instanton and anti-instanton parts, on the other hand, both provide two states in the spinor and its conjugate representations of $SO(2N_f)$. These states altogether fit into the $SO(2N_f)\times U(1)_I$ decomposition of the adjoint representation of $E_{N_f+1}$ at order $x^2$, and for this reason, they can be viewed as the states of the adjoint representation of $E_{N_f+1}$  for $N_f<6$.

For the $N_f=6, 7$ cases, the adjoint representations of $E_{N_f+1}$ are big enough to have rooms for other $SO(2N_f)$ representations than the spinor and adjoint representations.
More specifically, the adjoint representation of $E_{N_f+1}$  contains the extra states with the $U(1)_I$ charge of $\pm2$ in the decomposition to $SO(2N_f)\times U(1)_I$ which correspond to two instanton contributions.
If we believe the $E_{N_f}$ symmetry enhancement of the conformal theories, these extra states must be seen at order $x^2$ of the superconformal index. It seems, however, difficult to extract the extra states at order $x^2$ from two instanton contributions due to the pole structure of the integral formula \eqref{integral-formula1}. In other words, if we evaluate the integral by taking into account all poles inside the unit circle $|e^{i\phi_1}|=1$, the expansion of two instanton index in powers of $x$ starts  at order $x^4$
rather than at order $x^2$ regardless of the number of flavors $N_f$.
This is troublesome for $N_f=6, 7$, because two instanton contribution is supposed to appear at order $x^2$.
This means that naive pole prescription does not lead to the right result and thus it should be treated with care.

The subtlety arises because the unphysical pole at $e^{i\phi_1}=0$ appears in the contour integral when $N_f\ge 6$.
Unfortunately, we could not find a consistent prescription for this unphysical pole.
One can naively try to exclude the contribution from the pole at $e^{i\phi_1}=0$.
However this attempt yields two instanton states at order less than $x^2$.
There are no other states to be combined with these two instanton states in order to form an irreducible representation of $E_{N_f+1}$.

We here instead write our prediction for two instanton contributions to order $x^3$, based on the symmetry enhancement that naturally leads to the adjoint representation of $E_{N_f+1}$.
We see that the perturbative and one instanton contributions of $N_f=6,7$ cases give the right spectrum for the symmetry enhancement and thus two instanton contributions should arrange themselves to yield the adjoint representation of $E_{N_f+1}$.

It follows from the perturbative and one instanton contributions that the index for $N_f=6$ is given by
\begin{align}
I= &~ 1+ \big( 1+ \chi^{SO(12)}_{\bf 66} + q \,\chi^{SO(12)}_{\bf 32}+ q^{-1}\,\chi^{SO(12)}_{\bf 32} +\cdots \big)\,x^2 \cr
&+\chi_2(y) \Big[1+ (1+ \chi^{SO(12)}_{\bf 66}+ q \,\chi^{SO(12)}_{\bf 32}+ q^{-1}\,\chi^{SO(12)}_{\bf 32}+\cdots)\Big]x^{3} + \mathcal{O}(x^4),
\end{align}
where $\cdots$ represents two instanton contributions which we have not been able to compute.
An obvious guess for the index is that the superconformal index, taken into account the correct two instanton contribution, would be expressed as
\begin{align}\label{I4e7-1}
I&= 1+ \big( 1+ \chi^{SO(12)}_{\bf 66} + q \,\chi^{SO(12)}_{\bf 32}+ q^{-1}\,\chi^{SO(12)}_{\bf 32}
+q^2  + q^{-2}\big)\,x^2 \cr
&\quad+\chi_2(y) \Big[1+ (1+ \chi^{SO(12)}_{\bf 66}+ q \,\chi^{SO(12)}_{\bf 32}+ q^{-1}\,\chi^{SO(12)}_{\bf 32}+q^2  + q^{-2})\Big]x^{3} + \mathcal{O}(x^4).\\
&=1+ \chi^{E_7}_{{\bf 133}} \,x^2 +\chi_2(y) \big[1+  \chi^{E_7}_{{\bf 133}}\big]x^{3} + \mathcal{O}(x^4),
\end{align}
which is based on the branching rules
\begin{align}
E_7&\supset SO(12)\times U(1)_I\cr
{\bf 133}&={\bf 66}_0 + {\bf 32}_1+ {\bf 32}_{-1} + {\bf 1}_2+ {\bf 1}_0+{\bf 1}_{-2}.
\end{align}
It would then again exhibit global symmetry enhancement to $E_7$ with not only one instanton but also two instantons appearing at order $x^2$. 

In the same way, the superconformal index for $N_f=7$ up to one instanton contributions is given by
\begin{align}
I= &~ 1+ \big( 1+ \chi^{SO(14)}_{\bf 91} + q \,\chi^{SO(14)}_{\bf 64}+ q^{-1}\,\chi^{SO(14)}_{\overline{\bf 64}} +\cdots \big)\,x^2 \cr
&+\chi_2(y) \Big[1+ (1+ \chi^{SO(14)}_{\bf 91}+ q \,\chi^{SO(14)}_{\bf 64}+ q^{-1}\,\chi^{SO(14)}_{\overline{\bf 64}}+\cdots)\Big]x^{3} + \mathcal{O}(x^4).
\end{align}
When combined with correct two instanton contributions, this would exhibit symmetry enhancement to $E_8$   with the following form
\begin{align}\label{I4e8-2}
I&= ~ 1+ \big( 1+ \chi^{SO(14)}_{\bf 91} + q \,\chi^{SO(14)}_{\bf 64}+ q^{-1}\,\chi^{SO(14)}_{\overline{\bf 64}} + q^2 \,\chi^{SO(14)}_{\bf 14}+ q^{-2}\,\chi^{SO(14)}_{\overline{\bf 14}}\big)\,x^2 \cr
&+\chi_2(y) \Big[1+ (1+ \chi^{SO(14)}_{\bf 91}+ q \,\chi^{SO(14)}_{\bf 64}+ q^{-1}\,\chi^{SO(14)}_{\overline{\bf 64}}+q^2 \,\chi^{SO(14)}_{\bf 14}+ q^{-2}\,\chi^{SO(14)}_{\overline{\bf 14}})\Big]x^{3} + \mathcal{O}(x^4)\cr
&= 1+ \chi^{E_8}_{\bf 248} \,x^2 + \chi_2(y) \big[1+  \chi^{E_8}_{\bf 248}\big]x^{3} +\mathcal{O}(x^5),
\end{align}
where the branching rule is
\begin{align}
E_8 &\supset SO(14)\times U(1)_I\cr
{\bf 248}&={\bf 1}_0 +{\bf 91}_0 + {\bf  64}_1+\overline{\bf 64}_{-1}+{\bf 14}_{2}+{\bf 14}_{-2}.
\end{align}
Just as the $N_f=6$ case, both one and two instanton contributions appear at order $x^2$.\footnote{String origin of the massless two instanton states ($\mathbf{1}_{\pm2}$ for $E_7$ and $\mathbf{14}_{\pm2}$ for $E_8$) is discussed in \cite{Bergman:1997py}.}

We close this section by reporting an observation.
It is clear that not all representations of $E_{N_f+1}$ appear in the superconformal index for $N_f$. Singlets and  only representations associated with the adjoint representation show up. One may notice that the representations of $SO(2N_f)$ combine themselves to yields the products of the adjoint representations of $E_{N_f+1}$. 

To make this observation concrete, let us call the adjoint representation $adj$.
The tensor product of two $adj$'s splits into symmetric and antisymmetric parts
\begin{align}
adj\otimes adj = (adj\otimes adj )_A + (adj\otimes adj )_S.
\end{align}
The 2nd symmetrized tensor products are generically written as
\begin{align}
(adj\otimes adj )_A&=    adj\oplus R_A^{(2)},\cr
(adj\otimes adj )_S &= 1  \oplus R_S^{(2)}  \oplus adj^2 ,
\end{align}
where the notation $adj^k$ to denote the irreducible representations whose weight is given as $k$ times that for the adjoint representation, and $R_A^{(2)}$ and $R_S^{(2)}$ represent remaining other representations. 
The superconformal indices that we have computed for $N_f=2,\ldots, 5$ then take the form
\begin{align}\label{master}
I&= 1+ \chi_{\bf adj} \,x^2 + \chi_2(y) \big[1+\chi_{\bf adj}\big]x^{3} + \Big(\chi_3(y) \big[1+ \chi_{\bf adj}\big] +1+ \chi_{\bf adj^2}\Big)x^4 \cr
&+\Big(\chi_4(y) \big[1+ \chi_{\bf adj}\big] +\chi_2(y)\big[1+ \chi_{\bf adj^2} + \chi_{({\bf adj\otimes adj})_A}\big]\Big)x^{5} \cr
&+\Big(\chi_5(y) \big[1+ \chi_{\bf adj}\big] +\chi_3(y)\big[1+\chi_{\bf adj}+ \chi_{\bf adj^2}+ \chi_{\bf adj\otimes adj}\big]
+ \chi_{\bf adj}+ \chi_{\bf adj^3}+ \chi_{({\bf adj\otimes adj })_A}\Big)x^{6} \cr
&+ {\cal O}(x^{7}),
\end{align}
where $\chi_{{\bf adj}^k} \, (= \chi^{E_{N_f+1}}_{{\bf adj}^k}) $ is the character of $adj^k$ of the enhanced global symmetry group $E_{N_f+1}$, and $\chi_{({\bf adj\otimes adj})}=\chi_{\bf adj} \chi_{\bf adj}$. 
Recall that as $adj$ is associated with one instanton contribution, its second order tensor products contain two instanton contributions. In particular, $adj^2$ is the representation that precisely possesses the two instanton contributions  which appears from order $x^4$. In a similar way, the third order tensor products contain three instanton contributions, and $adj^3$ is the representation first appears at order $x^6$ and possesses three instantons. This pattern is expected to proceed to higher orders in $x$: $adj^k$ appear at $x^{2k}$ of the superconformal index and are those which first show $k$ instanton contributions for $N_f=2,\ldots, 5$.
The products of these $N_f$ are given in Appendix \ref{branching}.

Even though we do not have enough data for higher order terms in $x$ for $N_f=6, 7$, one may anticipate that the index for them would follow  the pattern of the index \eqref{master} as well, irrespective of how the instanton contributes the enhancements.
If so, the superconformal index for $N_f=6$ would take the form
\begin{align}\label{I4e7-1}
I&= 1+ \chi^{E_7}_{133} \,x^2 + \chi_2(y) \big[1+  \chi^{E_7}_{133}\big]x^{3} + \Big(\chi_3(y) \big[1+  \chi^{E_7}_{133}\big] +1+  \chi^{E_7}_{7371}\Big)x^4 \cr
&+\Big(\chi_4(y) \big[1+ \chi^{E_7}_{133}\big] +\chi_2(y)\big[1+  \chi^{E_7}_{7371} +  \chi^{E_7}_{(133\otimes 133 )_A}\big]\Big)x^{5} \cr
&+\Big(\chi_5(y) \big[1+  \chi^{E_7}_{133}\big] +\chi_3(y)\big[1+ \chi^{E_7}_{133}+ \chi^{E_7}_{7371} + \chi^{E_7}_{133\otimes 133} \big]
+ \chi^{E_7}_{133}+ \chi^{E_7}_{238602} + \chi^{E_7}_{(133\otimes 133)_A}\Big)x^{6} \cr
&+ {\cal O}(x^{7}),
\end{align}
where $adj\sim {\bf 133}$, $adj^2\sim {\bf 7371}$, and $adj^3\sim {\bf 238602}$.
A few relevant tensor products of the adjoint representation of $E_7$ are given as
\begin{align}
({\bf 133}\times{\bf 133})_S &= {\bf 1}+ {\bf 1539}+ {\bf 7371},\cr
({\bf 133}\times{\bf 133})_A &= {\bf 133}+ {\bf 8645}.
\end{align}
For $N_f=7$, the superconformal index would take the form
\begin{align}\label{I4e8-1}
I&= 1+ \chi^{E_8}_{248} \,x^2 + \chi_2(y) \big[1+  \chi^{E_8}_{248}\big]x^{3} + \Big(\chi_3(y) \big[1+  \chi^{E_8}_{248}\big] +1+  \chi^{E_8}_{27000}\Big)x^4 \cr
&+\Big(\chi_4(y) \big[1+ \chi^{E_8}_{248}\big] +\chi_2(y)\big[1+  \chi^{E_8}_{27000} +  \chi^{E_8}_{(248\otimes 248)_A}\big]\Big)x^{5} \cr
&+\Big(\chi_5(y) \big[1+  \chi^{E_8}_{248}\big] +\chi_3(y)\big[1+ \chi^{E_8}_{248}+ \chi^{E_8}_{27000} + \chi^{E_8}_{248\otimes 248} \big]
+ \chi^{E_8}_{248}+ \chi^{E_8}_{1763125} + \chi^{E_8}_{(248\otimes 248)_A}\Big)x^{6} \cr
&+ {\cal O}(x^{7}),
\end{align}
where $adj\sim {\bf 248}$, $adj^2\sim {\bf 27000}$, and $adj^3\sim {\bf 1763125}$, and relevant tensor products of the adjoint representation of $E_8$ are
\begin{align}
({\bf 248}\times{\bf 248})_S &= {\bf 1}+ {\bf 3875}+ {\bf 27000},\cr
({\bf 248}\times{\bf 248})_A &= {\bf 248}+ {\bf 30380}.
\end{align}
It would be interesting to see whether the index respects the pattern and whether there is a closed form for the result like a sort of Plethystic expansion.


\subsection{Superconformal index for $Sp(2)$}
When $N\ge 2$, as discussed in Sec \ref{sec:sp-gaugegr}, an additional antisymmetric hypermultiplet does not decouple and contribute to symmetry enhancement. In this case, we have the chemical potential $m$ for $SU(2)$ global symmetry of the antisymmetric hypermultiplet. Similar difficult pole structures as the case for $N=1$ still reside when we evaluate two instanton contributions. We believe that correct prescription will take care of irrelevant poles.

We computed the superconformal index for various $N_f$ and found that there is a universal expression for $N_f\le 6$. We see that the global symmetry $SO(2N_f)$ is enhancement to $E_{N_f+1}$, as expected, and the index is given by
\begin{align}
I&= 1+ \chi_2(e^{im})\,x + \Big(\chi_2(y)\chi_2(e^{im})  +2\chi_3(e^{im})+ \chi_{\bf adj}  \Big) x^2\cr
&+\Big(\chi_3(y) \chi_2(e^{im}) + \chi_2(y) [2+2\chi_3(e^{im})+\chi_{\bf adj}]
+ 2\chi_4(e^{im}) + \chi_2(e^{im})+2 \chi_2(e^{im})\chi_{\bf adj}
\Big) x^3\cr
&+\Big( \chi_4(y)\chi_2(e^{im})+\chi_3(y) [2+ 3\chi_3(e^{im})+\chi_{\bf adj}]
+ \chi_2(y) [3\chi_2(e^{im}) \chi_{\bf adj} + 3\chi_4(e^{im})+ 5\chi_2(e^{im}) ]\cr
&+ 3\chi_5(e^{im})+ \chi_3(e^{im}) + 3\chi_3(e^{im})\chi_{\bf adj} + 3 +\chi_{\bf (adj\otimes adj)_S} +\chi_{\bf adj}\Big)x^4 + \mathcal{O}(x^5),
\end{align}
where $\chi_\mathbf{adj}$ stands for the character of the adjoint representation for $E_{N_f+1}$ and their tensor products are given in Appendix \ref{branching}.
$\chi_{\rm dim}(y)$ is the $SU(2)$ character for the fugacity $y$ and $\chi_{\rm dim}(e^{im})$ is the $SU(2)$ character for the chemical potential $m$. For instance, the 2-dim representation is $\chi_2(e^{im}) = e^{im}+e^{-im}$ and the 3-dim representation is $\chi_3(e^{im}) = e^{2im}+1+e^{-2im}$.

As a representative the index for such hypermultiplets, we write the superconformal result for $N_f=5$
\begin{align}
I&= 1+ \chi_2(e^{im})\,x + \Big(\chi_2(y)\chi_2(e^{im})  +2\chi_3(e^{im})+ \chi^{E_6}_{\bf 78}  \Big) x^2\cr
&+\Big(\chi_3(y) \chi_2(e^{im}) + \chi_2(y) [2+2\chi_3(e^{im})+\chi^{E_6}_{\bf 78}]
+ 2\chi_4(e^{im}) + \chi_2(e^{im})+2 \chi_2(e^{im})\chi_{\bf 78}^{E_6}
\Big) x^3\cr
&+\Big( \chi_4(y)\chi_2(e^{im})+\chi_3(y) [2+ 3\chi_3(e^{im})+\chi^{E6}_{\bf 78}]
+ \chi_2(y) [3\chi_2(e^{im}) \chi^{E_6}_{\bf 78} + 3\chi_4(e^{im})+ 5\chi_2(e^{im}) ]\cr
&+ 3\chi_5(e^{im})+ \chi_3(e^{im}) + 3\chi_3(e^{im})\chi^{E_6}_{\bf 78} + 4 +\chi^{E6}_{\bf 2430} +\chi^{E6}_{\bf 650} +\chi^{E6}_{\bf 78}\Big)x^4 + \mathcal{O}(x^5).
\end{align}
Here, $\chi^{E_6}_{\bf 78}=\chi_{\bf adj}$ and $1+\chi^{E6}_{\bf 2430} +\chi^{E6}_{\bf 650}=\chi_{\bf (adj\otimes adj)_S}$.  As for the $Sp(1)$ cases, one instanton start to contribute at $x^2$ through $\chi_{\bf adj}$  while two instanton contributions would first appear at  $x^4$ through $\chi_{\bf adj^2}$ which is contained in $\chi_{\bf (adj\otimes adj)_S}$. We note that we have encountered similar obstacles in evaluating pole integral arising when one deals with two instanton contribution as for $Sp(1)$ case. So we leave this issue as it is, and here our expression for the two instanton contribution should be understood as an educed guess. All terms are well organized as above except for terms proportional to $q^{\pm 4}$ that is supposed to be taken care of from the pole integration associated with two instanton contributions.  For this reason, the index for $N_f=6,7$ are to be resolved as one finds the correct pole prescriptions.


\section{Conclusion and Discussions}\label{sec:discussions}

In this work we have set up and calculated the superconformal index for 5-dim gauge theories. The index has both perturbative and nonperturbative instanton contributions.  For $Sp(1)$ gauge group and $N_f\le 5$ fundamental hypermultiplets, we explicitly computed superconformal index up to four-instanton contributions. 
Our result for the superconformal index precisely shows the enhancement of the global symmetry from $SO(2N_f)\times U(1)$ to $E_{N_f+1}$ for $N_f\le 5$, which is expected to exist at the conformal fixed point. 
The characters of $SO(2N_f)\times U(1)$ appearing in the index reorganize themselves so that they belong to the characters of $E_{N_f+1}$. For lower powers of $x$ which is an expansion parameter related to energy, there is a universal pattern on how the $E_{N_f+1}$ characters appears as forms of  (anti-)symmetric products of the adjoint representation of $E_{N_f+1}$. 
The $N_f=0,1$ cases show fermionic contributions in the process of the symmetry enhancement to $E_1, E_2$, in terms of a singlet and the fundamental representation, respectively. The $N_f\ge2$ cases, however, no fermionic contributions appear at least to order $x^8$ that we have explicitly computed. It would be interesting to explore further how these fermionic contributions arise as instanton contributions. For $N_f=6,7$, we obtained one-instanton contribution only which also shows the symmetry enhancement to that order. Two or higher instanton contributions give rise difficult pole structures in the contour integral which we do not have a clear prescription to avoid irrelevant poles.

For $N\ge2$, we expect the superconformal index also exhibits the symmetry enhancement. As a simple example, we computed the superconformal index for $Sp(2)$ to the order which contains one-instanton contributions. To this order, we see that the index indeed shows the symmetry enhancement to $E_{N_f+1}$ as well. 
To include two or higher instanton contributions, just as for the  $N_f=6,7$ of $Sp(1)$, we have encountered similar obstacle related to pole structures. 

There are several directions to pursue from this point. We would like to resolve the obstacles that are mentioned above.   The investigation of  other 5-dim conformal field theories by our method may shed some new light on these theories. It would be interesting to evaluate the partition function on $S^5$ which also contains the information about the symmetry enhancement  and the degrees of freedom in large $N$ limit. See the recent work \cite{Kallen:2012cs,Kallen:2012va,kim:2012av} for the perturbative part of the partition function of 5-dim supersymmetric YM theory on $S^5$.

An interesting application with our result would be the dimensional reduction along ${S}^1$ to 4d theories on ${S}^4$ type manifold.
The 4d reduction then leads to the 4d theories whose base manifold ${S}^4$ is squashed by
the background gauge fields coupled to both the KK modes and the internal symmetries due to the chemical potentials we have turned on in 5d theories.
Concretely, we turned on the chemical potentials for the generators $2(j_1+R)$ and $2j_2$, which obviously break the isometry on the base ${S}^4$ into $U(1)_1\times U(1)_2$.
Upon the dimensional reduction, the resulting 4-manifold at which 4d theories are defined becomes the ellipsoid with $U(1)^2$ isometries.
As we can trade the generator $2(j_1+R)$ to other combinations of $\epsilon_0$, $j_1$, and $R$ using the BPS relation,
there is indeed a family of 4d ellipsoids with $U(1)^2$ isometries. The 4d ellipsoid given in \cite{Hama:2012bg} would probably be one of these ellipsoids
after correctly identifying the chemical potentials with the squashing parameters in the ellipsoid.
The simplest case would occurs when the generator $2(j_1+R)$ is replaced by $\epsilon_0-R$.
In this case the base manifold is deformed only by $j_2$ and thus it becomes 4d manifold preserving $SU(2)_1\times U(1)_2$
whose local geometry at the equator is the squashed  ${\rm S}^3$ investigated in \cite{Hama:2011ea, Imamura:2011wg}.
The chemical potential $\gamma_2$ is identified with the squashing parameter $u\sim (l^{-1}-\tilde{l}^{-1})^{1/2}$ on the {\it squashed} four-sphere.

We expect that our superconformal index reduces to 4d partition function on the squashed ${S}^4$ after the reduction along the time circle.
The reduction of the index can be easily achieved by removing the nonzero KK modes from our perturbative and instanton indices.
One may already notice that the one-loop determinants \eqref{1-loopvec} and \eqref{1-loophyp}, and the instanton indices \eqref{contour-integral} and so on are identical to the corresponding 4d results
upon the KK reduction on ${S}^1$ and the identification of the chemical potentials to the 4d parameters.
However, we may not be able to get the classical contribution in 4d partition function, proportional to the square of scalar vev of the vector multiplet, from our index,
which is analogous to the cases of 4d to 3d reduction considered in \cite{Dolan:2011rp,Gadde:2011ia,Imamura:2011uw,Benini:2011nc} where 3d Chern-Simons terms cannot be obtained under the reduction.
This would imply that 4d reduction of our indices leads to 4d partition functions on the generalized four-spheres
apart from the classical contributions.

Our index may count some M theoretic objects wrapping degenerate del Pezzo surfaces which also appear naturally in the (p,q)-web description of our theory with enhanced exceptional groups \cite{DeWolfe:1999hj}. The detail identification would be interesting.  As our Higgs phase is the moduli space of $E_{N_f+1}$ instantons, our calculation for $Sp(1)$ may have something to say the moduli space of a single $E_{N_f+1}$ instanton. While our calculation is done for small $N=1$, one can imagine the large $N$ limit and  compare to the gravity calculation which is also needed to be done.

We finally make remarks on the AdS/CFT correspondence of the 5d conformal theories.
The gravity dual of the $Sp(N)$ gauge theory with $N_f$ fundamental flavors is 
a warped product of $AdS_6 \times S^4$ whose gauge/gravity duality is studied in various literatures \cite{Ferrara:1998gv,Brandhuber:1999np,Bergman:2012kr}.
The spectrum of the gauge invariant operators in the boundary field theory we counted here
amounts to the gravity spectrum in the bulk and the KK spectrum of the $E_{N_f+1}$ twisted sector living on the boundary of the gravity theory on wrapped $AdS_6\times S^4$ background. The duality maps the $E_{N_f+1}$ neutral operators in the field theory to the bulk gravitons, and the $E_{N_f+1}$ charged operators maps to the twisted sectors. It would be interesting to see how our index matches to the index on the gravity side in the large $N$ limit.


\vskip 1.0cm
\hspace*{-0.8cm} {\bf\large Acknowledgements}
\vskip 0.2cm
\hspace*{-0.75cm} We are grateful to  Dongmin Gang, Eunkyung Koh, Noppadol Mekareeya, Hiroaki Nakajima, Sangmin Lee, Sungjay Lee, Vasily Pestun, Nathan Seiberg, and Jaewon Song for discussions and encouraging comments. 
We especially thank Seok Kim and Takuda Okuda for valuable discussions.
K.L. is grateful for Benasque Center for Science and Newton Institute for Mathematical Science where the part of this work is done. This work is supported   by   the National Research Foundation of Korea (NRF) Grants No. 2010-0007512 (HK),  2006-0093850, 2009-0084601 (KL) and 2005-0049409 through the Center for Quantum Spacetime(CQUeST) of Sogang University (KL).
\vskip 1.5cm


\centerline{\Large \bf Appendix}

\appendix

\section{Notation}\label{appendixA}

In 5d Lorentzian flat spacetime, we can choose $4\times 4$ gamma matrices using $2\times 2$ Pauli matrices $\sigma_{1,2,3}$ as follows:
\be
\gamma^0 =-i{\bf 1}\otimes\sigma_3\,,\quad \gamma^1 = \sigma_1\otimes\sigma_1 \,, \quad \gamma^2=\sigma_2\otimes\sigma_1\,,\quad \gamma^3 = \sigma_3\otimes\sigma_1\,, \quad \gamma^4 ={\bf 1}\otimes\sigma_2 \ .
\ee
They satisfy the Clifford algebra $\{\gamma_\mu,\gamma_\nu\} =\eta_{\mu\nu}$ and
\be
&&\gamma^{01234} = i\,,\quad \gamma^{\mu\nu\lambda\rho} =i\epsilon^{\mu\nu\lambda\rho\sigma}\gamma_\sigma \ \ (\epsilon^{012345}=1)\,, \nn \\
&&\gamma^{\mu\nu\lambda} = \gamma^\mu\gamma^{\nu\lambda} +2 \gamma^{[\nu}\eta^{\lambda]\mu} = \gamma^{\nu\lambda}\gamma^\mu +2\eta^{\mu[\nu}\gamma^{\lambda]} \ ,
\ee
with a flat metric $\eta = {\rm diag}(-1,+1,+1,+1,+1)$.
In five dimensions, the spinor representation is the fundamental of $Sp(2)\cong SO(1,4)$ Lorentz rotation, so it is pseudo-real.
This implies that the ordinary Majorana condition cannot be imposed on 5d spinors.
Instead, we can impose symplectic-Majorana reality condition on the spinors if they carry $SU(2)_R$ symmetry charges.
When a spinor $\lambda^A$ is a doublet of $SU(2)_R$, the symplectic-Majorana condition is given by
\be
\bar\lambda_A = (\lambda^T)^B\varepsilon_{BA}\Omega \ ,
\ee
where $A,B=1,2$, the R-symmetry indices, and $\varepsilon=i\sigma^2$ is the $SU(2)_R$ invariant tensor. $\Omega$ is the symplectic form of $Sp(2)$ defined as
\be
\Omega = \gamma^{24} = i\sigma_2\otimes\sigma_3 \ .
\ee
The 5d gamma matrices satisfy the following Fierz identities:
\be
\delta^{\;\;q}_p\delta^{\;\;n}_m &=& \frac{1}{4}\delta^{\;\;n}_p\delta^{\;\;q}_m + \frac{1}{4}(\gamma^\mu)^{\;\;n}_{p}(\gamma_\mu)^{\;\;q}_m -\frac{1}{8}(\gamma^{\mu\nu})^{\;\;n}_p(\gamma_{\mu\nu})^{\;\;q}_m \,,\nn \\
(\gamma^\mu)^{\;\;q}_p(\gamma_\mu)_m^{\;\;n} &=& \delta_p^q\delta_m^n+\delta_p^n\delta_m^q-(\gamma^\mu)_p^{\;\;n}(\gamma_\mu)_m^{\;\;q} \ ,
\ee
where $m,n=1,2,3,4$ are the $Sp(2)$ indices.


\section{Theories on $S^1\times S^4$ and $S^5$}\label{appendixB}

We consider the Euclidean version of the Chern-Simons Lagrangian (\ref{conformal-Lagrangian}).
The flat space Lagrangian can easily be obtained by the Wick rotation, $x^0 = -i\tau$.
Then the Lagrangian on the Riemannian curved manifold can be derived from the flat Lagrangian by the conformal mapping,
whenever the curved manifold is related to the flat space by the conformal transformation.
Using this fact, the Chern-Simons Lagrangian on the curved space is derived from the Lagrangian (\ref{conformal-Lagrangian})
\be
    \mathcal{L} \!\!&\!\!=\!\!&\!\! \mathcal{L}_{cs} +\mathcal{L}_\kappa \,, \nn \\
    \mathcal{L}_{cs} \!\!&\!\!=\!\!&\!\! -i\frac{\kappa}{24\pi^2}{\rm tr}\,\bigg[A\wedge F\wedge F+\frac{i}{2}A\wedge A\wedge A \wedge F -\frac{1}{10}A\wedge A\wedge A\wedge A\wedge A  \nn \\
&& \qquad \qquad  +3i\bar\lambda\gamma^{\mu\nu}\lambda F_{\mu\nu}+6 \bar\lambda{\rm D}\lambda \bigg] \,, \nn \\
\mathcal{L}_\kappa \!\!&\!\!=\!\!&\!\! \frac{\kappa}{2\pi^2}{\rm tr}\,\phi\bigg[\frac{1}{2}F_{\mu\nu}F^{\mu\nu} +\nabla_\mu \phi \nabla^\mu\phi+\frac{R}{12}\phi^2-\frac{i}{2}\nabla_\mu\bar\lambda\gamma^\mu\lambda+\frac{i}{2}\bar\lambda \gamma^\mu \nabla_\mu \lambda - {\rm D}^I {\rm D}^I -i\bar\lambda[\phi,\lambda] \bigg], \qquad
\ee
where $R$ is the Ricci curvature for the curved manifold and $\bar\lambda\equiv\lambda^\dagger$.
The covariant derivative $\nabla_\mu$ includes the connection on the curved space as well as the gauge connection.
For example, the covariant derivative acting on a spinor field is $\nabla_\mu = D_\mu +\frac{1}{4}\omega_{\mu ab}\gamma^{ab}$
where $D_\mu=\partial_\mu -i A_\mu$ and  $\omega_{\mu ab}$ is the spin connection.
As the 5d rotational symmetry group is pseudo-real $SO(5)$,
the $SU(2)_R$ doublet spinors $\lambda_A$ is restricted to be the symplectic-Majorana spinor satisfying $\bar\lambda_A = (\lambda^T)^B\varepsilon_{BA}\Omega$.
The action on the curved space is invariant under the following supersymmetry transformation:
\be
    \delta A_\mu &=& i\bar\lambda\gamma_\mu \epsilon \,, \nn \\
    \delta \phi &=& \bar\lambda\epsilon \,, \nn \\
    \delta \lambda &=& \frac{1}{2}F_{\mu\nu}\gamma^{\mu\nu}\epsilon -i \nabla_\mu\phi\gamma^\mu \epsilon  +i{\rm D}^I\sigma^I\epsilon -\frac{2i}{5}\phi\gamma^\mu \nabla_\mu\epsilon\,, \nn \\
    \delta \bar\lambda &=& -\frac{1}{2}F_{\mu\nu}\bar\epsilon\gamma^{\mu\nu} -i \bar\epsilon\gamma^\mu \nabla_\mu \phi -i\bar\epsilon\sigma^I {\rm D}^I -\frac{2i}{5}\nabla_\mu\bar\epsilon\gamma^\mu\phi\,, \nn \\
    \delta {\rm D}^I &=& \nabla_\mu\bar\lambda \gamma^\mu\sigma^I\epsilon -[\phi,\bar\lambda]\sigma^I\epsilon-\frac{1}{5}\bar\lambda\sigma^I\gamma^\mu \nabla_\mu\epsilon\ ,
\ee
where $\epsilon$ is the Killing spinor satisfying the Killing spinor equation
\be
    \nabla_\mu\epsilon = \gamma_\mu\tilde{\epsilon} \ ,
\ee
with an arbitrary spinor $\tilde\epsilon$.

The Lagrangian for the hypermultiplet can also be obtained by the conformal mapping from the flat space Lagrangian (\ref{matter-lagrangian}).
It is almost the same as the flat space Lagrangian but the scalar fields acquire the conformal mass term proportional to the scalar curvature $R$,
which reflects that the fields on the curved space non-trivially couples to the curvature.
The matter Lagrangian is then
\be\label{hyper-actionB}
    \mathcal{L}_{matter}=|\nabla_\mu q|^2 -i\bar\psi\gamma^\mu\nabla_\mu\psi +\frac{3}{16}R\,\bar{q}q +\bar{q}\phi^2q - q\sigma^I\bar{q} {\rm D}^I  -\sqrt{2}\bar\psi\lambda q +\sqrt{2}\bar{q}\bar\lambda \psi -i\bar\psi\phi\psi
\ee
This Lagrangian is invariant under the supersymmetry transformation of the matter fields
\be\label{hyper-SUSYB}
    \delta q^A &=& \sqrt{2}i\bar\epsilon^A\psi \ ,\nn \\
    \delta \bar{q}_A &=& \sqrt{2}i\bar\psi\epsilon_A \ , \nn \\
    \delta\psi &=& \sqrt{2}(-\nabla_\mu q^A \gamma^\mu\epsilon_A +\phi q^A\epsilon_A-\frac{3}{5}q^A\gamma^\mu\nabla_\mu\epsilon_A) \ ,\nn \\
    \delta\bar\psi &=& \sqrt{2}(\bar\epsilon^A\gamma^\mu\nabla_\mu\bar{q}_A+\bar\epsilon^A\bar{q}_A\phi+\frac{3}{5}\nabla_\mu\bar\epsilon^A\gamma^\mu\bar{q}_A) \ .
\ee


\subsection{Killing spinors on ${S}^1\times {S}^4$}
The explicit forms of the conformal Killing spinors on ${S}^1\times {S}^4$ depend on the choice of the metric and the vielbein basis.
We choose the ${S}^1\times {S}^4$ metric as follows:
\begin{eqnarray}
    ds^2 \!\!\!&=&\!\!\! d\tau^2 + ds_{S^4}^2 \,, \nonumber \\
    ds^2_{S^4} \!\!\!&=&\!\!\! d\theta_1^2+ \sin^2\theta_1d\theta_2^2
    +\sin^2\theta_1\sin^2\theta_2d\theta_3^2+ \sin^2\theta_1\sin^2\theta_2\sin^2\theta_3d\theta_4^2 \ ,
\end{eqnarray}
where the radii of the ${S}^1$ and ${S}^4$ are set to unit radius.
We choose the vielbein basis as the following standard form:
\begin{eqnarray}
    \!\!e^1 \!=\! d\theta_1 \,, \quad e^2 \!=\!\sin\theta_1 d\theta_2 \,, \quad e^3 \!=\! \sin\theta_1\sin\theta_2d\theta_3 \,,
    \quad e^4 \!=\! \sin\theta_1\sin\theta_2\sin\theta_3d\theta_4\,,\quad e^5 \!=\! d\tau \ .
\end{eqnarray}
With this choice, one can easily compute the spin connection whose components are given by
\begin{eqnarray}
    &&\hspace{-1.3cm}\omega^1_{\;\;2} = -\cos\theta_1 d\theta_2\,, \quad \omega^2_{\;\;3} = -\cos\theta_2 d\theta_3 \,, \quad \omega^3_{\;\;4} = -\cos\theta_3 d\theta_4 \,, \nonumber \\
    &&\hspace{-1.3cm}\omega^1_{\;\;3} = -\cos\theta_1\sin\theta_2d\theta_3 \,, \quad \omega^2_{\;\;4} = -\cos\theta_2\sin\theta_3d\theta_4 \,, \quad \omega^3_{\;\;4} = -\cos\theta_1\sin\theta_2\sin\theta_3d\theta_4 \ .
\end{eqnarray}
The manifold ${S}^1\times {S}^4$ admits 8 independent Killing spinors taking of the forms
\begin{eqnarray}
    \epsilon^q = e^{-\frac{1}{2}\tau} e^{\frac{1}{2}\theta_1\gamma^{51}}e^{\frac{1}{2}\theta_2\gamma^{12}}e^{\frac{1}{2}\theta_3\gamma^{23}}e^{\frac{1}{2}\theta_4\gamma^{34}}\epsilon_0^q \ , \quad
\epsilon^s = \gamma^5e^{\frac{1}{2}\tau} e^{\frac{1}{2}\theta_1\gamma^{51}}e^{\frac{1}{2}\theta_2\gamma^{12}}e^{\frac{1}{2}\theta_3\gamma^{23}}e^{\frac{1}{2}\theta_4\gamma^{34}}\epsilon_0^s\ .
\end{eqnarray}
where $\epsilon_0^{q,s}$ are constant spinors. They satisfy the Killing spinor equations
\begin{eqnarray}
    \nabla_\mu\epsilon^q =- \frac{1}{2}\gamma_\mu\gamma^5\epsilon^q , \quad \nabla_\mu\epsilon^s = \frac{1}{2}\gamma_\mu\gamma^5\epsilon^s.
\end{eqnarray}
The supercharges $Q_m^A$ and $S_A^m$ appearing in the superconformal algebra are parametrized by these Killing spinors.
The superconformal algebra requires the symplectic-Majorana condition on the constant spinor parameters
\be
    (\epsilon^q_0)^* = \varepsilon\,\Omega\,\epsilon^s_0
\ee


\section{$Q$-exact deformation}
In Section 3.1, we localize the path integral by deforming the Lagrangian with $Q$-exact terms
which respect the supercharge $Q+S$.
We now present the $Q$-exact terms used for the localization explicitly.

Firstly, we introduce the following $Q$-exact terms for the vector multiplet:
\be
    \delta \mathcal{L} = {\rm tr}\Big[t\delta_\epsilon((\delta_\epsilon\lambda)^\dagger \lambda)\Big] \ .
\ee
Here $\delta_\epsilon$ is the SUSY transformation with respect to the SUSY parameter $\epsilon$ defined in \eqref{SUSY-parameter}.
The parameter $\epsilon$ is a Grassmann-even spinor and normalized as $\bar\epsilon\epsilon=1$ throughout the localization procedure.
Then the bosonic terms become
\be
    \hspace{-.8cm}(\delta_\epsilon\lambda)^\dagger\delta_\epsilon\lambda \!\!\!&=&\!\!\!\!\!
     \left(\!\!-\frac{1}{2}\bar\epsilon\gamma^{\mu\nu}F_{\mu\nu}+i\bar\epsilon\gamma^\mu\nabla_\mu\phi+i\bar\epsilon\sigma^I{\rm D}^I+\frac{2i}{5}\phi\nabla_\mu\bar\epsilon\gamma^\mu\!\right)
      \!\!\left(\!\frac{1}{2}F_{\mu\nu}\gamma^{\mu\nu}\epsilon-i\nabla_\mu\phi\gamma^\mu\epsilon+i{\rm D}^I\sigma^I\epsilon-\frac{2i}{5}\phi\gamma^\mu\nabla_\mu\epsilon\!\right) \nn \\
    \!\!\!&=&\!\!\! \bar\epsilon\epsilon\frac{1}{2}F_{\mu\nu}F^{\mu\nu}-\frac{1}{4}\bar\epsilon\gamma^{\mu\nu\lambda\rho}\epsilon F_{\mu\nu}F_{\lambda\rho}+\bar\epsilon\epsilon(\nabla_\mu\phi)^2
    +\bar{\tilde\epsilon}\tilde\epsilon\phi^2-\frac{i}{2}(\bar{\tilde{\epsilon}}\gamma^5\gamma^{\mu\nu}\epsilon+\bar\epsilon\gamma^{\mu\nu}\gamma^5\tilde\epsilon)\phi F_{\mu\nu}-\bar\epsilon\epsilon{\rm D}^2\nn \\
    \!\!\!&=&\!\!\!F_{\tau\mu}F^{\tau\mu} + \frac{1}{2}F_{ij}F^{kl} + \frac{1}{4}cos\theta_1 \epsilon^{ijkl5}F_{ij}F_{kl} -i\bar{\tilde\epsilon}\gamma^5\gamma^{ij}\epsilon\phi F_{ij} +(\nabla_\mu\phi)^2 + \phi^2-{\rm D}^2 \nn \\
    \!\!\!&=&\!\!\!F_{\tau\mu}F^{\tau\mu} + \cos^2\frac{\theta_1}{2}(F^-_{ij}-\omega^-_{ij}\phi)^2+ \sin^2\frac{\theta_1}{2}(F^+_{ij}- \omega^+_{ij}\phi)^2  +(\nabla_\mu\phi)^2-{\rm D}^2 ,
\ee
with the following definitions
\be
    &&F^\pm_{ij} = \frac{1}{2}\left[F_{ij} \mp *F_{ij}\right] \ ,\nn \\
    &&\omega^+_{ij} = \frac{i}{2\sin^2\frac{\theta_1}{2}}\bar{\tilde\epsilon}^R\gamma^5\gamma^{ij}\epsilon^R \ , \quad \omega^-_{ij} = \frac{i}{2\cos^2\frac{\theta_1}{2}}\bar{\tilde\epsilon}^L\gamma^5\gamma^{ij}\epsilon^L
    \ , \quad \omega^+_{ij}\omega^{+ij} =\omega^-_{ij}\omega^{-ij}=1 ,
\ee
where $\gamma^5\epsilon^R = \epsilon^R ,\, \gamma^5\epsilon^L = -\epsilon^L$.
The fermionic terms are obtained by
\be
    \delta_\epsilon(\delta_\epsilon\lambda)^\dagger\lambda \!\!&\!\!=\!\!&\!\! -i\nabla_\mu(\bar\lambda\gamma_\nu\epsilon)\bar\epsilon\gamma^{\mu\nu}\lambda+i\nabla_\mu(\bar\lambda\epsilon)\bar\epsilon\gamma^\mu\lambda-\frac{2i}{5}\nabla_\mu\bar\epsilon\gamma^\mu\lambda (\bar\lambda\epsilon) +i[\bar\lambda\gamma_\mu\epsilon,\phi]\bar\epsilon\gamma_\mu\lambda \nn \\
    &&+i\left(D_\mu\bar\lambda\gamma^\mu\sigma^I\epsilon - [\phi,\bar\lambda]\sigma^I\epsilon-\frac{1}{5}\bar\lambda\sigma^I\gamma^\mu\nabla_\mu\epsilon\right)\bar\epsilon\sigma^I\lambda \nn \\
    \!\!&\!\!=\!\!&\!\! -i(\bar\lambda\gamma_\nu\epsilon)\bar\epsilon\gamma^\nu\gamma^\mu\nabla_\mu\lambda+
    3i(\bar\lambda\gamma_\nu\epsilon)\bar{\tilde{\epsilon}}\gamma^\nu\lambda-2i\bar{\tilde\epsilon}\lambda(\bar\lambda\epsilon)-i(\bar\lambda\sigma^A\epsilon)\bar\epsilon\sigma^I\gamma^\mu\nabla_\mu\lambda -i\bar{\tilde\epsilon}\sigma^I\lambda(\bar\lambda\sigma^I\epsilon) \nn \\
    &&-i[\phi,\bar\lambda\gamma^\mu\epsilon]\bar\epsilon\gamma_\mu\lambda -i[\phi,\bar\lambda\sigma^A\epsilon]\bar\epsilon\sigma^I\lambda \nn \\
    \!\!&\!\!=\!\!&\!\!  -i\bar\lambda\gamma^\mu\nabla_\mu\lambda -i[\phi,\bar\lambda]\lambda -2i\bar{\epsilon}\sigma^I\tilde\epsilon\bar\lambda\sigma^I\lambda
    +i\bar{\epsilon}\gamma^\mu\sigma^I\tilde\epsilon\bar\lambda\gamma_\mu\sigma^I\lambda+\frac{i}{2}\bar{\epsilon}\gamma^{\mu\nu}\tilde\epsilon\bar\lambda\gamma_{\mu\nu}\lambda\ .
\ee
Here we used for the last equality the 5d Fierz identities given in Appendix \ref{appendixA}.

For the localization of the hypermultiplet parts, we need to realize the off-shell supersymmetry for the hypermultiplet.
We do not have the off-shell extension of the action \eqref{hyper-actionB} and the SUSY transformation \eqref{hyper-SUSYB}
which exhibits the full superconformal algebra of the theory.
Instead one can find the off-shell formalism for any chosen supercharge, which partially breaks the superconformal algebra, using the method suggested in \cite{Hosomichi:2012ek}.
For any symplectic-Majorana spinor $\epsilon$, we can always find a pair of $\hat\epsilon^{A'}$ which satisfy \cite{Hosomichi:2012ek}
\be
    \bar\epsilon\epsilon = \bar{\hat\epsilon}\hat\epsilon \ , \quad \bar\epsilon_A \hat\epsilon^{A'} = 0 \ , \quad \bar\epsilon\gamma^\mu \epsilon + \bar{\hat\epsilon}\gamma^\mu\hat\epsilon=0\ .
\ee
where $\bar{\hat\epsilon}_{A'} = (\hat\epsilon^T)^{B'}\varepsilon_{B'A'}\Omega$ and $\varepsilon_{A'B'}$ is the invariant tensor of the new $SU(2)'$ symmetry under which $\hat\epsilon^{A'}$ transforms as a doublet.
The 4-component spinors $\epsilon^A$ and $\hat\epsilon^{A'}$ are all Grassmann-even spinors.
Let us introduce a pair of complex auxiliary scalars $F^{A'}$ and add the following Lagrangian
\be
    \Delta\mathcal{L} = -\bar{F}_{A'}F^{A'} \ .
\ee
Then this Lagrangian is invariant under the following off-shell supersymmetry transformation
with respect to any spinor $\epsilon$ and the corresponding $\hat\epsilon$
\be
    \delta q^A &=& \sqrt{2}i\bar\epsilon^A\psi \ ,\nn \\
    \delta \bar{q}_A &=& -\sqrt{2}i\bar\psi\epsilon_A  \ ,\nn \\
    \delta\psi &=& \sqrt{2}(-\nabla_\mu q^A \gamma^\mu\epsilon_A +\phi q^A\epsilon_A-\frac{3}{5}q^A\gamma^\mu\nabla_\mu\epsilon_A +i \hat\epsilon_{A'}F^{A'}) \ ,\nn \\
    \delta\bar\psi &=& \sqrt{2}(\bar\epsilon^A\gamma^\mu\nabla_\mu\bar{q}_A+\bar\epsilon^A\bar{q}_A\phi+\frac{3}{5}\nabla_\mu\bar\epsilon^A\gamma^\mu\bar{q}_A + i\bar{F}_{A'}(\hat\epsilon^\dagger)^{A'}) \ ,\nn \\
    \delta F^{A'} &=& \sqrt{2}(\hat\epsilon^\dagger)^{A'}(\gamma^\mu\nabla_\mu\psi+\phi\psi-\sqrt{2}\lambda_A q^A) \ , \nn \\
    \delta \bar{F}_{A'} &=& \sqrt{2}(\nabla_\mu \bar\psi\gamma^\mu-\bar\psi\phi - \sqrt{2}\bar{q}_A\bar\lambda^A)\hat\epsilon_{A'} \ .
\ee
One can easily check that the off-shell SUSY algebra closes such as
\be
    \delta^2q^A &=& \xi^\mu\partial_\mu q^A +i\Lambda q^A +\frac{3}{2}\rho q^A +\frac{3}{4}R^{IJ}(\sigma^{IJ}q)^A \ , \nn \\
    \delta^2\psi &=& \xi^\mu\partial_\mu\psi +\frac{1}{4}\Theta_{\mu\nu}\gamma^{\mu\nu}+i\Lambda\psi + 2\rho\psi \ ,\ \nn \\
    \delta^2 F^{A'} &=& \xi^\mu\partial_\mu F^{A'} +i\Lambda F^{A'}+\frac{5}{2}\rho F^{A'}+\frac{5}{4}\hat{R}^{IJ}(\hat\sigma^{IJ} F)^{A'}\ ,
\ee
where
\be
    \xi^\mu &=& -i\bar\epsilon\gamma^\mu \epsilon \ , \nn \\
    \Lambda &=& i\bar\epsilon\gamma^\mu\epsilon A_\mu +\bar\epsilon\epsilon \phi \ , \nn \\
    \Theta^{\mu\nu} &=& D^{[\mu}\xi^{\nu]} +\xi^\lambda \omega_\lambda^{\mu\nu}\ , \nn \\
    \rho &=& -\frac{i}{5}\nabla_\mu(\bar\epsilon\gamma^\mu\epsilon) \ , \nn \\
    R^{IJ} &=&- \frac{2i}{5}\bar\epsilon\gamma^\mu \sigma^{IJ}\nabla_\mu\epsilon \ , \nn \\
    \hat{R}^{IJ} &=& \frac{2i}{5}\bar{\hat\epsilon}\gamma^\mu \hat\sigma^{IJ}\nabla_\mu\hat\epsilon\ .
\ee
Hence, the square of the supercharge gives rise to the bosonic transformations in the superconformal algebra and the new $SU(2)'$ transformation
generated by $\hat{R}^{IJ}$.

The localization of the hypermultiplets is straightforward.
Since the superconformal index does not depend on the continuous deformation of the theory,
it is possible to deform the original action \eqref{hyper-actionB} with an continuous parameter $t$ in front and take $t\rightarrow +\infty$ limit without
altering the index.
As a result, the path integral for the hypermultiplets localizes to the saddle points at which the scalars in the hypermultiplets become trivial: $q^A=0$ and $F^{A'}=0$.

\section{$Sp(N)$ instanton quantum mechanics}\label{instanton-QM}
The instanton moduli space of 5d $Sp(N)$ gauge theories with $N_f$ fundamental flavors can be described by the 1d quantum mechanics of the $k$ D0-branes
with $N$ D4- and $N_f$ D8-branes near the orientifold plane.
The 1d quantum mechanics is then $O(k)$ gauge theory preserving $4$ real supersymmetries and its Higgs branch is equivalent to the instanton moduli space of the 5d gauge theory.
The quantum mechanics has the global symmetry
\be
    SO(4)_E\times SO(4)_R \times Sp(N) \times SO(2N_f)\ .
\ee
The $SO(4)_E$ is identified with the spatial rotation along the D4-branes and the $SO(4)_R$ is identified with the rotation transverse to the D4-branes but on D8-branes.
The $Sp(N)$ is the flavor symmetry coming from the D0-D4 connecting string modes and the $SO(2N_f)$ is the flavor symmetry from D0-D8 connecting string modes.
The Lagrangian and the field content are known in \cite{Douglas:1996uz, Aharony:1997pm}.
The field content is given by
$$
\begin{array}{c|cccc|ccc|c|c|cc}
    \hline &X_{1,2,3,4} & X_{5,6,7,8} & \lambda & \bar\lambda & J & \psi & \bar\psi&\xi & A_t,X_9 & \chi & \bar\chi \\
    \hline SO(4)_E & ({\bf 2,2}) & (\bf{1,1}) & ({\bf 1,2}) & ({\bf 2,1})  & ({\bf 2,1}) & ({\bf 1,1}) & ({\bf 1,1})& ({\bf 1,1})& ({\bf 1,1})& ({\bf 2,1})& ({\bf 1,2})\\
    SO(4)_R& (\bf{1,1}) & ({\bf 2,2}) & ({\bf 2,1}) & ({\bf 1,2}) & ({\bf 1,1})  & ({\bf 2,1}) & ({\bf 1,2})& ({\bf 1,1})& ({\bf 1,1})& ({\bf 2,1})& ({\bf 1,2})\\
    \hline Sp(N) & {\bf 1} & {\bf 1}& {\bf 1}&{\bf 1} & {\bf 2N} &{\bf 2N}&{\bf 2N}& {\bf 1} & {\bf 1}& {\bf 1}&{\bf 1}  \\
    SO(2N_f) & {\bf 1}& {\bf 1}& {\bf 1}& {\bf 1}& {\bf 1}& {\bf 1}& {\bf 1} & {\bf 2N_f}& {\bf 1}& {\bf 1}& {\bf 1} \\
    \hline O(k) & {\bf \frac{k(k+1)}{2}} & {\bf \frac{k(k+1)}{2}} & {\bf \frac{k(k+1)}{2}} &{\bf \frac{k(k+1)}{2}} & {\bf \bar{k}} & {\bf \bar{k}}  & {\bf \bar{k}}  & {\bf k} &{\rm adj}&{\rm adj}&{\rm adj}  \\ \hline
\end{array}
$$

It is convenient to divide the field content into three groups.
The first group consists of the bosonic fields $A_t,X_9,X_{1,2,3,4},J$ and the fermionic fields $\lambda,\psi, \chi$.
The $k\times k$ symmetric bosonic fields $X_{1,2,3,4}$ parametrize the positions of D0-branes along D4-branes
and the $2N\times k$ bosonic field $J$ represents the $Sp(N)$ gauge orientation modes.
The fermionic fields $\lambda$ and $\psi$ are their superpartners.
If we restrict ourselves to Higgs branch in which $X_{1,2,3,4}$ and $J$ take nonzero expectation values, the moduli space made of the first group is identical to the instanton moduli space in the 5d pure $Sp(N)$ gauge theory.
The adjoint fields $A_t$ and $X_9$ are lifted in the Higgs branch.
In fact, the matrices $X_{1,2,3,4}$ and $J$ coincide with the ADHM fields for the ADHM construction of the instanton moduli space.
The Higgs branch constraints on those matrices are equivalent to the ADHM constraints which are given by \cite{Nekrasov:2004vw,Atiyah:1978ri}
\be
    \mu^{12}\!\!&\!\!=\!\!&\!\![B_1,B_1^*] + [B_2,B_2^*] + J^\dagger_1J^1-J_2^\dagger J^2 = 0 ,\nn \\
    \mu^{11}\!\!&\!\!=\!\!&\!\![B_1,B_2] + J^\dagger_2 J^1=0,
\ee
where $B_1\equiv X_1+iX_2$ and $B_2\equiv X_3+iX_4$.
Here the field $J$ is subject to the reality condition $(J^A_i)^* = \varepsilon_{AB}\Omega^{ij}J^B_j$ where $i,j$ are fundamental indices and $\Omega^{ij}$
is the antisymmetric invariant tensor of $Sp(N)$ group.
The ADHM constraint $\mu^{AB}(\sigma^I)_{AB}$ is a triplet under $SU(2)_{1R}$ symmetry of $SU(2)_{1R}\times SU(2)_{2R}\in SO(4)_R$.

The second group consists of the bosonic $X_{5,6,7,8}$ and the fermionic $\bar\lambda,\bar\psi,\bar\chi$.
The matrices $\tilde{B}_1\equiv X_5+iX_6$ and $\tilde{B}_2 \equiv X_7+iX_8$ describe the positions of D0-branes perpendicular to D4-branes and they are lifted in the Higgs branch.
The degrees of freedom from this group in the Higgs branch describes the moduli space of the antisymmetric hypermultiplet on the instanton background in the 5d theory.
The $SU(2)$ global symmetry of the hypermultiplet is identified with the $SU(2)_{2R}$.
The last group is formed by the fermionic field $\xi$ which represents the fermionic zero modes of $N_f$ fundamental hypermultiplets of the 5d theory.


\subsection{Equivariant Chern character}\label{App:ECc}
The instanton index of 5d $Sp(N)$ gauge theories can be obtained from the index of the 1d quantum mechanics with the above matter content as $U(N)$ instanton case done in \cite{Kim:2011mv}.
Equivalently we can also use the localization technique used in \cite{Moore:1998et,Nekrasov:2002qd} by regarding the fields in the quantum mechanics as
the ADHM data of the instanton moduli space.
We then first construct a cohomological formulation of the above field content with a twisted supercharge $Q$ by identifying $SU(2)_{1E}\subset SO(4)_E$ and $SU(2)_{1R}$,
and evaluate the index through the localization procedure.
This allows us to easily read off the instanton part of the equivariant index from the weights of the torus actions on the ADHM data.
The conversion from the equivariant index to the instanton index is also very easy and we present it below.

Let us first compute the equivariant index (or the equivariant Chern character) for the $Sp(N)$ gauge multiplet, which gets contributions from the fields in the first group.
In the cohomological formulation, the BRST-like charge $Q$ acts on the fields as
\be\label{BRST-cohomology}
    Q\phi \!\!&\!\!=\!\!&\!\! 0\ , \quad Q\bar\phi = \eta \equiv \epsilon_{\dot\alpha\dot\beta}\chi^{\dot\alpha\dot\beta} \ , \quad Q\eta = [\phi,\bar\phi] \ ,\nn \\
    QJ^{\dot{\alpha}} \!\!&\!\!=\!\!&\!\! \psi^{\dot\alpha} \ , \quad Q\psi^{\dot\alpha} =  - J^{\dot\alpha} \phi+ a J^{\dot\alpha}+2i\gamma_1j_1J^{\dot\alpha} \ ,\nn \\
    QB^{\dot\alpha \beta} \!\!&\!\!=\!\!&\!\! \lambda^{\dot\alpha\beta} \ , \quad Q \lambda^{\dot\alpha\beta} = [\phi,B^{\dot\alpha\beta}] +2i(\gamma_1j_1+\gamma_2j_2)B^{\dot\alpha\beta} \ ,\nn \\
    Q\chi^{\dot\alpha\dot\beta}  \!\!&\!\!=\!\!&\!\! \mu^{\dot\alpha\dot\beta} \ , \quad Q\mu^{\dot\alpha\dot\beta} = [\phi,\chi^{\dot\alpha\dot\beta}] + 2i\gamma_1j_1\chi^{\dot\alpha\dot\beta} \ ,
\ee
with the equivariant parameters $\gamma_1,\gamma_2$ and
\be
    a = {\rm diag} (\alpha_1,\alpha_2,\cdots,\alpha_{N} )\otimes \sigma^3.
\ee
The indices $\dot\alpha,\dot\beta$ are for the diagonal subgroup of $\ SU(2)_{1E}\times SU(2)_{1R}$ and the indices $\alpha,\beta$ are for the $SU(2)_{2E}$.
The $k\times k$ scalar $\phi$ denotes the combination $A_t+ X_9$.
The localization lifts the off-diagonal components of the $\phi$ and leaves only the diagonal components $\phi_\pm$ defined in \eqref{gauge-parameter1} and \eqref{gauge-parameter2}
which play the role of the equivariant gauge parameter of $O(k)$ dual gauge group.
We note that the dual gauge group is divided into two components $O(k)_+$ and $O(k)_-$.
It is sufficient for the equivariant index to know the torus action $T_{\gamma_1}\times T_{\gamma_2}\times T_a\times T_\phi$ only for the $J,B$ and $\chi$
since the other fields are $Q$-exact from \eqref{BRST-cohomology}.
\be
    &&J^{\dot1} \rightarrow e^{ia}J^{\dot1}e^{-i\phi_\pm}e^{-\gamma_1} \ , \nn \\
    &&B^{\dot11} \rightarrow e^{i\phi_\pm}B^{\dot11}e^{-i\phi_\pm}e^{- \gamma_1-\gamma_2} \ , \quad B^{\dot12} \rightarrow e^{i\phi_\pm}B^{\dot12}e^{-i\phi_\pm}e^{- \gamma_1+\gamma_2} \ ,\nn \\
    &&\chi^{\dot1\dot1} \rightarrow e^{i\phi_\pm}\chi^{\dot1\dot1}e^{-i\phi_\pm}e^{-2\gamma_1}.
\ee
We need to compute the equivariant index for two distinct actions by $e^{i\phi_+}$ and $e^{i\phi_-}$ independently.
For $O(k)_+$, the equivariant index of each field is \cite{Nekrasov:2004vw,Shadchin:2005mx}
\be\label{equiv-index-vector}
    J^{\dot\alpha} &\rightarrow& e^{-\gamma_1}\Big[\sum_{I=1}^n\sum_{i=1}^N\left(e^{i\alpha_i+i\phi_I}+e^{-i\alpha_i+i\phi_I}+e^{i\alpha_i-i\phi_I}+e^{-i\alpha_i-i\phi_I}\right)
    +\chi \sum_{i=1}^N\left(e^{i\alpha_i}+e^{-i\alpha_i}\right)\!\!\Big], \\
    B^{\dot\alpha\beta}&\rightarrow&(e^{-\gamma_1-\gamma_2}+e^{-\gamma_1+\gamma_2})\Big[\sum_{I<J}^n\left(e^{i\phi_I+i\phi_J}+e^{-i\phi_I+i\phi_J}+e^{i\phi_I-i\phi_J}+e^{-i\phi_I-i\phi_J}\right)\nn \\
    &&\hspace{4cm}+\sum_{I=1}^n\left(e^{2i\phi_I}+e^{-2i\phi_I}\right) + \chi \sum_{I=1}^n\left(e^{i\phi_I}+e^{-i\phi_I}\right)+n+\chi\Big], \nn \\
    \chi^{\dot\alpha\dot\beta} &\rightarrow& -e^{-2\gamma_1}\Big[\sum_{I<J}^n\left(e^{i\phi_I+i\phi_J}+e^{-i\phi_I+i\phi_J}+e^{i\phi_I-i\phi_J}+e^{-i\phi_I-i\phi_J}\right)
    +\chi \sum_{I=1}^n\left(e^{i\phi_I}+e^{-i\phi_I}\right)+n\Big] .\nn
\ee
The overall minus sign of the last index comes from the consideration of the fermionic statistic of the $\chi^{\dot\alpha\dot\beta}$.
For $O(k)_-$, the last entries of $e^{i\phi_-}$ actions for odd and even $k$ are different, which are $-1$ for odd $k$ and $\sigma^3$ for even $k$.
The element $-1$ in the $O(k)_-$ action here has to be regarded as $e^{i\pi}$.
We need to carefully handle them in the $e^{i\phi_-}$ action.
Then the equivariant index with the $O(k)_-$ action for odd $k$ is given by
\be\label{equiv-index-vector2}
    J^{\dot\alpha} &\rightarrow& e^{-\gamma_1}\Big[\sum_{I=1}^n\sum_{i=1}^N\left(e^{i\alpha_i+i\phi_I}+e^{-i\alpha_i+i\phi_I}+e^{i\alpha_i-i\phi_I}+e^{-i\alpha_i-i\phi_I}\right)
    +e^{i\pi} \sum_{i=1}^N\left(e^{i\alpha_i}+e^{-i\alpha_i}\right)\!\!\Big] \\
    B^{\dot\alpha\beta}&\rightarrow&(e^{-\gamma_1-\gamma_2}+e^{-\gamma_1+\gamma_2})\Big[\sum_{I<J}^n\left(e^{i\phi_I+i\phi_J}+e^{-i\phi_I+i\phi_J}+e^{i\phi_I-i\phi_J}+e^{-i\phi_I-i\phi_J}\right)\nn \\
    &&\hspace{4cm}+\sum_{I=1}^n\left(e^{2i\phi_I}+e^{-2i\phi_I}\right) + e^{i\pi} \sum_{I=1}^n\left(e^{i\phi_I}+e^{-i\phi_I}\right)+n+1\Big] \nn \\
    \chi^{\dot\alpha\dot\beta} &\rightarrow& -e^{-2\gamma_1}\Big[\sum_{I<J}^n\left(e^{i\phi_I+i\phi_J}+e^{-i\phi_I+i\phi_J}+e^{i\phi_I-i\phi_J}+e^{-i\phi_I-i\phi_J}\right)
    +e^{i\pi} \sum_{I=1}^n\left(e^{i\phi_I}+e^{-i\phi_I}\right)+n\!\Big] \nn
\ee
and the equivariant index for even $k$ is given by
\be\label{equiv-index-vector3}
    \hspace{-.6cm}J^{\dot\alpha} \!&\!\rightarrow\!&\! e^{-\gamma_1}\Big[\sum_{I=1}^{n-1}\sum_{i=1}^N\left(e^{i\alpha_i+i\phi_I}+e^{-i\alpha_i+i\phi_I}+e^{i\alpha_i-i\phi_I}+e^{-i\alpha_i-i\phi_I}\right)
    +(1+e^{i\pi}) \sum_{i=1}^N\left(e^{i\alpha_i}+e^{-i\alpha_i}\right)\!\!\Big], \\
   \hspace{-.6cm} B^{\dot\alpha\beta}\!&\!\rightarrow\!&\!(e^{-\gamma_1-\gamma_2}+e^{-\gamma_1+\gamma_2})\Big[\sum_{I<J}^{n-1}\left(e^{i\phi_I+i\phi_J}+e^{-i\phi_I+i\phi_J}+e^{i\phi_I-i\phi_J}+e^{-i\phi_I-i\phi_J}\right)\nn \\
   \hspace{-.6cm} &&\hspace{3.5cm}+\sum_{I=1}^{n-1}\left(e^{2i\phi_I}+e^{-2i\phi_I}\right) + (1+e^{i\pi}) \sum_{I=1}^{n-1}\left(e^{i\phi_I}+e^{-i\phi_I}\right)+(n+1+e^{i\pi})\Big], \nn \\
   \hspace{-.6cm} \chi^{\dot\alpha\dot\beta} \!&\!\rightarrow\!&\! -e^{-2\gamma_1}\!\Big[\sum_{I<J}^{n-1}\!\left(e^{i\phi_I+i\phi_J}\!+\!e^{-i\phi_I+i\phi_J}\!+\!e^{i\phi_I-i\phi_J}\!+\!e^{-i\phi_I-i\phi_J}\right)
    \!+\!(1\!+\!e^{i\pi}) \sum_{I=1}^{n-1}\left(e^{i\phi_I}\!+\!e^{-i\phi_I}\right)\!+\!n\!-\!1\!+\!e^{i\pi}\!\Big].\nn
\ee

We use a conversion rule from the equivariant index to the Euler class
\be\label{conversion-rule}
    \sum_i\epsilon_ie^{iw_i} \rightarrow \prod_i\left(\sin \frac{w_i}{2}\right)^{-\epsilon_i} ,
\ee
which follows from the conversion rule explained in Section 3.3 after considering the momentum modes along the time circle.
This is analogous to the Plethystic exponential of a single letter index if we regard the above equivariant index as single letter index.
The instanton index is the $O(k)$ gauge invariant projection of the Euler class.
Therefore, inserting the proper Haar measure of $O(k)$ gauge group, we derive from \eqref{equiv-index-vector},\eqref{equiv-index-vector2},\eqref{equiv-index-vector3} the vector multiplet contribution
to the $Sp(N)$ instanton index in \eqref{integral-formula1},\eqref{integral-formula2},\eqref{integral-formula3}, respectively.

We now turn to the equivariant index for the antisymmetric hypermultiplet on the instanton background, which gets contributions from the fields in the second group:
$\tilde{B}^{\dot\alpha a},\bar\lambda^{\dot\alpha a},\bar\psi^{a}$, and $\bar\chi^{\alpha a}$ where the superscript $a=\pm$ denotes the $SU(2)_{2R}$ doublet index of the global symmetry of the hypermultiplet.
The equivariant transformations of these fields under the torus action are given by
\be\label{torus-action-anti}
    \bar\psi^+ &\rightarrow& e^{ia}\bar\psi^+e^{-i\phi_\pm}e^{im},\nn\\
    \tilde{B}^{\dot1+} &\rightarrow& e^{i\phi_\pm}\tilde{B}^{\dot1+}e^{-i\phi_\pm}e^{im-\gamma_1}  , \quad \tilde{B}^{\dot2+} \rightarrow e^{-i\phi_\pm}\tilde{B}^{\dot2+}e^{-i\phi_\pm}e^{im+\gamma_1}, \nn \\
    \bar\chi^{1+} &\rightarrow& e^{i\phi_\pm}\bar\chi^{1+}e^{-i\phi_\pm}e^{im-\gamma_2} , \quad \bar\chi^{2+} \rightarrow e^{i\phi_\pm}\bar\chi^{2+}e^{-i\phi_\pm}e^{im+\gamma_2}.
\ee
Here we do not consider the contribution from $\bar\lambda$ as it is $Q$-exact.
One can then easily compute the equivariant index from \eqref{torus-action-anti}.
For $O(k)_+$ action, the equivariant index of an antisymmetric hypermultiplet is given by
\be
    \bar\psi^a \!&\!\rightarrow\!&\! -e^{im}\Big[\sum_{I=1}^n\sum_{i=1}^N\left(e^{i\phi_I+i\alpha_i}+e^{-i\phi_I+i\alpha_i}+e^{i\phi_I-i\alpha_i}+e^{-i\phi_I-i\alpha_i}\right) +\chi\sum_{i=1}^N\left(e^{i\alpha_i}+e^{-i\alpha_i}\right)\!\!\Big], \\
    \tilde{B}^{\dot\alpha a} \!&\!\rightarrow\!&\! e^{im}(e^{\gamma_1}+e^{-\gamma_1})\Big[\sum_{I<J}^n\left(e^{i\phi_I+i\phi_J}+e^{-i\phi_I+i\phi_J}+e^{i\phi_I-i\phi_J}+e^{-i\phi_I-i\phi_J}\right) \nn \\
    && \hspace{4cm}+ \sum_{I=1}^n\left(e^{2i\phi_I}+e^{-2i\phi_I}\right)+\chi\sum_{I=1}^n\left(e^{i\phi_I}+e^{-i\phi_I}\right)+n+\chi\Big] ,\nn \\
    \bar\chi^{\alpha a} \!&\!\rightarrow\!&\! -e^{im}(e^{\gamma_2}\!+\!e^{-\gamma_2})\!\Big[\!\sum_{I<J}^n\!\left(e^{i\phi_I+i\phi_J}\!+\!e^{-i\phi_I+i\phi_J}\!+\!e^{i\phi_I-i\phi_J}\!+\!e^{-i\phi_I-i\phi_J}\right)
    \!+\!\chi \sum_{I=1}^n\left(e^{i\phi_I}\!+\!e^{-i\phi_I}\right)\!+\!n\!\Big]. \nn
\ee
Similarly, for $O(k)_-$ with odd $k$, the equivariant index is given by
\be
    \bar\psi^a \!&\!\rightarrow\!&\! -e^{im}\Big[\sum_{I=1}^n\sum_{i=1}^N\left(e^{i\phi_I+i\alpha_i}+e^{-i\phi_I+i\alpha_i}+e^{i\phi_I-i\alpha_i}+e^{-i\phi_I-i\alpha_i}\right) +e^{i\pi}\sum_{i=1}^N\left(e^{i\alpha_i}+e^{-i\alpha_i}\right)\!\!\Big], \\
    \tilde{B}^{\dot\alpha a} \!&\!\rightarrow\!&\! e^{im}(e^{\gamma_1}+e^{-\gamma_1})\Big[\sum_{I<J}^n\left(e^{i\phi_I+i\phi_J}+e^{-i\phi_I+i\phi_J}+e^{i\phi_I-i\phi_J}+e^{-i\phi_I-i\phi_J}\right) \nn \\
    && \hspace{4cm} +\sum_{I=1}^n\left(e^{2i\phi_I}+e^{-2i\phi_I}\right)+e^{i\pi}\sum_{I=1}^n\left(e^{i\phi_I}+e^{-i\phi_I}\right)+n+1\Big], \nn \\
    \bar\chi^{\alpha a} \!&\!\rightarrow\!&\! -e^{im}(e^{\gamma_2}\!+\!e^{-\gamma_2})\!\Big[\!\sum_{I<J}^n\!\left(e^{i\phi_I+i\phi_J}\!+\!e^{-i\phi_I+i\phi_J}\!+\!e^{i\phi_I-i\phi_J}\!+\!e^{-i\phi_I-i\phi_J}\right)
    \!+\!e^{i\pi} \sum_{I=1}^n\left(e^{i\phi_I}\!+\!e^{-i\phi_I}\right)\!+\!n\!\Big], \nn
\ee
and, for $O(k)_-$ with even $k$, it is given by 
\be
    \hspace{-.8cm}\bar\psi^a \!&\!\rightarrow\!&\! -e^{im}\Big[\sum_{I=1}^{n-1}\sum_{i=1}^N\left(e^{i\phi_I+i\alpha_i}+e^{-i\phi_I+i\alpha_i}+e^{i\phi_I-i\alpha_i}+e^{-i\phi_I-i\alpha_i}\right) +(1+e^{i\pi})\sum_{i=1}^N\left(e^{i\alpha_i}+e^{-i\alpha_i}\right)\!\!\Big], \\
    \hspace{-.8cm}\tilde{B}^{\dot\alpha a} \!&\!\rightarrow\!&\! e^{im}(e^{\gamma_1}+e^{-\gamma_1})\Big[\sum_{I<J}^{n-1}\left(e^{i\phi_I+i\phi_J}+e^{-i\phi_I+i\phi_J}+e^{i\phi_I-i\phi_J}+e^{-i\phi_I-i\phi_J}\right) \nn \\
    \hspace{-.8cm}&& \hspace{4cm} +\sum_{I=1}^{n-1}\left(e^{2i\phi_I}+e^{-2i\phi_I}\right)+(1+e^{i\pi})\sum_{I=1}^{n-1}\left(e^{i\phi_I}+e^{-i\phi_I}\right)+(n+1+e^{i\pi})\Big], \nn \\
    \hspace{-.8cm}\bar\chi^{\alpha a} \!&\!\rightarrow\!&\!\!\! -e^{im}\!(e^{\gamma_2}\!+\!e^{-\gamma_2})\!\Big[\!\sum_{I<J}^{n-1}\!\left(e^{i\phi_I+i\phi_J}\!+\!e^{-i\phi_I+i\phi_J}\!+\!e^{i\phi_I-i\phi_J}\!+\!e^{-i\phi_I-i\phi_J}\right)
    \!+\!(1\!+\!e^{i\pi})\!\! \sum_{I=1}^{n-1}\!\!\left(e^{i\phi_I}\!+\!e^{-i\phi_I}\right)\!+\!n\!-\!1\!+\!e^{i\pi}\!\Big]. \nn
\ee
Using the conversion rule \eqref{conversion-rule} from the equivariant index,
we can derive the antisymmetric matter part of the $Sp(N)$ instanton index \eqref{formula-anti1}, \eqref{formula-anti2}, and \eqref{formula-anti3}.

Finally, we compute the equivariant index for the fermion zero modes $\xi$ corresponding to 0-8 string modes.
The field $\xi$ rotates under the torus action as
\begin{equation}
    \xi_l \rightarrow e^{i\phi_\pm} \xi_l e^{im_l}
\end{equation}
Then equivariant index for the $\xi$ is
\be
    -\sum_{l=1}^{N_f}e^{im_l}\Big[\sum_{I=1}^n(e^{i\phi_I}+e^{-i\phi_I})+\chi\Big]&& \quad {\rm for} \ O(k)_+ \, \nn \\
    -\sum_{l=1}^{N_f}e^{im_l}\Big[\sum_{I=1}^n(e^{i\phi_I}+e^{-i\phi_I})+e^{i\pi}\Big]&& \quad {\rm for} \ O(k)_- \ {\rm with\ odd\ } k  , \nn \\
    -\sum_{l=1}^{N_f}e^{im_l}\Big[\sum_{I=1}^{n-1}(e^{i\phi_I}+e^{-i\phi_I})+1+e^{i\pi}\Big]&& \quad {\rm for}  \ O(k)_- \ {\rm with\ even\ } k,
\ee
which also yield the fundamental matter part of the instanton index in \eqref{integral-formula1},\eqref{integral-formula2}, and \eqref{integral-formula3}.


\section{Haar measure of $O_{\pm}(N)$}
In the main text, we have used the Haar measure to obtain gauge invariant quantities in the path integral. Here we list the Haar measure $[d\alpha]$ for the classical groups:\\
For $U(N)$,
\begin{align}
[d\alpha]
&=\frac{1}{N!} \left[\prod^N_{k=1} \frac{d\alpha_k}{2\pi}\right] \prod_{ j <k }^N \left[2\sin \Big(\frac{\alpha_i-\alpha_j}{2}\Big)\right]^2.
\end{align}
For $O_+(2N)$ ($=SO(2N)$),
\begin{align}
[d\alpha] 
&=\frac{1}{2^{N-1}N!} \left[\prod^N_{k=1} \frac{d\alpha_k}{2\pi}\right] \prod_{ i<j }^N \left[2\sin \Big(\frac{\alpha_i-\alpha_j}{2}\Big)\right]^2\left[2\sin \Big(\frac{\alpha_i+\alpha_j}{2}\Big)\right]^2.
\end{align}
For $O_+(2N+1)$ $(= SO(2N+1))$,
\begin{align}
[d\alpha]
&=\frac{2^N}{N!} \left[\prod^N_{k=1} \frac{d\alpha_k}{2\pi}\sin^2\frac{\alpha_k}{2}\right] \prod_{ i <j }^N \left[2\sin \Big(\frac{\alpha_i-\alpha_j}{2}\Big)\right]^2\left[2\sin \Big(\frac{\alpha_i+\alpha_j}{2}\Big)\right]^2.
\end{align}
For $O_-(2N+2)$ and $Sp(N)$,
\begin{align}
[d\alpha]
&= \frac{2^{N}}{N!}  \left[\prod^N_{k=1} \frac{d\alpha_k}{2\pi}\sin^2\alpha_k\right] \prod_{ i <j }^N \left[2\sin \Big(\frac{\alpha_i-\alpha_j}{2}\Big)\right]^2\left[2\sin \Big(\frac{\alpha_i+\alpha_j}{2}\Big)\right]^2.
\end{align}
For $O_-(2N+1)$,
\begin{align}
[d\alpha] 
&= \frac{2^N}{N!}\left[\prod^N_{k=1} \frac{d\alpha_k}{2\pi}\cos^2\frac{\alpha_k}{2}\right] \prod_{ i <j }^N \left[2\sin \Big(\frac{\alpha_i-\alpha_j}{2}\Big)\right]^2\left[2\sin \Big(\frac{\alpha_i+\alpha_j}{2}\Big)\right]^2.
\end{align}
The normalization constants are chosen such that
\begin{align}
\int_0^{2\pi}\cdots\int_0^{2\pi}[d\alpha] =1.\nn
\end{align}


\section{Characters and Branching Rules}\label{app:character}
\subsection{Characters}
$\bullet$ $SO(N)$\\
The superconformal index can be represented in terms of the character of the representations of $SO(2N_f)$ together with the $U(1)_I$ factor which we will denote by the powers of $q$. The Weyl character formula for $SO(2N_f)$ is given by \cite{Choi:2008za}
\begin{align}
\chi(h, \mu) = \frac{\det[\sinh(\mu_i(h_j + N_f-j))] + \det[\cosh(\mu_i(h_j + N_f-j))]}{\det[\cosh(\mu_i(N_f-j))]},
\end{align}
where $h$ denotes the highest weight with $(h_1, h_2, \cdots, h_{N_f-1}, h_{N_f})$ subject to the condition that $h_1\ge h_2\ge \cdots \ge h_{N_f-1}\ge |h_{N_f}|\ge 0$, $\mu_i$ denotes the chemical potentials, and $i, j = 1,\ldots, N_f$.

\noindent$\bullet$ $SU(N)$\\
The $SU(2)$ character is given by
\begin{align}
\chi^{SU(2)}[m] = \frac{e^{i(2m+1)r}-e^{-i(2m+1)r}}{e^{ir}-e^{-ir}}.
\end{align}
For example, we used in Section \ref{sec:Indexless5} that $\chi^{SU(2)}_{\bf 3}= e^{i \frac{m}{2}}+e^{-i \frac{m}{2}}$ and $\chi^{SU(2)}_{\bf 3}= e^{i m}+1+e^{-i m}$ with a chemical potential $m$.\\
  The  $SU(3)$ part is given by
\begin{align}
\chi^{SU(3)}[m,n] = \frac{\left| \begin{array}{ccc}
y_1^{m+n+2} &y_2^{-m-n-2}& (y_2/y_1)^{m+n+2}\\
y_1^{n+1} &y_2^{-n-1}& (y_2/y_1)^{n+1}\\
1 &1&1
\end{array}
 \right|}{\left| \begin{array}{ccc}
y_1^{2} &y_2^{-2}& (y_2/y_1)^{2}\\
y_1^{1} &y_2^{-1}& (y_2/y_1)^{1}\\
1 &1&1
\end{array}
 \right|}.
\end{align}
In this way, one can easily obtain character formulas for $SU(N)$.


\subsection{Branching rules}\label{branching}
Here we list branching rule associated with non-semi-simple embedding
 $E_{N_f+1} \supset SO(2N_f)\times U(1)_I$ that is discussed in Section \ref{enhancement}.

\noindent$\bullet$ $E_3=SU(3)\times SU(2) \supset SO(4) \times U(1)_I$\\
With the embedding
\begin{align}
E_3 = SU(3)\times SU(2) \supset SO(4)\times U(1)_I\cong SU(2)\times SU(2)\times U(1)_I,
\end{align}
to yield
\begin{align}
SU(3)&\supset   SU(2)\times U(1)_I\cr
{\bf 8}&=({\bf 1},{\bf 1})_0+({\bf 3},{\bf 1})_0+({\bf 2},{\bf 1})_1+({\bf 2},{\bf 1})_{-1}.
\end{align}
The adjoint representation of $E_3 = SU(3)\times SU(2)$ is expressed as $({\bf 8,1})+({\bf 1,3})$, and its products are given as follows:
\begin{align}
adj&=({\bf 8},{\bf 1})+({\bf 1},{\bf 3}),\cr
adj^2&=({\bf 27},{\bf 1})+({\bf 1},{\bf 5}),\cr
adj^3&=({\bf 64},{\bf 1})+({\bf 1},{\bf 7}),\cr
adj^4&=({\bf 125},{\bf 1})+({\bf 1},{\bf 9}),\cr
(adj \times adj)_S &= ({\bf 27},{\bf 1})+({\bf 1},{\bf 5})+({\bf 8},{\bf 3})+({\bf 8},{\bf 1})+2({\bf 1},{\bf 1}),\cr
(adj\times adj)_A & = ({\bf 8},{\bf 1})+({\bf 1},{\bf 3})+({\bf 8},{\bf 3})+({\bf 10},{\bf 1})+(\overline{\bf 10},{\bf 1}).
\end{align}

\noindent$\bullet$ $E_4=SU(5)$
 \begin{align}
SU(5)&\supset SO(6)\times U(1)_I\\
{\bf 24}&={\bf 1}_{0}+{\bf 4}_{1}+ \overline{\bf 4}_{-1} + {\bf 15}_{0}\cr
{\bf 75}&={\bf 15}_{0}+{\bf 20}_{-1}+ \overline{\bf 20}_{1} + {\bf 20}'_{0}\cr
{\bf 126}&={\bf \overline4}_{-1}+{\bf 6}_{-2} +{\bf 15}_{0}+{\bf 20}_{-1}+{\bf 36}_{1}+{\bf 45}_{0}\cr
{\bf \overline{126}}&={\bf 4}_{1}+{\bf 6}_{2} +{\bf 15}_{0}+{\bf \overline{20}}_{1}+{\bf\overline{36}}_{-1}+{\bf \overline{45}}_{0}\cr
{\bf 200}&= {\bf 1}_{0}+{\bf 4}_{1}+ \overline{\bf 4}_{-1} +{\bf 10}_{2}+ \overline{\bf 10}_{-2}+{\bf 15}_{0}+{\bf 36}_{1}+ \overline {\bf 36}_{-1}+{\bf 84}_{0}\cr
{\bf 224}&= {\bf 4}_{-3}+{\bf 6}_{-2}+{\bf 10}_{-2}+{\bf 20}''_{-1}+{\bf 20}_{-1}+{\bf 35}_{0}+{\bf 45}_{0}+{\bf 84}'_{1}\cr
\overline{\bf 224}&= \overline{\bf 4}_{3}+\overline{\bf 6}_{2}+\overline{\bf 10}_{2}+\overline{\bf 20}''_{1}+\overline{\bf 20}_{1}+\overline{\bf 35}_{0}+\overline{\bf 45}_{0}+\overline{\bf 84}'_{-1}\cr
{\bf 1000}&={\bf 1}_{0}+{\bf 4}_{1}+ \overline{\bf 4}_{-1}+{\bf 15}_{0} +{\bf 10}_{2}+ \overline{\bf 10}_{-2}+{\bf 36}_{1}+ \overline {\bf 36}_{-1}+{\bf 20}_{3}+{\bf \overline{20}}_{-3}\cr
&\quad+{\bf 70}_2+{\bf \overline{70}}_{-2}+{\bf 84}_{0}+{\bf {\bf 160}_{1}+\overline{160}}_{-1}+{\bf 300}_0\cr
{\bf 1024}&={\bf 15}_0+{\bf 20}_{-1}+{\bf 20}_0+\overline{\bf 20}_1+{\bf 36}_1+\overline{\bf 36}_{-1}+{\bf 45}_0 +\overline{\bf 45}_0+{\bf 60}_1+\overline{\bf 60}_{-1}\cr
&\quad +{\bf 64}_{-2}+{\bf 64}_2+{\bf 84}_0+{\bf 140}_{-1}+\overline{\bf 140}_{1}+{\bf 175}_0
\cr
{\bf 1050}&=
\overline{\bf 4}_{-1}+{\bf 6}_{-2}+\overline{\bf10}_{-2}+{\bf 15}_0+{\bf 20}_{-1}+\overline{\bf 20}_{-3} + {\bf 36}_{1}+\overline{\bf 36}_{-1} + {\bf 45}_{0}\cr
&
\quad +{\bf 64}_{-2}+{\bf 70}_{2}+{\bf 84}_{0} +{\bf 84}'_{1}+{\bf 140}_{-1}+{\bf 160}_{1} +{\bf 256}_{0}
\cr
\overline{\bf 1050}&=
{\bf 4}_{1}+ {\bf 6}_{2}+{\bf 10}_{2}+ {\bf 15}_{0}+\overline{\bf 20}_{1}+{\bf 20}_{3} + \overline{\bf 36}_{-1}+ {\bf 36}_{1} \cr
&\quad
+ \overline{\bf 45}_{0}+ {\bf 64}_{2}+\overline{\bf 70}_{-2}+{\bf 84}_{0} +\overline{\bf 84}'_{-1}+\overline{\bf 140}_{1}+\overline{\bf 160}_{-1} +\overline{\bf 256}_{0}
\cr
{\bf 3765}&={\bf 1}_{0}+{\bf 4}_{1}+ \overline{\bf 4}_{-1}+{\bf 15}_{0} +{\bf 10}_{2}+ \overline{\bf 10}_{-2}
+{\bf 20}_{3}+{\bf \overline{20}}_{-3}
+ {\bf 35}_{4}+ \overline{\bf 35}_{-4}
\cr
&\quad
+{\bf 36}_{1}+ \overline {\bf 36}_{-1}
+{\bf 70}_2+{\bf \overline{70}}_{-2}
+{\bf 84}_{0}+ {\bf 120}_{3}+ \overline{\bf 120}_{-3}
+{\bf {\bf 160}_{1}+\overline{160}}_{-1}\cr
&\quad
+{\bf 300}_0
+ {\bf 270}_{2}+ \overline{\bf 270}_{-2}
+ {\bf 500}_{1}+ \overline{\bf 500}_{-1}
+ {\bf 825}_{0},
\nn
\end{align}
where the $\bf 4$ of $SO(6)$ is associated with $\chi^{SO(6)}_{[\frac12,\frac12,\frac12]}$, and the $\bf \overline 4$ of $SO(6)$ is with $\chi^{SO(6)}_{[\frac12,\frac12,-\frac12]}$. Besides,
\begin{align*}
{\bf 20}_{SO(6)}=\chi^{SO(6)}_{[\frac32,\frac12,\frac12]}, &&
{\bf 20}'_{SO(6)}=\chi^{SO(6)}_{[2,0,0]}, &&
{\bf 20}''_{SO(6)}=\chi^{SO(6)}_{[\frac32,\frac32,\frac32]},&&\\
{\bf 84}_{SO(6)}=\chi^{SO(6)}_{[2,2,0]}, &&
{\bf 84}'_{SO(6)}=\chi^{SO(6)}_{[\frac52,\frac32,\frac32]}, &&
{\bf 84}''_{SO(6)}=\chi^{SO(6)}_{[3,3,3]},&&
\end{align*}
For $E_4$ case, $adj$ is 24-dimensional, and some relevant tensor products are
\begin{align}
({\bf 24}\times {\bf 24})_S&= {\bf 1}+ {\bf 24}+ {\bf 75} +{\bf 200}, \cr
({\bf 24}\times {\bf 24})_A&= {\bf 24}+{\bf 126}+\overline{{\bf 126}}, \cr
({\bf 24} \times {\bf 24}\times {\bf 24})_S&={\bf 1}+ 2\times {\bf 24} + {\bf 75}+{\bf 126}+\overline{{\bf 126}}+{\bf 200}+ {\bf 1024}+ {\bf 1000},\cr
({\bf 24} \times {\bf 24}\times {\bf 24})_A&={\bf 1}+ {\bf 24} + {\bf 75}+{\bf 126}+\overline{{\bf 126}}+{\bf 200}+{\bf 224}+\overline{{\bf 224}}+ {\bf 1024}.
\end{align}

\noindent$\bullet$ $E_5=Spin(10)$
 \begin{align}
SO(10)&\supset SO(8)\times U(1)_I\\
{\bf 45}&= {\bf 1}_{0}+{\bf 8_s}_{-1}+ {\bf 8_s}_{1}+{\bf 28}_{0} \cr
{\bf 54}&={\bf 1}_{-2}+{\bf 1}_0+{\bf 1}_2 +{\bf 8_s}_{-1}+{\bf 35_s}_{\text{0}}+{\bf 8_s}_{\text{1}}\cr
{\bf 210}&= {\bf 28}_0 +{\bf 35_c}_{\text{0}}+{\bf 35_v}_{0}+{\bf 56_c}_{-1}+{\bf 56_c}_{\text{1}}\cr
{\bf 770}&= {\bf1}_0+{\bf 8_s}_{1}+{\bf 8_s}_{-1} +{\bf 28}_{0} + {\bf 35_s}_{2}+{\bf 35_s}_{0}+{\bf 35_s}_{-2} + {\bf 160_s}_{1} + {\bf 160_s}_{-1}+{\bf 300_s}_{0}  \cr
{\bf 945}&= {\bf 8_s}_{-1}+ {\bf 8_s}_{1}+{\bf 28}_{2} + 2\times {\bf 28}_{0}+{\bf 28}_{-2} +{\bf 35_s}_{0}+{\bf 56_c}_{1}+{\bf 56_c}_{-1}+{\bf 160_s}_{1}+{\bf 160_s}_{-1} +{\bf 350}_{0} \nn\\
{\bf 1386} &={\bf 1}_{2} + {\bf 1}_{-2} +{\bf 1}_{0} +{\bf 8_s}_{3}+ 2\times {\bf 8_s}_{1}+ 2\times {\bf 8_s}_{-1} +{\bf 8_s}_{-3} + {\bf 28}_{2} + {\bf 28}_{0}+ {\bf 28}_{-2}\cr
&\quad
+ {\bf 35_s}_{2}  + 2\times {\bf 35_s}_{0}  + {\bf 35_s}_{-2}  + {\bf 112_s}_{1}  + {\bf 112}_{-1}  + {\bf 160_s}_{1}  + {\bf 160_s}_{-1}    + {\bf 567_s}_{0}
\cr
{\bf 4125} &={\bf 35_s}_{0} +{\bf 160_s}_{1} +{\bf 160_s}_{-1}+{\bf 300}_{2}+{\bf 300}_{0}+{\bf 300}_{-2}+{\bf 350}_{0} +{\bf 840'_c}_{1}+{\bf 840_c}_{0} +{\bf 840'_c}_{-1}
\cr
{\bf 5940} &={\bf 28}_{0}+ {\bf 35_c}_{0}+{\bf 35_v}_{0}+{\bf 56_c}_{1}+{\bf 56_c}_{-1}+{\bf 160_s}_{1}+{\bf 160_s}_{-1}+{\bf 224_s}_{1}+{\bf 224_s}_{-1}+{\bf 224'_c}_{1}\cr
&\quad
+{\bf 224'_c}_{-1}+{\bf 300}_{0}+{\bf 350}_{2}+2\times {\bf 350}_{0}+{\bf 350}_{-2}+{\bf 567_v}_{0}+{\bf 567_c}_{0}+{\bf 840'_c}_{1}+{\bf 840'_c}_{-1}
\cr
{\bf 7644}& = {\bf 1}_0+{\bf 8_s}_{-1}+{\bf 8_s}_1+{\bf 28}_0+{\bf 35_s}_{-2}+{\bf 35_s}_0+{\bf 35_s}_2+{\bf 112_s}_{-3}+{\bf 112_s}_{-1}+{\bf 112_s}_1+{\bf 112_s}_3\cr
&\quad+{\bf 160_s}_{-1}+{\bf 160_s}_1+{\bf 300}_0+{\bf 567_s}_{-2}+{\bf 567_s}_0+{\bf 567_s}_2+{\bf 1400_s}_{-1}+{\bf 1400_s}_1+{\bf 1925}_0\cr
{\bf 8085} &={\bf 28}_{2}+{\bf 28}_{0}+{\bf 28}_{-2}+{\bf 35_v}_{2}+{\bf 35_v}_{0}+{\bf 35_v}_{-2}+{\bf 35_c}_{2}+{\bf 35_c}_{0}+{\bf 35_c}_{-2}+{\bf 56_c}_{3}+2\times{\bf 56_c}_{1}\cr
&\quad+2\times{\bf 56_c}_{-1}+{\bf 56_c}_{-3}
+{\bf 160_s}_{1}+{\bf 160_s}_{-1}+{\bf 224_s}_{1}+{\bf 224_s}_{-1}+{\bf 224'_c}_{1}+{\bf 224'_c}_{-1}\cr
&\quad+{\bf 350}_{2}+2\times {\bf 350}_{0}+{\bf 350}_{-2}+{\bf 567_s}_{0}+{\bf 840_s}_{0}+{\bf 840_v}_{0}+ {\bf 1296_s}_{1}+ {\bf 1296_s}_{-1}
\cr
{\bf 17920} &={\bf 8_s}_{1}+{\bf 8_s}_{-1}+{\bf 28}_{2}+2\times{\bf 28}_{0}+{\bf 28}_{-2}+{\bf 35_s}_{2}+2\times{\bf 35_s}_{0}+{\bf 35_s}_{-2}+{\bf 56_c}_{1}+{\bf 56_c}_{-1}\cr
&\quad +{\bf 112_s}_{1}+{\bf 112_s}_{-1}+{\bf 160_s}_{3}+3\times{\bf 160_s}_{1}+3\times{\bf 160_s}_{-1}+{\bf 160_s}_{-3}+{\bf 300}_{2}+2\times {\bf 300}_{0}\cr
&\quad
+{\bf 300}_{-2}
+{\bf 350}_{2}+2\times {\bf 350}_{0}+{\bf 350}_{-2}+{\bf 567_s}_{2}+2\times{\bf 567_s}_{0}+{\bf 567_s}_{-2}\cr
&\quad
+{\bf 840_s}_{1}+{\bf 840_s}_{-1}+{\bf 1296_s}_{1}+{\bf 1296_s}_{-1}+{\bf 1400_s}_{1}+{\bf 1400_s}_{-1}
+{\bf 4096}_{0}
\cr
{\bf 52920}& = {\bf 1}_0+{\bf 8_s}_{1}+{\bf 8_s}_{-1}+{\bf 28}_{0}+{\bf 35_s}_{2}+{\bf 35_s}_{0}+{\bf 35_s}_{-2}
+{\bf 112_s}_{3}+{\bf 112_s}_{1}+{\bf 112_s}_{-1}+{\bf 112_s}_{-3}\cr
&\quad
+{\bf 160_s}_{1}+{\bf 160_s}_{-1}
+{\bf 294_s}_{4}+ {\bf 294_s}_{2}+{\bf 294_s}_{0}+{\bf 294_s}_{-2}+{\bf 294_s}_{-4}
+{\bf 300}_{0}+{\bf 567_s}_{2}\cr
&\quad
+{\bf 567_s}_{0}+{\bf 567_s}_{-2}+{\bf 1400_s}_{1}+{\bf 1400_s}_{-1}
+ {\bf 1568_s}_{3} +{\bf 1568_s}_{1}+{\bf 1568_s}_{-1}+{\bf 1568_s}_{-3}
\cr
&\quad
+{\bf 1925}_{0}
+{\bf 4312_s}_{2}+{\bf 4312_s}_{0}+{\bf 4312_s}_{-2}
+{\bf 7840_s}_{1}+{\bf 7840_s}_{-1}
+{\bf 8918}_{0}.\nn
\end{align}
Our convention for representations of $SO(8)$ with the same dimensions is as follows:
\begin{align}
&{\bf 8_v}=\chi^{SO(8)}_{[1,0,0,0]},\qquad
{\bf 8_s} =\chi^{SO(8)}_{[\frac12,\frac12,\frac12,\frac12]},\qquad
{\bf 8_c} =\chi^{SO(8)}_{[\frac12,\frac12,\frac12,-\frac12]},
\end{align}
where spinor and conjugate spinor representations differ by the the sign for the last weight, and in addition,
\begin{align}
&{\bf 224_s}=\chi^{SO(8)}_{[\frac52,\frac12,\frac12,\frac12]}, \qquad {\bf 224'_c}=\chi^{SO(8)}_{[\frac32,\frac32,\frac32,-\frac12]},\cr
&{\bf 840_v}=\chi^{SO(8)}_{[2,2,2,0]},\qquad
{\bf 840_c}=\chi^{SO(8)}_{[3,1,1,-1]},\qquad
{\bf 840'_c} =\chi^{SO(8)}_{[\frac52,\frac32,\frac12,-\frac12]}.
\end{align}
Some relevant tensor products are 
\begin{align}
({\bf 45}\times {\bf 45})_S&= {\bf 1}+ {\bf 54}+{\bf 210}+{\bf 770}, \cr
({\bf 45}\times {\bf 45})_A&= {\bf 45}+{\bf 945}, \cr
({\bf 45} \times {\bf 45}\times {\bf 45})_S&=2\times{\bf 45}+ {\bf 210} +{\bf 945}+{\bf 1386}+{\bf 5940}+{\bf 7644},\cr
({\bf 45} \times {\bf 45}\times {\bf 45})_A&={\bf 1}+{\bf 54}+ {\bf 210}+{\bf 770} +{\bf 945}+{\bf 4125}+{\bf 8085}.
\end{align}

\noindent$\bullet$ $E_6$
\begin{align}
E_6&\supset SO(10)\times U(1)_I\\
{\bf 78}&={\bf 1}_0 +{\bf 16}_{1}+ \overline{\bf 16}_{-1}+{\bf 45}_{0}\cr
{\bf 650}&={\bf 1}_0+{\bf 10}_{2} +{\bf 10}_{-2}+{\bf 16}_{1}+ \overline{\bf 16}_{-1}+{\bf 45}_{0}+{\bf 54}_{0}+{\bf 144}_{1}+\overline{\bf 144}_{-1}+ {\bf 210}_{0}\cr
{\bf 2430}&={\bf 1}_0+{\bf 16}_{1}+ \overline{\bf 16}_{-1}+{\bf 45}_{0}+{\bf 126}_{-2}+ \overline{\bf 126}_{2}+{\bf 210}_{0}+{\bf 560}_{1}+\overline{\bf 560}_{-1}+ {\bf 770}_{0}\cr
{\bf 2925}&={\bf 16}_{1}+ \overline{\bf 16}_{-1}+{\bf 45}_{0}+{\bf 45}_{0}+{\bf 120}_{2}+\overline{\bf 120}_{-2}+{\bf 144}_{1}+\overline{\bf 144}_{-1}+ {\bf 210}_{0}+ {\bf 560}_{1}+ \overline{\bf 560}_{-1}+ {\bf 945}_{0}\cr
{\bf 34749}&=
\mathbf{1}_0+\mathbf{10}_3+\mathbf{10}_{-3}+2\times\mathbf{16}_1+2\times \overline{\mathbf{16}}_{-1}+2
\times \mathbf{45}_0+\mathbf{54}_0+\mathbf{120}_2+\mathbf{120}_{-2}+\mathbf{126}_{-2}+\overline{\mathbf{126}}_2\cr
&+\overline{\mathbf{144}}_3+2\times \mathbf{144}_1+2\times \overline{\mathbf{144}}_{-1}+\mathbf{144}_{-3}+3\times \mathbf{210}_0+\mathbf{320}_2+\mathbf{320}_{-2}+2\times \mathbf{560}_1+2\times \overline{\mathbf{560}}_{-1}\cr
&+\mathbf{720}_1+\overline{\mathbf{720}}_{-1}+\mathbf{770}_{0}+2\times\mathbf{945}_0+\mathbf{1050}_0+\overline{\mathbf{1050}}_0+\mathbf{1200}_1+\overline{\mathbf{1200}}_{-1}+\mathbf{1386}_0\cr
&
+\mathbf{1440}_1+\overline{\mathbf{1440}}_{-1}+\mathbf{1728}_2+\mathbf{1728}_{-2}+\mathbf{3696}_1+\overline{\mathbf{3696}}_{-1}+\mathbf{5940}_0\cr
{\bf 43758}&={\bf 1}_0+{\bf 16}_{1}+ \overline{\bf 16}_{-1}+{\bf 45}_{0}+{\bf 126}_{-2}+ \overline{\bf 126}_{2}+{\bf 210}_{0}
+{\bf 560}_{1}+\overline{\bf 560}_{-1}+{\bf 672}_{-3}+ \overline{\bf 672}_{3}\cr
&+ {\bf 770}_{0}+{\bf 1440}_{1}+\overline{\bf 1440}_{-1}+{\bf 3696}'_{-2}+\overline{\bf 3696}'_{2}+{\bf 5940}_{0}+{\bf 7644}_{0}+{\bf 8064}_{1}+\overline{\bf 8064}_{-1},\cr
{\bf 70070}&=
\mathbf{45}_0+\mathbf{54}_0+\mathbf{120}_2+\mathbf{120}_{-2}+\mathbf{126}_2+\overline{\mathbf{126}}_{-2}+2\times\mathbf{144}_1+2\times\overline{\mathbf{144}}_{-1}+2\times \mathbf{210}_0\cr
&+\mathbf{320}_2+\mathbf{320}_{-2}+\overline{\mathbf{560}}_3+2\times \mathbf{560}_1+2\times\overline{\mathbf{560}}_{-1}+\mathbf{560}_{-3}+\mathbf{720}_1+\overline{\mathbf{720}}_{-1}+\mathbf{770}_0\cr
&+3\times\mathbf{945}_0+\mathbf{1050}_0+\overline{\mathbf{1050}}_0+2\times\mathbf{1200}_1+2\times\overline{\mathbf{1200}}_{-1}+\mathbf{1728}_2+\mathbf{1728}_{-2}\cr
&+\mathbf{2970}_2+\mathbf{2970}_{-2}+\mathbf{3696}_1+\overline{\mathbf{3696}}_{-1}+\mathbf{4125}_0+\mathbf{5940}_0+\mathbf{8085}_0+\mathbf{8800}_1+\overline{\mathbf{8800}}_{-1}
\cr
{\bf 105600}&=
\mathbf{16}_1+\overline{\mathbf{16}}_{-1}+2\times\mathbf{45}_0+\mathbf{120}_2+\mathbf{120}_{-2}+\mathbf{126}_{-2}+\overline{\mathbf{126}}_2+\mathbf{144}_1+\overline{\mathbf{144}}_{-1}+2\times\mathbf{210}_0\cr
&+3\times\mathbf{560}_1+3\times\overline{\mathbf{560}}_{-1}+2\times\mathbf{770}_0+2\times\mathbf{945}_0+\mathbf{1050}_0+\overline{\mathbf{1050}}_0+\overline{\mathbf{1200}}_3+\mathbf{1200}_1\cr
&+\overline{\mathbf{1200}}_{-1}+\mathbf{1200}_{-3}+\mathbf{1440}_1+\overline{\mathbf{1440}}_{-1}+\mathbf{1728}_{2}+\mathbf{1728}_{-2}+\mathbf{2970}_2+\mathbf{2970}_{-2}+\mathbf{3696}_{1}\cr
&+\overline{\mathbf{3696}}_{-1}+\mathbf{3696}'_{-2}+\overline{\mathbf{3696}}'_2+2\times\mathbf{5940}_0+\mathbf{8064}_1+\overline{\mathbf{8064}}_{-1}+\mathbf{8800}_{1}+\overline{\mathbf{8800}}_{-1}+\mathbf{17920}_0
\cr
{\bf 537966}&= {\bf 1}_{0}+{\bf 16}_{1}+\overline{\bf 16}_{-1}+{\bf 45}_{0}+{\bf 126}_{-2}+\overline{\bf 126}_{2}+{\bf 210}_{0}+{\bf 560}_{1}+\overline{\bf 560}_{-1}\cr
&+\overline{\bf 672}_{3}+{\bf 672}_{-3}+{\bf 770}_{0}+{\bf 1440}_{1}+\overline{\bf 1440}_{-1}+{\bf 2772}_{-4}+\overline{\bf 2772}_{4}+{\bf 3696}'_{-2}+\overline{\bf 3696}'_{2}\cr
&
+{\bf 5940}_{0}+{\bf 6930}'_{-2}+\overline{\bf 6930}'_{2}+{\bf 7644}_{0}+{\bf 8064}_{1}+\overline{\bf 8064}_{-1}+{\bf 8910}_{0}
\cr
&
+\overline{\bf 17280}_{3}+{\bf 17280}_{-3}+{\bf 34992}_{1}+\overline{\bf 34992}_{-1}+{\bf 46800}_{-2}+\overline{\bf 46800}_{2}+{\bf 52920}_{0}
\cr
&
+{\bf 70560}_{1}+\overline{\bf 70560}_{-1}+{\bf 73710}_{0},
\end{align}
where we used a Mathematica application LieART (ver 1.0.1) \cite{2012arXiv1206.6379F} to obtain the branching rules above.
For $E_6$ case, $adj$ is 78-dimensional and relevant tensor products of the adjoint representation of $E_6$ are as follows:x
\begin{align}
({\bf 78}\times {\bf 78})_S&= {\bf 1}+ {\bf 650}+{\bf 2430}, \cr
({\bf 78}\times {\bf 78})_A&= {\bf 78}+{\bf 2925}, \cr
({\bf 78} \times {\bf 78}\times {\bf 78})_S&={\bf 78}+ {\bf 650} +{\bf 2925}+{\bf 34749}+{\bf 43758},\cr
({\bf 78} \times {\bf 78}\times {\bf 78})_A&={\bf 1}+ {\bf 650}+{\bf 2430} +{\bf 2925}+{\bf 70070}.
\end{align}

\noindent$\bullet$ $E_7$
\begin{align}
E_7&\supset SO(12)\times U(1)_I\cr
{\bf 133}&={\bf 66}_0 + {\bf 32}_1+ {\bf 32}_{-1} + {\bf 1}_2+ {\bf 1}_0+{\bf 1}_{-2},\cr
{\bf 912}&={\bf 12}_{-1}+ {\bf 12}_1+ {\bf 32}_{-2}+ {\bf 32}_{0}+ {\bf 32}_{2}+ {\bf 220}_{-1}+ {\bf 220}_{1}++ {\bf 332}_{0}\cr
{\bf 1463}&={\bf 66}_0 + {\bf 77}_{2}+{\bf 77}_{0}+{\bf 77}_{-2}+ {\bf 352}'_{1} + {\bf 352}'_{-1} + {\bf 462}_0\cr
{\bf 1539}&={\bf 1}_0+ {\bf 32}'_{1}+ {\bf 32}'_{-1} + {\bf 66}_{2}+ {\bf 66}_{0}+ {\bf 66}_{-2}+{\bf 77}_{0}+ {\bf 352}'_{1}+ {\bf 352}'_{-1}+{\bf 495}_{0}\cr
{\bf 7371}&={\bf 1}_4+{\bf 1}_2+{\bf 1}_0+{\bf 1}_0 +{\bf 1}_{-2}+{\bf 1}_{-4}+
  {\bf 32}_{3}+ {\bf 32}_{1}+ {\bf 32}_{1}+ {\bf 32}_{-1} + {\bf 32}_{-1} + {\bf 32}_{-3} \cr
&\quad+ {\bf 66}_{2}+ {\bf 66}_{0}+ {\bf 66}_{-2}+{\bf 462}_{2}+{\bf 462}_{0}+{\bf 462}_{-2}+ {\bf 495}_{0}+ {\bf 1638}_{0}+{\bf 1728}_{1}+{\bf 1728}_{-1} \nn
\end{align}

$adj\sim {\bf 133}$, $adj^2\sim {\bf 7371}$, and $adj^3\sim {\bf 238602}$ and 
some relevant tensor products of the adjoint representation of $E_7$ are as follows:
\begin{align}
({\bf 133}\times{\bf 133})_S &= {\bf 1}+ {\bf 1539}+ {\bf 7371},\cr
({\bf 133}\times{\bf 133})_A &= {\bf 133}+ {\bf 8645},\cr
({\bf 133}\times{\bf 133}\times{\bf 133})_S &={\bf 133}+{\bf 1463}+{\bf 8645}+{\bf 152152}+{\bf 238602},\cr
({\bf 133}\times{\bf 133}\times{\bf 133})_A &={\bf 1}+{\bf 1539}+{\bf 7371}+{\bf 8645}+{\bf 365750}.
\end{align}

\noindent$\bullet$ $E_8$ \\
\begin{align}
E_8 &\supset SO(14)\times U(1)_I\cr
{\bf 248}&={\bf 91}_0+{\bf 14}_{2}+{\bf 14}_{-2} + {\bf 1}_0 +{\bf  64}_1+\overline{\bf 64}_{-1}.
\end{align}
 $adj\sim {\bf 248}$, $adj^2\sim {\bf 27000}$, and $adj^3\sim {\bf 1763125}$, and relevant tensor products of the adjoint representation of $E_8$ are as follows:
\begin{align}
({\bf 248}\times{\bf 248})_S &= {\bf 1}+ {\bf 3875}+ {\bf 27000},\cr
({\bf 248}\times{\bf 248})_A &= {\bf 248}+ {\bf 30380},\cr
({\bf 248}\times{\bf 248}\times{\bf 248})_S &={\bf 248}+{\bf 30380}+{\bf 779247}+{\bf 1763125},\cr
({\bf 248}\times{\bf 248}\times{\bf 248})_A &={\bf 1}+{\bf 3875}+{\bf 30380}+{\bf 27000}+{\bf 2450240}.
\end{align}


\begin{thebibliography}{10}

\bibitem{Seiberg:1996bd}
N.~Seiberg, ``{Five-dimensional SUSY field theories, nontrivial fixed points
  and string dynamics},''
  \href{http://dx.doi.org/10.1016/S0370-2693(96)01215-4}{{\em Phys.Lett.} {\bf
  B388} (1996)  753--760},
\href{http://arxiv.org/abs/hep-th/9608111}{{\tt arXiv:hep-th/9608111
  [hep-th]}}.

\bibitem{Morrison:1996xf}
D.~R. Morrison and N.~Seiberg, ``{Extremal transitions and five-dimensional
  supersymmetric field theories},''
  \href{http://dx.doi.org/10.1016/S0550-3213(96)00592-5}{{\em Nucl.Phys.} {\bf
  B483} (1997)  229--247},
\href{http://arxiv.org/abs/hep-th/9609070}{{\tt arXiv:hep-th/9609070
  [hep-th]}}.

\bibitem{Douglas:1996xp}
M.~R. Douglas, S.~H. Katz, and C.~Vafa, ``{Small instantons, Del Pezzo surfaces
  and type I-prime theory},''
  \href{http://dx.doi.org/10.1016/S0550-3213(97)00281-2}{{\em Nucl.Phys.} {\bf
  B497} (1997)  155--172},
\href{http://arxiv.org/abs/hep-th/9609071}{{\tt arXiv:hep-th/9609071
  [hep-th]}}.

\bibitem{Ganor:1996pc}
O.~J. Ganor, D.~R. Morrison, and N.~Seiberg, ``Branes, calabi-yau spaces, and
  toroidal compactification of the n=1 six-dimensional e(8) theory,''
  \href{http://dx.doi.org/10.1016/S0550-3213(96)00690-6}{{\em Nucl.Phys.} {\bf
  B487} (1997)  93--127},
\href{http://arxiv.org/abs/hep-th/9610251}{{\tt hep-th/9610251 [hep-th]}}.

\bibitem{Intriligator:1997pq}
K.~A. Intriligator, D.~R. Morrison, and N.~Seiberg, ``{Five-dimensional
  supersymmetric gauge theories and degenerations of Calabi-Yau spaces},''
  \href{http://dx.doi.org/10.1016/S0550-3213(97)00279-4}{{\em Nucl.Phys.} {\bf
  B497} (1997)  56--100},
\href{http://arxiv.org/abs/hep-th/9702198}{{\tt arXiv:hep-th/9702198
  [hep-th]}}.

\bibitem{Danielsson:1997kt}
U.~H. Danielsson, G.~Ferretti, J.~Kalkkinen, and P.~Stjernberg, ``{Notes on
  supersymmetric gauge theories in five-dimensions and six-dimensions},''
  \href{http://dx.doi.org/10.1016/S0370-2693(97)00645-X}{{\em Phys.Lett.} {\bf
  B405} (1997)  265--270},
\href{http://arxiv.org/abs/hep-th/9703098}{{\tt arXiv:hep-th/9703098
  [hep-th]}}.

\bibitem{Kugo:2000af}
T.~Kugo and K.~Ohashi, ``{Off-shell D = 5 supergravity coupled to matter
  Yang-Mills system},'' \href{http://dx.doi.org/10.1143/PTP.105.323}{{\em
  Prog.Theor.Phys.} {\bf 105} (2001)  323--353},
\href{http://arxiv.org/abs/hep-ph/0010288}{{\tt arXiv:hep-ph/0010288
  [hep-ph]}}.

\bibitem{Bergshoeff:2001hc}
E.~Bergshoeff, S.~Cucu, M.~Derix, T.~de~Wit, R.~Halbersma, {\em et al.},
  ``{Weyl multiplets of N=2 conformal supergravity in five-dimensions},'' {\em
  JHEP} {\bf 0106} (2001)  051,
\href{http://arxiv.org/abs/hep-th/0104113}{{\tt arXiv:hep-th/0104113
  [hep-th]}}.

\bibitem{Bergshoeff:2002qk}
E.~Bergshoeff, S.~Cucu, T.~De~Wit, J.~Gheerardyn, R.~Halbersma, {\em et al.},
  ``{Superconformal N=2, D = 5 matter with and without actions},'' {\em JHEP}
  {\bf 0210} (2002)  045,
\href{http://arxiv.org/abs/hep-th/0205230}{{\tt arXiv:hep-th/0205230
  [hep-th]}}.

\bibitem{Witten:1996qb}
E.~Witten, ``{Phase transitions in M theory and F theory},''
  \href{http://dx.doi.org/10.1016/0550-3213(96)00212-X}{{\em Nucl.Phys.} {\bf
  B471} (1996)  195--216},
\href{http://arxiv.org/abs/hep-th/9603150}{{\tt arXiv:hep-th/9603150
  [hep-th]}}.

\bibitem{Aharony:1997an}
O.~Aharony, M.~Berkooz, and N.~Seiberg, ``{Light cone description of (2,0)
  superconformal theories in six-dimensions},'' {\em Adv.Theor.Math.Phys.} {\bf
  2} (1998)  119--153,
\href{http://arxiv.org/abs/hep-th/9712117}{{\tt arXiv:hep-th/9712117
  [hep-th]}}.

\bibitem{Douglas:2010iu}
M.~R. Douglas, ``{On D=5 super Yang-Mills theory and (2,0) theory},''
  \href{http://dx.doi.org/10.1007/JHEP02(2011)011}{{\em JHEP} {\bf 1102} (2011)
   011},
\href{http://arxiv.org/abs/1012.2880}{{\tt arXiv:1012.2880 [hep-th]}}.

\bibitem{Lambert:2010iw}
N.~Lambert, C.~Papageorgakis, and M.~Schmidt-Sommerfeld, ``{M5-Branes,
  D4-Branes and Quantum 5D super-Yang-Mills},''
  \href{http://dx.doi.org/10.1007/JHEP01(2011)083}{{\em JHEP} {\bf 1101} (2011)
   083},
\href{http://arxiv.org/abs/1012.2882}{{\tt arXiv:1012.2882 [hep-th]}}.

\bibitem{Lambert:2011gb}
N.~Lambert and P.~Richmond, ``{(2,0) Supersymmetry and the Light-Cone
  Description of M5-branes},''
  \href{http://dx.doi.org/10.1007/JHEP02(2012)013}{{\em JHEP} {\bf 1202} (2012)
   013},
\href{http://arxiv.org/abs/1109.6454}{{\tt arXiv:1109.6454 [hep-th]}}.

\bibitem{Kim:2011mv}
H.-C. Kim, S.~Kim, E.~Koh, K.~Lee, and S.~Lee, ``{On instantons as Kaluza-Klein
  modes of M5-branes},'' \href{http://dx.doi.org/10.1007/JHEP12(2011)031}{{\em
  JHEP} {\bf 1112} (2011)  031},
\href{http://arxiv.org/abs/1110.2175}{{\tt arXiv:1110.2175 [hep-th]}}.

\bibitem{Nahm:1977tg}
W.~Nahm, ``{Supersymmetries and their Representations},''
\href{http://dx.doi.org/10.1016/0550-3213(78)90218-3}{{\em Nucl.Phys.} {\bf
  B135} (1978)  149}.

\bibitem{Romans:1985tw}
L.~Romans, ``{The F(4) Gauged Supergravity in Six-Dimensions},''
\href{http://dx.doi.org/10.1016/0550-3213(86)90517-1}{{\em Nucl.Phys.} {\bf
  B269} (1986)  691}.

\bibitem{Minwalla:1997ka}
S.~Minwalla, ``{Restrictions imposed by superconformal invariance on quantum
  field theories},'' {\em Adv.Theor.Math.Phys.} {\bf 2} (1998)  781--846,
\href{http://arxiv.org/abs/hep-th/9712074}{{\tt arXiv:hep-th/9712074
  [hep-th]}}.

\bibitem{Bhattacharya:2008zy}
J.~Bhattacharya, S.~Bhattacharyya, S.~Minwalla, and S.~Raju, ``{Indices for
  Superconformal Field Theories in 3,5 and 6 Dimensions},''
  \href{http://dx.doi.org/10.1088/1126-6708/2008/02/064}{{\em JHEP} {\bf 0802}
  (2008)  064},
\href{http://arxiv.org/abs/0801.1435}{{\tt arXiv:0801.1435 [hep-th]}}.

\bibitem{Ganor:1996mu}
O.~J. Ganor and A.~Hanany, ``{Small E(8) instantons and tensionless noncritical
  strings},'' \href{http://dx.doi.org/10.1016/0550-3213(96)00243-X}{{\em
  Nucl.Phys.} {\bf B474} (1996)  122--140},
\href{http://arxiv.org/abs/hep-th/9602120}{{\tt arXiv:hep-th/9602120
  [hep-th]}}.

\bibitem{Horava:1995qa}
P.~Horava and E.~Witten, ``{Heterotic and type I string dynamics from
  eleven-dimensions},''
  \href{http://dx.doi.org/10.1016/0550-3213(95)00621-4}{{\em Nucl.Phys.} {\bf
  B460} (1996)  506--524},
\href{http://arxiv.org/abs/hep-th/9510209}{{\tt arXiv:hep-th/9510209
  [hep-th]}}.

\bibitem{Horava:1996ma}
P.~Horava and E.~Witten, ``{Eleven-dimensional supergravity on a manifold with
  boundary},'' \href{http://dx.doi.org/10.1016/0550-3213(96)00308-2}{{\em
  Nucl.Phys.} {\bf B475} (1996)  94--114},
\href{http://arxiv.org/abs/hep-th/9603142}{{\tt arXiv:hep-th/9603142
  [hep-th]}}.

\bibitem{Polchinski:1995df}
J.~Polchinski and E.~Witten, ``{Evidence for heterotic - type I string
  duality},'' \href{http://dx.doi.org/10.1016/0550-3213(95)00614-1}{{\em
  Nucl.Phys.} {\bf B460} (1996)  525--540},
\href{http://arxiv.org/abs/hep-th/9510169}{{\tt arXiv:hep-th/9510169
  [hep-th]}}.

\bibitem{Polchinski:1996fm}
J.~Polchinski, S.~Chaudhuri, and C.~V. Johnson, ``{Notes on D-branes},''
\href{http://arxiv.org/abs/hep-th/9602052}{{\tt arXiv:hep-th/9602052
  [hep-th]}}.

\bibitem{Pestun:2007rz}
V.~Pestun, ``{Localization of gauge theory on a four-sphere and supersymmetric
  Wilson loops},'' {\em Commun.Math.Phys.} {\bf 313} (2012)  71--129,
\href{http://arxiv.org/abs/0712.2824}{{\tt arXiv:0712.2824 [hep-th]}}.

\bibitem{Gomis:2011pf}
J.~Gomis, T.~Okuda, and V.~Pestun, ``{Exact Results for 't Hooft Loops in Gauge
  Theories on $S^4$},'' {\em JHEP} {\bf 1205} (2012)  141,
\href{http://arxiv.org/abs/1105.2568}{{\tt arXiv:1105.2568 [hep-th]}}.

\bibitem{Atiyah}
M.~F. Atiyah, {\em Elliptic operators and compact groups}, vol.~401.
\newblock Springer-Verlag, Lecture Notes in Mathematics, 1974.

\bibitem{Aharony:2003sx}
O.~Aharony, J.~Marsano, S.~Minwalla, K.~Papadodimas, and M.~Van~Raamsdonk,
  ``{The Hagedorn - deconfinement phase transition in weakly coupled large N
  gauge theories},'' {\em Adv.Theor.Math.Phys.} {\bf 8} (2004)  603--696,
\href{http://arxiv.org/abs/hep-th/0310285}{{\tt arXiv:hep-th/0310285
  [hep-th]}}.

\bibitem{Kim:2009wb}
S.~Kim, ``{The Complete superconformal index for N=6 Chern-Simons theory},''
  \href{http://dx.doi.org/10.1016/j.nuclphysb.2009.06.025}{{\em Nucl.Phys.}
  {\bf B821} (2009)  241--284},
\href{http://arxiv.org/abs/0903.4172}{{\tt arXiv:0903.4172 [hep-th]}}.

\bibitem{Nekrasov:2002qd}
N.~A. Nekrasov, ``{Seiberg-Witten prepotential from instanton counting},'' {\em
  Adv.Theor.Math.Phys.} {\bf 7} (2004)  831--864,
\href{http://arxiv.org/abs/hep-th/0206161}{{\tt arXiv:hep-th/0206161
  [hep-th]}}.

\bibitem{Nekrasov:2003rj}
N.~Nekrasov and A.~Okounkov, ``{Seiberg-Witten theory and random partitions},''
\href{http://arxiv.org/abs/hep-th/0306238}{{\tt arXiv:hep-th/0306238
  [hep-th]}}.

\bibitem{Nakajima:2003pg}
H.~Nakajima and K.~Yoshioka, ``{Instanton counting on blowup. 1.4-dimensional
  pure gauge theory},''
\href{http://arxiv.org/abs/math/0306198}{{\tt arXiv:math/0306198 [math-ag]}}.

\bibitem{Nekrasov:2004vw}
N.~Nekrasov and S.~Shadchin, ``{ABCD of instantons},''
  \href{http://dx.doi.org/10.1007/s00220-004-1189-1}{{\em Commun.Math.Phys.}
  {\bf 252} (2004)  359--391},
\href{http://arxiv.org/abs/hep-th/0404225}{{\tt arXiv:hep-th/0404225
  [hep-th]}}.

\bibitem{Shadchin:2005mx}
S.~Shadchin, ``{On certain aspects of string theory/gauge theory
  correspondence},''
\href{http://arxiv.org/abs/hep-th/0502180}{{\tt arXiv:hep-th/0502180
  [hep-th]}}.

\bibitem{Kim:2008kn}
S.~Kim, K.-M. Lee, and S.~Lee, ``{Dyonic Instantons in 5-dim Yang-Mills
  Chern-Simons Theories},''
  \href{http://dx.doi.org/10.1088/1126-6708/2008/08/064}{{\em JHEP} {\bf 0808}
  (2008)  064},
\href{http://arxiv.org/abs/0804.1207}{{\tt arXiv:0804.1207 [hep-th]}}.

\bibitem{Collie:2008vc}
B.~Collie and D.~Tong, ``{Instantons, Fermions and Chern-Simons Terms},''
  \href{http://dx.doi.org/10.1088/1126-6708/2008/07/015}{{\em JHEP} {\bf 0807}
  (2008)  015},
\href{http://arxiv.org/abs/0804.1772}{{\tt arXiv:0804.1772 [hep-th]}}.

\bibitem{Tachikawa:2004ur}
Y.~Tachikawa, ``{Five-dimensional Chern-Simons terms and Nekrasov's instanton
  counting},'' \href{http://dx.doi.org/10.1088/1126-6708/2004/02/050}{{\em
  JHEP} {\bf 0402} (2004)  050},
\href{http://arxiv.org/abs/hep-th/0401184}{{\tt arXiv:hep-th/0401184
  [hep-th]}}.

\bibitem{Bruzzo:2002xf}
U.~Bruzzo, F.~Fucito, J.~F. Morales, and A.~Tanzini, ``{Multiinstanton calculus
  and equivariant cohomology},'' {\em JHEP} {\bf 0305} (2003)  054,
\href{http://arxiv.org/abs/hep-th/0211108}{{\tt arXiv:hep-th/0211108
  [hep-th]}}.

\bibitem{Marino:2004cn}
M.~Marino and N.~Wyllard, ``{A Note on instanton counting for N=2 gauge
  theories with classical gauge groups},''
  \href{http://dx.doi.org/10.1088/1126-6708/2004/05/021}{{\em JHEP} {\bf 0405}
  (2004)  021},
\href{http://arxiv.org/abs/hep-th/0404125}{{\tt arXiv:hep-th/0404125
  [hep-th]}}.

\bibitem{Fucito:2004gi}
F.~Fucito, J.~F. Morales, and R.~Poghossian, ``{Instantons on quivers and
  orientifolds},'' \href{http://dx.doi.org/10.1088/1126-6708/2004/10/037}{{\em
  JHEP} {\bf 0410} (2004)  037},
\href{http://arxiv.org/abs/hep-th/0408090}{{\tt arXiv:hep-th/0408090
  [hep-th]}}.

\bibitem{Okuda:2010ke}
T.~Okuda and V.~Pestun, ``{On the instantons and the hypermultiplet mass of
  N=2* super Yang-Mills on $S^{4}$},''
  \href{http://dx.doi.org/10.1007/JHEP03(2012)017}{{\em JHEP} {\bf 1203} (2012)
   017},
\href{http://arxiv.org/abs/1004.1222}{{\tt arXiv:1004.1222 [hep-th]}}.

\bibitem{Douglas:1996uz}
M.~R. Douglas, ``{Gauge fields and D-branes},''
  \href{http://dx.doi.org/10.1016/S0393-0440(97)00024-7}{{\em J.Geom.Phys.}
  {\bf 28} (1998)  255--262},
\href{http://arxiv.org/abs/hep-th/9604198}{{\tt arXiv:hep-th/9604198
  [hep-th]}}.

\bibitem{Aharony:1997pm}
O.~Aharony, M.~Berkooz, S.~Kachru, and E.~Silverstein, ``{Matrix description of
  (1,0) theories in six-dimensions},'' {\em Phys.Lett.} {\bf B420} (1998)
  55--63,
\href{http://arxiv.org/abs/hep-th/9709118}{{\tt arXiv:hep-th/9709118
  [hep-th]}}.

\bibitem{Bergman:1997py}
O.~Bergman, M.~R. Gaberdiel, and G.~Lifschytz, ``{String creation and heterotic
  type I' duality},'' {\em Nucl.Phys.} {\bf B524} (1998)  524--544,
\href{http://arxiv.org/abs/hep-th/9711098}{{\tt arXiv:hep-th/9711098
  [hep-th]}}.

\bibitem{Kallen:2012cs}
J.~Kallen and M.~Zabzine, ``{Twisted supersymmetric 5D Yang-Mills theory and
  contact geometry},'' {\em JHEP} {\bf 1205} (2012)  125,
\href{http://arxiv.org/abs/1202.1956}{{\tt arXiv:1202.1956 [hep-th]}}.

\bibitem{Kallen:2012va}
J.~Kallen, J.~Qiu, and M.~Zabzine, ``{The perturbative partition function of
  supersymmetric 5D Yang-Mills theory with matter on the five-sphere},''
\href{http://arxiv.org/abs/1206.6008}{{\tt arXiv:1206.6008 [hep-th]}}.

\bibitem{kim:2012av}
H.-C. Kim and S.~Kim, ``{M5-branes from gauge theories on the 5-sphere},''
\href{http://arxiv.org/abs/1206.6339}{{\tt arXiv:1206.6339 [hep-th]}}.

\bibitem{Hama:2012bg}
N.~Hama and K.~Hosomichi, ``{Seiberg-Witten Theories on Ellipsoids},''
\href{http://arxiv.org/abs/1206.6359}{{\tt arXiv:1206.6359 [hep-th]}}.

\bibitem{Hama:2011ea}
N.~Hama, K.~Hosomichi, and S.~Lee, ``{SUSY Gauge Theories on Squashed
  Three-Spheres},'' {\em JHEP} {\bf 1105} (2011)  014,
\href{http://arxiv.org/abs/1102.4716}{{\tt arXiv:1102.4716 [hep-th]}}.

\bibitem{Imamura:2011wg}
Y.~Imamura and D.~Yokoyama, ``{N=2 supersymmetric theories on squashed
  three-sphere},'' {\em Phys.Rev.} {\bf D85} (2012)  025015,
\href{http://arxiv.org/abs/1109.4734}{{\tt arXiv:1109.4734 [hep-th]}}.

\bibitem{Dolan:2011rp}
F.~Dolan, V.~Spiridonov, and G.~Vartanov, ``{From 4d superconformal indices to
  3d partition functions},''
  \href{http://dx.doi.org/10.1016/j.physletb.2011.09.007}{{\em Phys.Lett.} {\bf
  B704} (2011)  234--241},
\href{http://arxiv.org/abs/1104.1787}{{\tt arXiv:1104.1787 [hep-th]}}.

\bibitem{Imamura:2011uw}
Y.~Imamura, ``{Relation between the 4d superconformal index and the $S^3$
  partition function},'' {\em JHEP} {\bf 1109} (2011)  133,
\href{http://arxiv.org/abs/1104.4482}{{\tt arXiv:1104.4482 [hep-th]}}.

\bibitem{Benini:2011nc}
F.~Benini, T.~Nishioka, and M.~Yamazaki, ``{4d Index to 3d Index and 2d
  TQFT},''
\href{http://arxiv.org/abs/1109.0283}{{\tt arXiv:1109.0283 [hep-th]}}.

\bibitem{Gadde:2011ia}
A.~Gadde and W.~Yan, ``{Reducing the 4d Index to the $S^3$ Partition
  Function},''
\href{http://arxiv.org/abs/1104.2592}{{\tt arXiv:1104.2592 [hep-th]}}.

\bibitem{DeWolfe:1999hj}
O.~DeWolfe, A.~Hanany, A.~Iqbal, and E.~Katz, ``{Five-branes, seven-branes and
  five-dimensional E(n) field theories},'' {\em JHEP} {\bf 9903} (1999)  006,
\href{http://arxiv.org/abs/hep-th/9902179}{{\tt arXiv:hep-th/9902179
  [hep-th]}}.

\bibitem{Ferrara:1998gv}
S.~Ferrara, A.~Kehagias, H.~Partouche, and A.~Zaffaroni, ``{AdS(6)
  interpretation of 5-D superconformal field theories},''
  \href{http://dx.doi.org/10.1016/S0370-2693(98)00560-7}{{\em Phys.Lett.} {\bf
  B431} (1998)  57--62},
\href{http://arxiv.org/abs/hep-th/9804006}{{\tt arXiv:hep-th/9804006
  [hep-th]}}.

\bibitem{Brandhuber:1999np}
A.~Brandhuber and Y.~Oz, ``{The D-4 - D-8 brane system and five-dimensional
  fixed points},'' \href{http://dx.doi.org/10.1016/S0370-2693(99)00763-7}{{\em
  Phys.Lett.} {\bf B460} (1999)  307--312},
\href{http://arxiv.org/abs/hep-th/9905148}{{\tt arXiv:hep-th/9905148
  [hep-th]}}.

\bibitem{Bergman:2012kr}
O.~Bergman and D.~Rodriguez-Gomez, ``{5d quivers and their AdS(6) duals},''
\href{http://arxiv.org/abs/1206.3503}{{\tt arXiv:1206.3503 [hep-th]}}.

\bibitem{Hosomichi:2012ek}
K.~Hosomichi, R.-K. Seong, and S.~Terashima, ``{Supersymmetric Gauge Theories
  on the Five-Sphere},''
\href{http://arxiv.org/abs/1203.0371}{{\tt arXiv:1203.0371 [hep-th]}}.

\bibitem{Atiyah:1978ri}
M.~Atiyah, N.~J. Hitchin, V.~Drinfeld, and Y.~Manin, ``{Construction of
  Instantons},''
{\em Phys.Lett.} {\bf A65} (1978)  185--187.

\bibitem{Moore:1998et}
G.~W. Moore, N.~Nekrasov, and S.~Shatashvili, ``{D particle bound states and
  generalized instantons},''
  \href{http://dx.doi.org/10.1007/s002200050016}{{\em Commun.Math.Phys.} {\bf
  209} (2000)  77--95},
\href{http://arxiv.org/abs/hep-th/9803265}{{\tt arXiv:hep-th/9803265
  [hep-th]}}.

\bibitem{Choi:2008za}
J.~Choi, S.~Lee, and J.~Song, ``{Superconformal Indices for Orbifold
  Chern-Simons Theories},''
  \href{http://dx.doi.org/10.1088/1126-6708/2009/03/099}{{\em JHEP} {\bf 0903}
  (2009)  099},
\href{http://arxiv.org/abs/0811.2855}{{\tt arXiv:0811.2855 [hep-th]}}.

\bibitem{2012arXiv1206.6379F}
R.~{Feger} and T.~W. {Kephart}, ``{LieART -- A Mathematica Application for Lie
  Algebras and Representation Theory},''
  \href{http://arxiv.org/abs/1206.6379}{{\tt arXiv:1206.6379 [math-ph]}};
LieART project home page: \href{http://lieart.hepforge.org}{\tt http://lieart.hepforge.org}.
  
\end{thebibliography}
\providecommand{\href}[2]{#2}\begingroup\raggedright\endgroup
\end{document}